\documentclass[acmsmall,screen]{acmart}

\AtBeginDocument{%
  }

\usepackage[T1]{fontenc}

\usepackage{tikz}
\usetikzlibrary{automata,arrows,positioning}
\usepackage{multirow}
\usepackage{subfigure}
\usepackage{makecell}
\usepackage{algorithmic}
\usepackage{xcolor,xspace}
\usepackage{mathtools}
\usepackage{colortbl}
\usepackage{enumitem}
\usepackage{hhline}
\usepackage{textcomp}
\usepackage{xcolor}
\usepackage{booktabs}
\usepackage{hyperref}
\usepackage{graphicx}
\usepackage{relsize}
\usepackage{makecell}
\usepackage{wrapfig}
\usepackage{bbding}
\usepackage{algorithmic}
\usepackage[algoruled,noend,linesnumbered]{algorithm2e}%

\definecolor{mygray}{gray}{.85}

\newcommand{\minor}[1]{\textcolor{black}{{#1}}}
\newcommand{\major}[1]{\textcolor{black}{{#1}}}

\newcommand{\tool}{{\sf BFAVerifier}\xspace}

\newcommand{\deepPoly}{{\sf DeepPoly}\xspace}
\newcommand{\symPoly}{{\sf SymPoly}\xspace}

\newcommand{\mI}{{\mathcal{I}}}

\newcommand{\mA}{{\mathcal{A}}}
\newcommand{\mN}{{\mathcal{N}}}
\newcommand{\bs}[1]{{\mathbf{#1}}}

\newcommand{\mm}{\mathfrak{m}}

\newcommand{\nn}{\mathfrak{n}}

\usepackage{booktabs}
\usepackage{multirow}
\newcommand{\tabincell}[2]{\begin{tabular}{@{}#1@{}}#2\end{tabular}}

\makeatletter
\newcommand{\Rmnum}[1]{\expandafter\@slowromancap\romannumeral #1@}
\makeatother

\begin{document}

\title{Verification of Bit-Flip Attacks against Quantized Neural Networks}


\author{Yedi Zhang}
\affiliation{%
  \institution{National University of Singapore}
  \city{Singapore}
  \country{Singapore}}
  \email{yd.zhang@nus.edu.sg}

 \author{Lei Huang}
 \affiliation{%
   \institution{ShanghaiTech University}
   \city{Shanghai}
   \country{China}}
   \email{huanglei@shanghaitech.edu.cn}

 \author{Pengfei Gao}
 \affiliation{%
  \institution{ByteDance Inc}
  \city{Beijing}
  \country{China}}
   \email{gaopengfei.se@bytedance.com}

 \author{Fu Song}
 \authornote{Corresponding author.}
 \affiliation{%
   \institution{Key Laboratory of System Software (Chinese Academy of Sciences) and State Key Laboratory of Computer Science, Institute of Software, Chinese Academy of Sciences}
   \city{Beijing}
   \country{China}}
\additionalaffiliation{
\institution{University of Chinese Academy of Sciences}
  \city{Beijing}
  \country{China}}
\additionalaffiliation{
\institution{Nanjing Institute of Software Technology}
  \city{Nanjing}
  \country{China}}
\email{songfu@ios.ac.cn}
   
\author{Jun Sun}
\affiliation{%
  \institution{Singapore Management University}
  \city{Singapore}
  \country{Singapore}}
\email{junsun@smu.edu.sg}

\author{Jin Song Dong}
\affiliation{%
  \institution{National University of Singapore}
  \city{Singapore}
  \country{Singapore}}
\email{dcsdjs@nus.edu.sg}



 \renewcommand{\shortauthors}{Zhang et al.}

\begin{abstract}
  In the rapidly evolving landscape of neural network security, the resilience of neural networks against bit-flip attacks (i.e., an attacker maliciously flips an extremely small amount of bits within its parameter storage memory system to induce harmful behavior), has emerged as a relevant area of research. Existing studies suggest that quantization may serve as a viable defense against such attacks. Recognizing the documented susceptibility of real-valued neural networks to such attacks and the comparative robustness of quantized neural networks (QNNs), in this work, we introduce \tool, the first verification framework designed to formally verify the absence of bit-flip attacks or to identify all vulnerable parameters in a sound and rigorous manner.
\tool comprises two integral components: an abstraction-based method and an MILP-based method.
Specifically, we first conduct a reachability analysis with respect to symbolic parameters that represent the potential bit-flip attacks, based on a novel abstract domain with a sound guarantee. If the reachability analysis fails to prove the resilience of such attacks, then we encode this verification problem into an equivalent MILP problem which can be solved by off-the-shelf solvers. 
Therefore, \tool is sound, complete, and reasonably efficient. We conduct extensive experiments, which demonstrate its effectiveness and efficiency across various network architectures, quantization bit-widths, and adversary capabilities. 
\end{abstract}




\maketitle

\section{Introduction}\label{sec:intro}
Neural networks have demonstrated their potential to achieve human-level performance in multiple domains~\cite{dong2021survey,MusaHLLY23}. 
However, they are fragile in many ways and can be easily manipulated through various attacks~\cite{khalid2021exploiting,LiuWLX17,yao2020deephammer,HONG_USENIX19,CCFDZSL20,SongLCFL21,DZBS21,ChenZZS23,ZhaoCLLSWS24,ChenZS24}.
Recently, bit-flip attacks (BFAs)~\cite{BFAICCV19,liyes,1bitallyouneed,SURVEY} have become a critical class of hardware-based adversarial threats that exploit the physical vulnerability of neural networks. These attacks involve maliciously flipping the bits in the memory cells that store the parameters of a neural network during the deployment stage or changing the real-time activation values during the inference stage. Such attacks have been demonstrated to be feasible in practice for altering the behavior of networks in multiple cases~\cite{walker2021dram,PRACTICALATTACK,tol2022toward}. For instance, RowHammer~\cite{DONTKNOCK,ROWHAMMER} is one of the most widely used BFA methods which exploits a vulnerability in DRAM by repeatedly accessing memory rows to induce unintended bit flips in adjacent rows, compromising data integrity and security for network parameters. 
Unlike traditional software-level adversarial attacks, which typically require modifications to input data, BFAs directly target the underlying hardware (e.g., memory), making them particularly effective and difficult to defend against. 

\begin{figure}
    \centering
    \includegraphics[width=.98\textwidth]{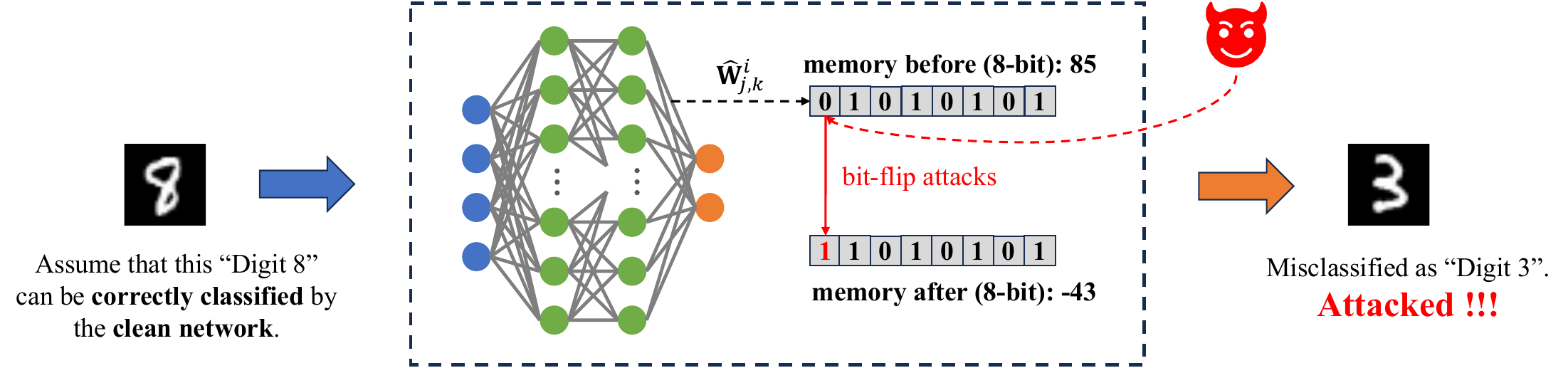}
    \caption{An illustration example of bit-flip attacks on an 8-bit quantized neural network. The attacker flips a single bit in the final/output layer, altering the value of parameter $\widehat{\bs{W}}^i_{j,k}$ from 85 to -43 (represented in two's complement) and misleading the network behavior.}
    \Description{An illustration of bit-flip attacks on an 8-bit quantized neural network.}
    \label{fig:illu}
\end{figure}

Modern DNNs, characterized by their large sizes and 32-bit floating-point parameters, face high computational and storage demands, hindering their deployment on resource-limited embedded devices. Quantization~\cite{DSQ,HanMD15,JacobKCZTHAK18}, reducing the precision of parameters and/or activation values, offers a promising solution to compress the network, and enable{s} the deployment of quantized neural networks (QNNs) on such devices. For example, the Tesla-FSD chip~\cite{FSDChip} employs an 8-bit integer format to store all network weights. On the other hand, QNNs have been demonstrated to exhibit greater resilience to BFAs compared to their real-valued counterparts. Specifically, DNNs are highly susceptible to BFAs, with successful attack rates reaching nearly 99\%~\cite{HONG_USENIX19}, particularly through the manipulation of the exponential bit of compromised parameters. In response, numerous defense strategies have been proposed \cite{advwa,HARDeNN,randomDNN}, leveraging parameter quantization to fortify network security against bit-flip attacks. Despite these measures, QNNs remain vulnerable to BFAs, as existing defense techniques fall short of providing formal security assurances against such attacks. This vulnerability underscores the critical need for developing a rigorous verification method to ascertain the absence of BFAs, ensuring the integrity and reliability of QNNs in security-sensitive applications. An illustration example of bit-flip attacks on an 8-bit QNN can be found in Figure~\ref{fig:illu}.

\smallskip
\noindent{\bf Main contributions.} 
In this work, we propose the first \textbf{B}it-\textbf{F}lip \textbf{A}ttacks \textbf{V}erification method (\tool) to efficiently and effectively verify if the bit-flip attacks are absent given a QNN, concerning a given input region, that is also sound and complete. It guarantees the safety of the QNN (such as robustness with respect to a specified input region) when facing potential bit-flip attacks. Given a QNN and an input region, \tool first conducts a novel reachability analysis to compute an overapproximation of the output range of the network under the potential attacks. Such an analysis generates two outcomes: i) \texttt{Proved}, meaning the absence of the potential BFAs, or ii) \texttt{Unknown}, meaning that it fails to prove the absence of successful attacks possibly due to a conservative approximation of the abstraction throughout the reachability analysis process. If the result is \texttt{Unknown}, we further encode this bit-flip attacks verification problem into an equivalent mixed-integer linear programming (MILP) problem, which can be solved by off-the-shelf solvers.

A key technical challenge is how to conduct the reachability analysis for QNNs, given the interested input region and the threat of potential bit-flip attacks (i.e., some network parameters become symbolic with unknown values). To tackle the challenge, we propose \symPoly, an advanced abstract domain that is built on \deepPoly and is equipped with new abstract transformers specifically designed for handling symbolic parameters. Initially, symbolic parameters are determined with specific parameter intervals for the QNN concerning the potential bit-flip attacks. Subsequent reachability analysis can then be conducted on the modified QNN, which is equipped with symbolic parameters, using \symPoly. To enhance the precision of our reachability analysis results, we also propose two optimization strategies, namely, \emph{sub-interval division} and \emph{binary search strategy}, to reduce the precision loss that arises from the abstract transformation concerning large value discrepancies with a single interval.

\smallskip
\noindent
{\bf Experimental results.}
We implement our method as an end-to-end tool that uses Gurobi~\cite{Gurobi} as the back-end MILP solver. We extensively evaluate it on a large set of verification tasks using multiple QNNs for the MNIST~\cite{MNIST} and ACAS Xu~\cite{julian2019deep} datasets, where the number of hidden neurons varies from 30 to 5120, the quantization bit-width of QNNs ranges from 4 to 8, and the number of bits for bit-flip attacks ranges from 1 to 4 bits. For the reachability analysis, we compare \tool with a naive method that iteratively generates a new QNN $\mN'$ for each possible bit-flip attack and verifies whether the $\mN'$ still preserves the robustness property within the given input region via \deepPoly. The experimental results show that our method is much more efficient than the naive method (up to 30x faster), successfully proving a similar number of verification tasks and even proving some tasks that return unknown by the naive method. Moreover, with the binary search strategy, we can prove even more tasks. The results also confirm the effectiveness of the MILP-based method, which can help verify many tasks that cannot be solved by \symPoly solely. The experimental results also show that \tool can verify the absence of BFAs for most of the benign neural networks in our benchmark.

Our contributions are summarized as follows.
\begin{itemize}
    \item We propose a novel abstract domain \symPoly to conduct reachability analysis for neural networks with symbolic parameters soundly and efficiently;
    \item We introduced the first sound, complete, and reasonably efficient bit-flip attacks verification method \tool for QNNs by combining \symPoly and an MILP-based method;
    \item We implement \tool as an end-to-end tool and conduct an extensive evaluation of various verification tasks, demonstrating its effectiveness and efficiency.
\end{itemize}

\smallskip
\noindent
{\bf Outline.} Section~\ref{sec:pre} presents the preliminary. Section~\ref{sec:pro} defines our problem and a naive method for solving the problem based on \deepPoly is given. We present our method in Section~\ref{sec:method}. Section~\ref{sec:exp} reports experimental results. Section~\ref{sec:relatedwork} discusses related work and finally, Section~\ref{sec:con} concludes this work. 
Missing proofs can be found in the appendix.

\section{Preliminary}\label{sec:pre}

We denote by $\mathbb{R}$ (resp. $\mathbb{N}$) the set of real (resp. integer) numbers.
Given a positive integer $n$,  we denote by $[n]$ the set of positive integers $\{1,2, \ldots, n\}$.
We use $x,x',\ldots$ to denote scalars, $\bs{x}, \bs{x}',\ldots$ to denote vectors, and $\bs{W}, \bs{W}',\ldots$ to denote matrices. We denote by $\bs{W}_{i,:}$ and $\bs{W}_{:,j}$ to denote the $i$-th row and $j$-th column of the matrix $\bs{W}$, and use $\bs{x}_i$ to denote the $i$-th entry of the vector $\bs{x}$.

\subsection{Neural Network and Quantization}\label{sec:quant}
In this section, we provide the minimal necessary background on neural networks and the quantization scheme considered in this work. Specifically, we focus on feedforward deep neural networks (DNNs) used for classification problems.

\smallskip
\noindent
{\bf Neural networks.}
%
A DNN consists of an input layer, multiple hidden layers, and an output layer. Each layer contains neurons connected via weighted edges to the neurons in the subsequent layer. Specifically, each neuron in a non-input layer is additionally linked with a bias term. Given an input, a DNN computes an output by propagating it through the network layer by layer and gets the classification result by identifying the dimension with the highest value in the output vector.

A DNN with $d$ layers can be represented by a non-linear multivariate function $\mN: \mathbb{R}^n \rightarrow \mathbb{R}^s$. For any input $\bs{x}\in\mathbb{R}^n$, let $\bs{x}=\bs{x}^1$, the output $\mN(\bs{x})=\bs{W}^d \bs{x}^{d-1}+\bs{b}^d$ can be obtained via the recursive definition $\bs{x}^{i}=\text{ReLU}(\bs{W}^{i} \bs{x}^{i-1}+\bs{b}^{i}) \text{ for } i\in\{2,3,\ldots, d-1\}$, where $\bs{W}^i$ and $\bs{b}^i$ (for $2\le i\le d$) are the weight matrix and bias vector of the $i$-th layer. We refer to $\bs{x}^i_j$ as $j$-th neuron in the $i$-th layer and use $n_i$ to denote the dimension of the $i$-th layer. $n=n_1$ and $s=n_d$.

\smallskip
\noindent
{\bf Quantization.}
%
Quantization is the process of converting high-precision floating-point values into a finite range of lower-precision ones, i.e., fixed-point numbers, without significant accuracy loss. 
A quantized neural network (QNN) is structurally similar to a DNN, except that the parameters and/or activation values are quantized into fixed-pointed numbers, e.g., 
4-bit or 8-bit integers. 
In this work, we adopt the symmetric quantization scheme widely utilized in prior research concerning bit-blip attack (BFA) strategies on QNNs~\cite{1bitallyouneed}, where only parameters are quantized to reduce the memory requirements~\cite{HanMD15,zhou2022incremental,zhang2023post}. During inference, we assume that the parameters are de-quantized and all operations within the quantized networks are executed using floating-point arithmetic.

Given the weight matrix $\bs{W}^i$ and the bias vector $\bs{b}^i$, their signed integer counterparts $\widehat{\bs{W}}^i$ and $\hat{\bs{b}}^i$ with respect to quantization bit-width $Q$ are respectively defined as follows. For each $j,k$,
\[
\widehat{\bs{W}}^i_{j,k} = \lfloor \bs{W}^i_{j,k}/\Delta w^i \rceil, \quad 
        \hat{\bs{b}}^i_{j} = \lfloor \bs{b}^i_{j}/\Delta w^i \rceil
\]
where $\Delta w^i=\text{maxAbs}(\bs{W}^i,\bs{b}^i)/(2^{Q-1}-1)$ is the quantization step size of the $i$-th layer and the \text{max} function returns the maximal value of $\bs{W}^i$ and $\bs{b}^i$. $\lfloor \cdot \rceil$ is the rounding operator, maxAbs$(\bs{W}^i,\bs{b}^i)$ means finding the maximum absolute value among all the entries from $\bs{W}^i$ and $\bs{b}^i$.

\begin{figure*}[t]
	\centering
	\subfigure[DNN.]{\label{fig:dnnDemo}
		\begin{minipage}[b]{0.43\textwidth}
			\includegraphics[width=1.0\textwidth]{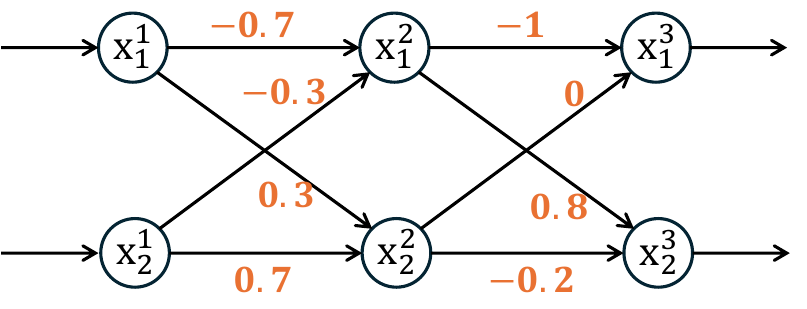}
		\end{minipage}	
	}\hspace{3mm}
	\subfigure[QNN.]{\label{fig:qnnDemo}
		\begin{minipage}[b]{0.43\textwidth}
			\includegraphics[width=1.0\textwidth]{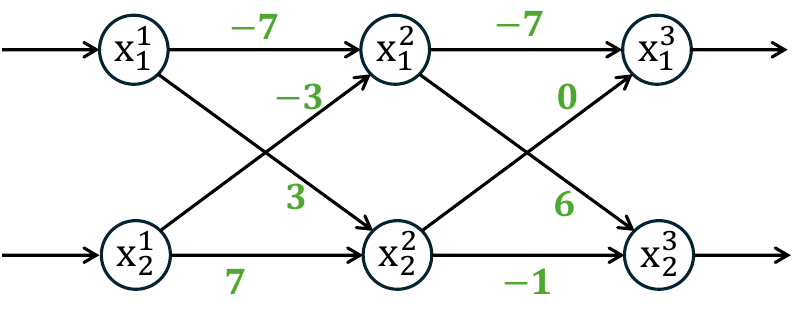}
		\end{minipage}	
	}
	\caption{A 3-layer DNN \minor{with ReLU activations} and its quantized version.}
	\Description{A 3-layer DNN and its quantized version.}
    \label{fig:nnDemo}
\end{figure*}

Once quantized into an integer, the parameter will be stored as the two's complement format in the memory. In the forward pass, the parameters will be de-quantized by multiplying the step size $\Delta w^i$. Taking a quantized parameter $\widehat{\bs{W}}^{i}_{j,k}$ as an example and let $\vec{v}(\cdot)$ denote the operation that converts an integer into its two's complement expressions. Assume that $\vec{v}(\widehat{\bs{W}}^i_{j,k})=[v_{Q};v_{Q-1};\cdots;v_1]$, then the de-quantized version $\widetilde{\bs{W}}^i_{j,k}$ can be calculated as follows with $\widetilde{\bs{W}}^i_{j,k}\approx \bs{W}^i_{j,k}$:
\[
\widetilde{\bs{W}}^i_{j,k}=\widehat{\bs{W}}^i_{j,k} \cdot \Delta w^i =\big( -2^{Q-1}\cdot v_{Q}+\sum_{q=1}^{Q-2} 2^{q-1}\cdot v_{q} \big) \cdot \Delta w^i
\]

\begin{example}\label{eg:dnnDemo}
    Consider the DNN shown in Figure~\ref{fig:dnnDemo}. It contains three layers: one input layer, one hidden layer, and one output layer. The weights are associated with the edges and all the biases are 0 and the quantization bit-width $Q=4$. Then, the step size of the parameter quantizer for each non-input layer is $\Delta w^2 = 0.7/(2^3-1) = 0.1$, $\Delta w^3 = 1/(2^3-1) = 1/7$. 
    The integer counterparts of weight parameters are associated with the edges in Figure~\ref{fig:qnnDemo}. 
    
    Take the hidden layer for instance, we obtain their quantized weights, two's complement counterparts, and de-quantized versions as follows: 
    \begin{itemize}
        \item $\widehat{\bs{W}}^2_{1,1}=\lfloor -0.7/\Delta w^2 \rceil=\lfloor -0.7*10 \rceil =-7$, $\vec{v}(\widehat{\bs{W}}^{2}_{1,1})=[1001]$, and $\widetilde{\bs{W}}^2_{1,1}=-0.7$;
        
        \item $\widehat{\bs{W}}^2_{1,2}=\lfloor -0.3/\Delta w^2  \rceil=\lfloor -0.3*10  \rceil=-3$, $\vec{v}(\widehat{\bs{W}}^{2}_{1,2})=[1101]$, and  $\widetilde{\bs{W}}^2_{1,2}=-0.3$;

        \item $\widehat{\bs{W}}^2_{2,1}=\lfloor 0.3/\Delta w^2  \rceil =\lfloor 0.3*10 \rceil =3$, $\vec{v}(\widehat{\bs{W}}^{2}_{2,1})=[0011]$, and $\widetilde{\bs{W}}^2_{2,1}=0.3$;
        
        \item $\widehat{\bs{W}}^2_{2,2}=\lfloor 0.5/\Delta w^2 \rceil=\lfloor 0.7*10 \rceil =7$, $\vec{v}(\widehat{\bs{W}}^{2}_{2,2})=[0111]$, and $\widetilde{\bs{W}}^2_{2,2}=0.7$;
    \end{itemize}
    
    Similarly, for the output layer, we have 
     \begin{itemize}
        \item $\widehat{\bs{W}}^3_{1,1}=\lfloor -1/\Delta w^3 \rceil=\lfloor -1*7 \rceil =-7$, $\vec{v}(\widehat{\bs{W}}^{3}_{1,1})=[1001]$, and $\widetilde{\bs{W}}^3_{1,1}=-1$;
        \item $\widehat{\bs{W}}^3_{1,2}=\lfloor 0/\Delta w^3  \rceil=\lfloor 0*7  \rceil=0$, $\vec{v}(\widehat{\bs{W}}^{3}_{1,2})=[0000]$, and $\widetilde{\bs{W}}^3_{1,2}=0$;
        \item $\widehat{\bs{W}}^3_{2,1}=\lfloor 0.8/\Delta w^3  \rceil=\lfloor 0.8*7 \rceil =6$, $\vec{v}(\widehat{\bs{W}}^{3}_{2,1})=[0110]$, and $\widetilde{\bs{W}}^2_{1,1}=0.8571$;
        \item $\widehat{\bs{W}}^3_{2,2}=\lfloor -0.2/\Delta w^3  \rceil=\lfloor -0.2*7 \rceil =-1$, $\vec{v}(\widehat{\bs{W}}^{3}_{2,2})=[1111]$, and $\widetilde{\bs{W}}^2_{1,1}=-0.1429$.
    \end{itemize}
    
\end{example}

\subsection{Bit-Flip Attacks}
\major{
Bit-flip attacks (BFAs) are a class of fault-injection attacks that were originally proposed to breach cryptographic primitives~\cite{BarenghiBKN12,BihamS97,BonehDL97}. Recently, BFAs have been ported to neural networks.}

\smallskip
\noindent
\major{{\bf Attack scenarios and threat model.}
Recent studies~\cite{kim2014flipping,yao2020deephammer,rakin2022deepsteal} have revealed vulnerabilities in DRAM chips, which act as a crucial memory component in hardware systems. 
Specifically, an adversary can induce bit-flips in memory by repeatedly accessing the adjacent memory rows in DRAM, without \emph{direct} access to the victim model's memory, known as Rowhammer attack~\cite{kim2014flipping}. Such attacks exploit an unintended side effect in DRAM, where memory cells interact electrically by leaking charges, potentially altering the contents of nearby memory rows that were not originally targeted in the memory access. 
Although such attacks do not grant adversaries full control over the number or precise location of bit-flips and the most prevalent BFA tools such as DeepHammer~\cite{yao2020deephammer} can typically induce only a single bit-flip,
the recent study~\cite{1bitallyouneed} has demonstrated that an adversary can effectively attack a QNN by flipping, on average, just one critical bit during the deployment stage. While indirectly flipping multiple bits is theoretically feasible, achieving this would require highly sophisticated techniques that are both extremely time-intensive and have a low likelihood of success in practice~\cite{rakin2022deepsteal}.
Therefore, in this study, we assume that the adversary can \emph{indirectly} manipulate only a minimal number of parameters in a QNN, by default 1. More powerful attacks that can \emph{directly} manipulate memory go beyond the scope of this work.
On the other hand, though most of the existing BFAs target weights only~\cite{liyes,1bitallyouneed,HONG_USENIX19,BFAICCV19}, in this work, we consider a more general setting where all parameters (weights and biases) of QNNs are vulnerable to BFAs~\cite{randomDNN}.}

\begin{example}
    Consider the QNN given in Example~\ref{eg:dnnDemo}. 
    Suppose a bit-flip attacker can alter any single bit in the memory cell storing parameters and we use two dots ``$\cdot\cdot$'' to represent a parameter that is targeted for such attacks.
    Take the parameter $\widehat{\bs{W}}^3_{2,2}$ with $\vec{v}(\widehat{\bs{W}}^3_{2,2})=[1111]$ for example. Its potential attacked representations are  $\vec{v}(\ddot{\widehat{\bs{W}}}^3_{2,2})\in\{[0111],[1011],[1101],[1110]\}$, thus we have $\ddot{\widehat{\bs{W}}}^3_{2,2}\in\{7,-5,-3,-2\}$ and $\ddot{\widetilde{\bs{W}}}^3_{2,2}\in\{1,-0.7143,-0.4286, -0.2857\}$. Given an input $\bs{x}^1=(1,1)$, after de-quantizing integer parameters during the inference, we can get the output of each non-input layer as $\bs{x}^2=(0,1)$ and $\bs{x}^3=(0,-0.1429)$.

    
    Now, suppose that the attacker flips the fourth bit (i.e., sign bit) of the parameter $\widehat{\bs{W}}^3_{2,2}$, 
    then we have $\ddot{\widehat{\bs{W}}}^3_{2,2}=7$ and $\ddot{\widetilde{\bs{W}}}^3_{2,2}=1$. Finally, the network output after the attack is $\bs{x}^3=(0,1)$, resulting in an altered classification outcome. 
\end{example}

\begin{definition}[Attack Vector]
    Given a QNN $\mN$ with quantization bit-width $Q$, and two integers $\mm$ and $\nn$ such that an adversary can attack any $\mm$ parameters by flipping $\nn$ bits at most within each parameter ($\nn\le Q$). 
    An $(\mm,\nn)$-attack vector $\rho$ is a set of pairs $\{(\alpha_i, P_i) \mid i\le \mm \}$ where $\alpha_i$ is a parameter (weight or bias) of $\mN$ and $P_i$ is a set of bit positions with $|P_i|\leq \nn$. 
    We use $\mN^\rho$ to denote the resulting network by invoking the attack vector $\rho$ on $\mN$. 
\end{definition}

An $(\mm,\nn)$-attack vector defines the vulnerable parameters and bits that are flipped by the adversary during a specific BFA. 
\begin{example}
    Consider the QNN given in Example~\ref{eg:dnnDemo}. Let $\mm=\nn=2$ and an attack vector $\rho=\{(\widehat{\bs{W}}^2_{1,1}, \{2,4\}),(\widehat{\bs{W}}^2_{1,2}, \{3\})\}$. Intuitively, $\rho$ describes a specific bit-flip attack that the 2nd and 4th bits in $\vec{v}(\widehat{\bs{W}}^2_{1,1})=[1001]$ and the 3rd bit in $\vec{v}(\widehat{\bs{W}}^2_{1,2})=[1101]$ are flipped. Then, we have the two's complement representations of attacked parameters as $\vec{v}(\ddot{\widehat{\bs{W}}}^2_{1,1})=[0011]$ and $\vec{v}(\ddot{\widehat{\bs{W}}}^2_{1,2})=[1001]$.
\end{example}

Note that for clarity and convenience, given a QNN, the de-quantized parameters before (resp. after) BFAs $\widetilde{\bs{W}}^i_{j,k}$ (resp. $\ddot{\widetilde{\bs{W}}}^i_{j,k}$) may be directly represented by $\bs{W}^i_{j,k}$ (resp. $\ddot{\bs{W}}^i_{j,k}$) when it is clear from the context in the subsequent sections.



\section{Bit-Flip Attack Verification Problem}\label{sec:pro}

In this section, we define the verification problem considered in this work and discuss a naive baseline solution based on \deepPoly.

\subsection{Problem Definition}\label{sec:proDef}

\begin{definition}[BFA-tolerance]
Let $\mN:\mathbb{R}^n\rightarrow\mathbb{R}^s$ be a QNN. Given a pre-condition $\phi$ over the input $\bs{x}\in\mathbb{R}^n$ and post-condition $\psi$ over the output $\mN(\bs{x})\in\mathbb{R}^s$. We use $\mN \models^\rho_{\mm,\nn} \langle \phi,\psi\rangle$ to denote that for any $(\mm,\nn)$-attack vector $\rho$, $\phi(\bs{x})\Rightarrow \psi(\mN^\rho(\bs{x}))$ always holds, where $\mN^\rho$ is the network obtained from $\mN$ given the attack vector $\rho$.
\end{definition}

If $\mN \models^\rho_{\mm,\nn} \langle \phi,\psi\rangle$ holds,  we say that $\mN$ is BFA-tolerant to the property $\langle \psi,\phi\rangle$. Note that, such a formulation of the problem is expressive enough to cover a range of desired neural network properties, including safety, robustness, (counterfactual) fairness, and backdoor-absence.

\begin{theorem}\label{the:npc}
Verifying whether $\mN \models^\rho_{\mm,\nn} \langle \phi,\psi\rangle$ holds is NP-complete. \hfill \qed
\end{theorem}

In the following, for the sake of readability, our discussion focuses on the following general BFA-tolerant robustness property.

\begin{definition}[BFA-tolerant Robustness]
    Let $\mN:\mathbb{R}^n\rightarrow\mathbb{R}^m$ be a QNN, $\mI \subset \mathbb{R}^n$ be an input region, and $g$ is a target class. $\mN$ is BFA-tolerant for robustness with respect to the region $\mI$ and the class $g$ if $\mN\models^\rho_{\mm,\nn}\langle \phi,\psi\rangle$ returns true, where $\phi(\bs{x}):= \bs{x}\in\mI$, $\psi(\bs{y}):= \text{argmax}(\bs{y})=g$.
\end{definition}

Intuitively, the BFA-tolerant robustness verification problem with $\mm=\nn=0$ is the vanilla robustness verification problem $\mN\models\langle \phi,\psi\rangle$ of neural networks, following the prior works~\cite{GuyKatz2017ReluplexAE}. 
In this work, we consider input regions that are expressible by polyhedra, \minor{following the literature, e.g., ~\cite{GMDTCV18,LiLYCHZ19,LiLHY0ZXH20,SGMPV18,SGPV19,TranBXJ20,TranLMYNXJ19,WangPWYJ18,YLLHWSXZ20,GiacobbeHL20,scaleQNN21,ZZCSZC22,ZhangCSSD24} to cite a few}. 

\subsection{A Naive Method by \deepPoly}\label{sec:naiveMethod}

Next, we present a baseline approach that reduces the BFA verification problem to a classic neural network verification problem so that the existing verifier, such as \deepPoly~\cite{SGPV19}, can be used to verify the above BFA-tolerant properties.


\smallskip
\noindent
{\bf Review of \deepPoly.}
The key idea of \deepPoly is to approximate the behavior of the neural network based on an abstract interpreter specifically tailored to the setting of neural networks. Specifically, the abstract domain $\mA$ is a combination of polyhedra, coupled with abstract transformers for neural network functions, including affine functions and activation functions. To achieve this, each neuron in the hidden layer $\bs{x}^i_j$ (the $j$-th neuron in the $i$-th layer) with $\bs{x}^i_j = \text{ReLU}(\bs{W}^i\bs{x}^{i-1}+\bs{b}^i)$ is seen into two nodes $\bs{x}^i_{j,1}$ and $\bs{x}^{i}_{j,2}$ such that $\bs{x}^{i}_{j,1}=\bs{W}^i_{j,:}\bs{x}^{i-1}+\bs{b}^i_j$ and $\bs{x}^i_{j,2}=\text{ReLU}(\bs{x}^i_{j,1})$, where $\bs{x}^{i-1}_{k}=\bs{x}^{i-1}_{k,2}$ for $k\in[n_{i-1}]$.
Formally, the abstract element $\bs{a}^i_{j,s}\in\mA$ for each neuron $\bs{x}^i_{j,s}$ ($s\in\{1,2\}$) is a tuple $\langle a^{i,\le}_{j,s}, a^{i,\ge}_{j,s}, l^i_{j,s}, u^i_{j,s}\rangle$, where $a^{i,\le}_{j,s}$ (resp. $a^{i,\ge}_{j,s}$) is a symbolic lower (resp. upper) bound in the form of a linear combination of variables which appear before it and $l^i_{j,s},u^i_{j,s}\in\mathbb{R}$. 
For an affine function $\bs{x}^i_{j,1}=\bs{W}^i_{j,:}\bs{x}^{i-1}+\bs{b}^i_j$, the abstract transformer sets $a^{i,\le}_{j,1}=a^{i,\ge}_{j,1}=\bs{W}^i_{j,:}\bs{x}^{i-1}+\bs{b}^i_j$. To compute the concrete lower (resp. upper) bound $l^i_{j,1}$ (resp. $u^i_{j,1}$), we first repeatedly substitute the variables in $a^{i,\le}_{j,1}$ (resp. $a^{i,\ge}_{j,1}$) with their symbolic bounds according to the coefficient until no further substitution is possible. Then, we can obtain a sound lower (resp. upper) bound in the form of the linear combination of input variables, and $l^i_{j,1}$ (resp. $u^i_{j,1}$) can be computed immediately from the input domain.
For an activation function $\bs{x}^{i}_{j,2}=\text{ReLU}(\bs{x}^i_{j,1})$, the abstract transformers set the abstract element $\bs{a}^i_{j,2}=\langle a^{i,\le}_{j,2},a^{i,\ge}_{j,2}, l^i_{j,2}, u^i_{j,2}\rangle$ as follows:
\begin{itemize}
    \item If $l^i_{j,1}\ge 0$: $a^{i,\le}_{j,2}= a^{i,\le}_{j,1}$, $a^{i,\ge}_{j,2}=a^{i,\ge}_{j,1}$, $l^i_{j,2}=l^i_{j,1}$, and $u^i_{j,2}=u^i_{j,1}$;
    \item If $u^i_{j,1}\le 0$: $a^{i,\le}_{j,2}=a^{i,\ge}_{j,2}=l^i_{j,2}=u^i_{j,2}=0$; 
    \item If $l^i_{j,1}< 0 < u^i_{j,1}$: $a^{i,\ge}_{j,2}=u^i_{j,1}(\bs{x}^{i}_{j,1}-l^i_{j,1})/(u^i_{j,1}-l^i_{j,1})$, $a^{i,\le}_{j,2}=\lambda\bs{x}^i_{j,1}$ where $\lambda\in\{0,1\}$ such that the area of resulting shape by $a^{i,\le}_{j,2}$ and $a^{i,\ge}_{j,2}$ is minimal, $l^i_{j,2}=\lambda l^i_{j,1}$ and $u^i_{j,2}=u^i_{j,1}$.
\end{itemize}

\smallskip
\noindent
{\bf A naive method.}
Given the problem of verifying whether $\mN \models^\rho_{\mm,\nn} \langle \phi,\psi \rangle$ holds, a naive solution is to iteratively create an attacked network $\mN^\rho$ for each possible $(\mm,\nn)$-attack vector $\rho$ and check the vanilla robustness verification problem $ \mN^\rho \models \langle \phi,\psi \rangle$ by \deepPoly, which conducts a reachability analysis and returns a sound and incomplete verification result. 
Following this method, the number of possible attack vectors increases quickly with $\mm$, $\nn$, and the number of parameters in $\mN$, causing the infamous combinatorial explosion problem. 
For instance, suppose the number of parameters of a QNN is $K$ and the quantization bit-width is $Q$, the number of possible attack vectors (or the number of attacked networks $\mN^\rho$) is 
$\binom{K}{\mm}\times (\sum_{i=1}^\nn\binom{Q}{i})^\mm$. 


\section{Methodology of \tool}\label{sec:method}

In this work, we operate under the assumption that the adversary is limited to attacking a small number of parameters in a QNN, specifically targeting only one parameter by default ($\mm=1$). 
Note that even if the adversary is limited to flipping only one parameter, the number of possible $(1,\nn)$-attack vectors is still $K\cdot\sum_{i=1}^\nn\binom{Q}{i}$.
Consider a QNN $\mN$ which is quantized by $Q$ and comprises $K$ parameters. The naive method introduced in Section~\ref{sec:proDef} can only verify each $(1,\nn)$-attack vector separately and invokes $K\cdot\sum_{i=1}^\nn\binom{Q}{i}$ times \deepPoly in total, which is highly inefficient. Our idea is to verify multiple attack vectors at one time.

\subsection{Overview of \tool}

The overall verification procedure is given in Algorithm~\ref{alg:overall}. 
Given a QNN $\mN$, an input region $\mI$, a target class $g$, and the maximum number of bits to flip $\nn$, we firstly traverse each parameter $w$, 
performing a reachability analysis via function {\sf BFA\_RA}$(\cdot)$ independently (lines 3-4) to compute a sound output range for $\mN$ considering all potential $\sum_{i=1}^\nn\binom{Q}{i}$ attack vectors with respect to parameter $w$, and subsequently identify all parameters potentially susceptible to bit-flip attacks (line 5). 
If the set $\xi$ is empty, we return \texttt{True} which means all parameters are safe to BFA and the network $\mN$ is BFA-tolerant with respect to the region $\mI$ and class $g$. Otherwise, it implies the existence of at least one parameter for which the reachability analysis fails to confirm safety against such attacks. 
In this case, we reformulate the verification problem into an equivalent MILP problem based on the intermediate results (i.e., all susceptible parameters $\xi$) derived before, which can then be solved using off-the-shelf solvers. Therefore, the whole verification process \tool is sound, complete yet reasonably efficient. We remark that the MILP-based verification method is often more time-consuming and thus the first step allows us to quickly verify many tasks first or identify all vulnerable parameters soundly and formally.

Below, we present the details of functions {\sf BFA\_RA} and {\sf BFA\_MILP}. We first introduce an abstract domain designed for networks with symbolic parameters, which will be utilized throughout our reachability analysis procedure.

\begin{algorithm}[t]
    \SetKwProg{myproc}{Proc}{}{}

    \myproc{{\sf BFA\_Verifier}{$(\mN, \mI,g, \nn)$}}{
        $\xi = \emptyset$\;
    
        \ForEach{parameter $w$ in $\mN$}{
            
            \If{{\sf BFA\_RA}$(\mN, \mI,g, w, \nn)$ = \texttt{Unknown}}{
                $\xi$.append($w$)\; 
            }
        }
        \If{$\xi==\emptyset$}{
            \Return{\texttt{True}};
        }
        \Else{
            \Return{{\sf BFA\_MILP$(\mN,\mI,g,\xi, \nn)$}};
        }
    }

    \caption{Overall Algorithm of {\sf BFAVerifier}}\label{alg:overall}

\end{algorithm}

\subsection{\symPoly: An Abstract Domain for Networks with Symbolic Parameters}\label{sec:deepPolyR}

In this section, we introduce a new abstract domain \symPoly designed for networks with symbolic parameters, equipped with abstract transformers tailored to our bit-flip attack setting to conduct a sound reachability analysis.

Let us consider the $(i+1)$-th layer with neuron function $\bs{x}^{i+1}_j=\text{ReLU}(\bs{W}^{i+1}_{j,:}\bs{x}^{i}+\bs{b}^{i+1}_j)$ in a QNN $\mN$ such that $\bs{W}^{i+1}_{j,k}$ (for some $k\in[n_i]$) or/and $\bs{b}^{i+1}_j$ may be replaced by symbolic parameters.
Following \deepPoly, we first split each neuron (e.g., $\bs{x}^{i+1}_j$) into two nodes (e.g., $\bs{x}^{i+1}_{j,1}$ and $\bs{x}^{i+1}_{j,2}$) and reformulate the neuron function as follows: 
\[
\bs{x}^{i+1}_{j,1}= \sum_{t\in[n_i]} \bs{W}^{i+1}_{j,t}\bs{x}^{i}_{t,2}+\bs{b}^{i+1}_j, \quad \bs{x}^{i+1}_{j,2}= \text{ReLU}(\bs{x}^{i+1}_{j,1})
\]

\subsubsection{Abstract domain.} We inherit the abstract domain $\mA$ introduced in \deepPoly which consists of a set of polyhedral constraints, each relating one variable to a linear combination of the variables from preceding layers. 
Formally, the abstract element of each neuron $\bs{x}^{i+1}_{j,s}$ ($s\in\{1,2\}$) in our abstract domain is represented as $\bs{a}^{i+1}_{j,s}=\langle a^{i+1,\le}_{j,s}, a^{i+1,\ge}_{j,s}, l^{i+1}_j, u^{i+1}_j\rangle \in \mA$, and it satisfy the following invariant: $\gamma(\bs{a}^{i+1}_{j,s})=\{x\in\mathbb{R}\mid a^{i+1,\le}_{j,s}\le x\le a^{i+1,\ge}_{j,s}\}\subseteq [l^{i+1}_{j,s},u^{i+1}_{j,s}]$. By repeatedly substituting each variable $\bs{x}^{i'}_{j'}$ in $a^{i+1,\le}_{j,s}$ (resp. $a^{i+1,\ge}_{j,s}$) using $a^{i',\le}_{j'}$ or $a^{i',\ge}_{j'}$ according to the coefficient of $\bs{x}^{i'}_{j'}$, until no further substitution is possible, $a^{i+1,\le}_{j,s}$ (resp. $a^{i+1,\ge}_{j,s}$) will be a linear combination over the input variables of the QNN.
%
We next introduce our abstract transformation for functions in $\mN$.

\subsubsection{Affine abstract transformer for symbolic weights.}\label{sec:symWeight}
Without loss of generality, we consider the transformer for the case where there is only one concrete parameter is replaced by a symbolic one, e.g., $\overrightarrow{\bs{W}}^{i+1}_{j,k}$ for some $k\in[n_i]$.
For all nodes other than $\bs{x}^{i+1}_{j,1}$, we directly inherit the abstract transformers from \deepPoly.

In this work, we need to abstract affine functions with symbolic parameters and the ReLU function, both of which contribute to precision loss. To improve accuracy, we abstract them jointly as a symbolic weighted ReLU function with ReLU applied internally, as shown in Figure 3b. For the very first affine layer, we abstract the affine function solely, since there is no preceding ReLU, as given later at the end of this section. We remark that our abstract transformations can be compositionally applied to settings involving multiple symbolic parameters. 

\smallskip
\noindent
\minor{\bf Symbolic weights on hidden neurons.}
Consider a symbolic weight parameter $\overrightarrow{\bs{W}}^{i+1}_{j,k}$ constrained by an interval range $[w_l,w_u]$. Then, the updated neuron function for $\bs{x}^{i+1}_{j,1}$ is as follows: 
\begin{equation}\label{eq:symWeight}
    \bs{x}^{i+1}_{j,1}=\sum_{t\in [n_{i}]\backslash k}\bs{W}^{i+1}_{j,t}\bs{x}^i_{t,2}+\overrightarrow{\bs{W}}^{i+1}_{j,k}\bs{x}^i_{k,2}+\bs{b}^{i+1}_j
\end{equation}

\begin{figure*}[t]
	\centering
        \subfigure[Areas obtained by abstracting $\overrightarrow{\bs{W}}^{i+1}_{j,k}\cdot \bs{a}^i_{k,2}$.]{\label{fig:TransAct}
		\begin{minipage}[b]{0.44\textwidth}
			\includegraphics[width=1.0\textwidth]{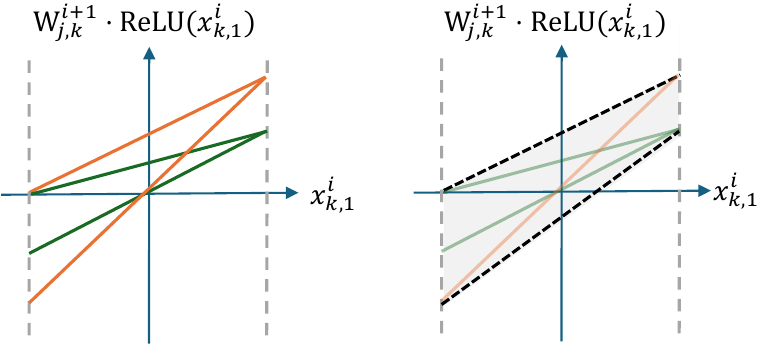}
		\end{minipage}	
	}\hspace{5mm}
        \subfigure[Areas obtained by abstracting $\overrightarrow{\bs{W}}^{i+1}_{j,k}\cdot\text{ReLU}(\bs{x}^i_{k,1})$.]{\label{fig:TransX}
		\begin{minipage}[b]{0.44\textwidth}
			\includegraphics[width=1.0\textwidth]{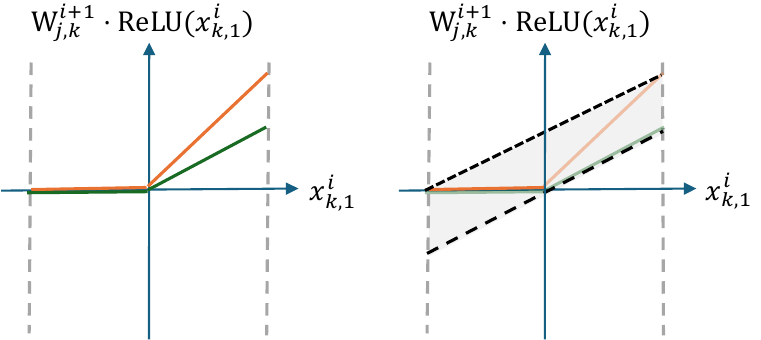}
		\end{minipage}	
	}\vspace{-2mm}
        \caption{Convex approximations of $\overrightarrow{\bs{W}}^{i+1}_{j,k}\bs{x}^i_{k,2}=\overrightarrow{\bs{W}}^{i+1}_{j,k}\cdot \text{ReLU}(\bs{x}^i_{k,1})$ via different abstract transformations: (a) depicts the approximation derived by abstracting $\overrightarrow{\bs{W}}^{i+1}_{j,k}\cdot \bs{a}^i_{k,2}$, where $\bs{a}^i_{k,2}$ is an over-approximation of $\text{ReLU}(\bs{x}^i_{k,1})$ obtained from \deepPoly, and the yellow and green lines gives the boundaries of $w_u\cdot \bs{a}^i_{k,2}$ and $w_u\cdot \bs{a}^i_{k,2}$. (b) directly give the approximation by linear boundaries with the minimal area in the input-output plane of the function $\overrightarrow{\bs{W}}^{i+1}_{j,k}\cdot \text{ReLU}(\bs{x}^i_{k,1})$.}
        \Description{Convex approximations for $\overrightarrow{\bs{W}}^{i+1}_{j,k}\bs{x}^i_{k,2}$ via different abstract transforms.}
    \label{fig:abstrDomains}
\end{figure*}

To perform abstract transformations on $\overrightarrow{\bs{W}}^{i+1}_{j,k}\bs{x}^i_{k,2}$, an intuitive idea is to directly make an affine transformation with respect to symbolic parameter $\overrightarrow{\bs{W}}^{i+1}_{j,k}$ on the abstract element of $\bs{x}^i_{k,2}$. However, it will lead to an over-approximate result compared to abstracting the symbolic-weighted ReLU function.
To illustrate it, let us consider the setting where $(l^i_{k,2}<0<u^i_{k,2})\wedge(l^i_{k,2}+u^i_{k,2}>0)$ and $w_l\ge 0$. As shown in Figure~\ref{fig:TransAct}, the areas within the yellow boundaries and the green boundaries are captured by the weighted abstract elements $w_u\cdot \bs{a}^{i}_{k,2}$ and $w_l\cdot\bs{a}^{i}_{k,2}$, respectively, with $\gamma(w_u\cdot\bs{a}^{i}_{k,2})=\{ w_u  \cdot x \in\mathbb{R} \mid a^{i,\le}_{k,2}\le  x\le a^{i,\ge}_{k,2} \}$ and $\gamma(w_l\cdot\bs{a}^{i}_{k,2})=\{ w_l \cdot x \in\mathbb{R} \mid a^{i,\le}_{k,2}\le  x\le a^{i,\ge}_{k,2} \}$. 
It is obvious that the area captured by the dotted polyhedra in Figure~\ref{fig:TransAct} is larger than that in Figure~\ref{fig:TransX}, whose area is captured by directly abstracting the weighted ReLU function $\overrightarrow{\bs{W}}^{i+1}_{j,k}\cdot \text{ReLU}(\bs{x}^i_{k,1})$.

\begin{figure*}
    \centering
    \includegraphics[width=0.95\textwidth]{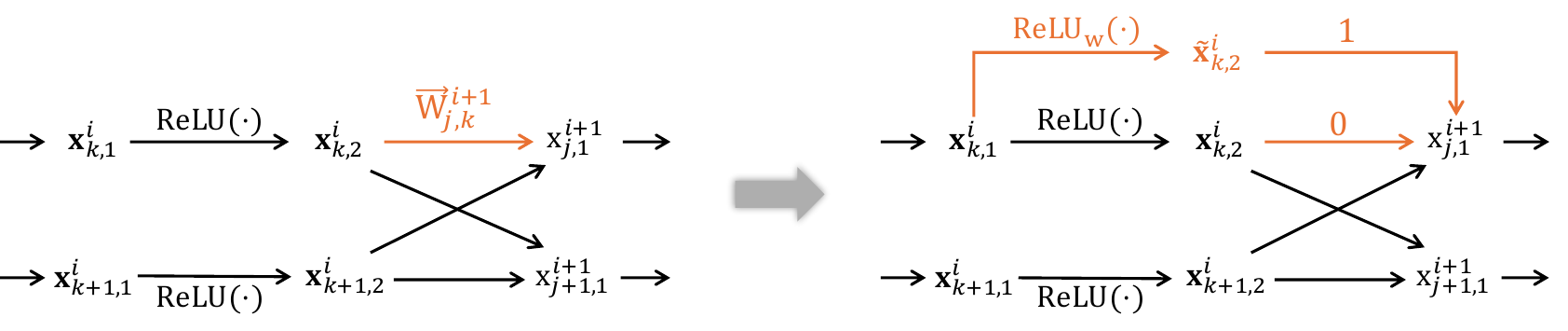}
    \caption{The construction of additional node $\tilde{\bs{x}}^{i}_{k,2}$ given a symbolic parameter $\overrightarrow{\bs{W}}^{i+1}_{j,k}$.}
    \Description{The construction of additional node $\tilde{\bs{x}}^{i}_{k,2}$ given a symbolic parameter $\overrightarrow{\bs{W}}^{i+1}_{j,k}$.}
    \label{fig:modQNN}
\end{figure*}

Therefore, our idea is to abstract the symbolic-weighted ReLU function directly. To achieve it, we initially introduce an additional node $\tilde{\bs{x}}^{i}_{k,2}$ for the symbolic parameter $\overrightarrow{\bs{W}}^{i+1}_{j,k}$ such that $\tilde{\bs{x}}^i_{k,2}=\text{ReLU}_w(\bs{x}^{i}_{k,1}) = \overrightarrow{\bs{W}}^{i+1}_{j,k} \cdot \text{ReLU}(\bs{x}^{i}_{k,1})$. After that, we set the weight as 1 between $\tilde{\bs{x}}^i_{k,2}$ and $\bs{x}^{i+1}_{j,1}$, and set the weight between $\bs{x}^i_{k,2}$ and $\bs{x}^{i+1}_{j,1}$ as 0. An illustration of the construction can be found in Figure~\ref{fig:modQNN}.
Then, given the abstract element of $\bs{x}^i_{k,2}$ as $\bs{a}^{i}_{k,2}=\langle a^{i,\le}_{k,2}, a^{i,\ge}_{k,2}, l^i_{k,2}, u^i_{k,2} \rangle$, we define the abstract element $\bs{a}^{i,\ast}_{k,2}=\langle \tilde{a}^{i,\le}_{k,2},\tilde{a}^{i,\ge}_{k,2}, \tilde{l}^i_{k,2}, \tilde{u}^i_{k,2} \rangle$ of neuron $\tilde{\bs{x}}^i_{k,2}$ as follows:
\begin{itemize}
    \item If $w_l\ge 0$: $\tilde{a}^{i,\le}_{k,2}=w_l\cdot a^{i,\le}_{k,2}$, $\tilde{a}^{i,\ge}_{k,2}=w_u\cdot a^{i,\ge}_{k,2}$, $\tilde{l}^i_{k,2}=w_l\cdot l^i_{k,2}$, and $\tilde{u}^i_{k,2}=w_u\cdot u^i_{k,2}$;
    
    \item If $w_u\le 0$: $\tilde{a}^{i,\le}_{k,2}=w_l\cdot a^{i,\ge}_{k,2}$, $\tilde{a}^{i,\ge}_{k,2}=w_u\cdot a^{i,\le}_{k,2}$, $\tilde{l}^i_{k,2}=w_l\cdot u^i_{k,2}$, and $\tilde{u}^i_{k,2}=w_u\cdot l^i_{k,2}$;
    
    \item If $w_l < 0 < w_u$: $\tilde{a}^{i,\le}_{k,2}=w_l\cdot a^{i,\ge}_{k,2}$, $\tilde{a}^{i,\ge}_{k,2}=w_u\cdot a^{i,\ge}_{k,2}$, $\tilde{l}^i_{k,2}=w_l\cdot u^i_{k,2}$, and $\tilde{u}^i_{k,2}=w_u\cdot u^i_{k,2}$.
\end{itemize}

\begin{figure*}[t]
    \centering
    \includegraphics[width=0.85\textwidth]{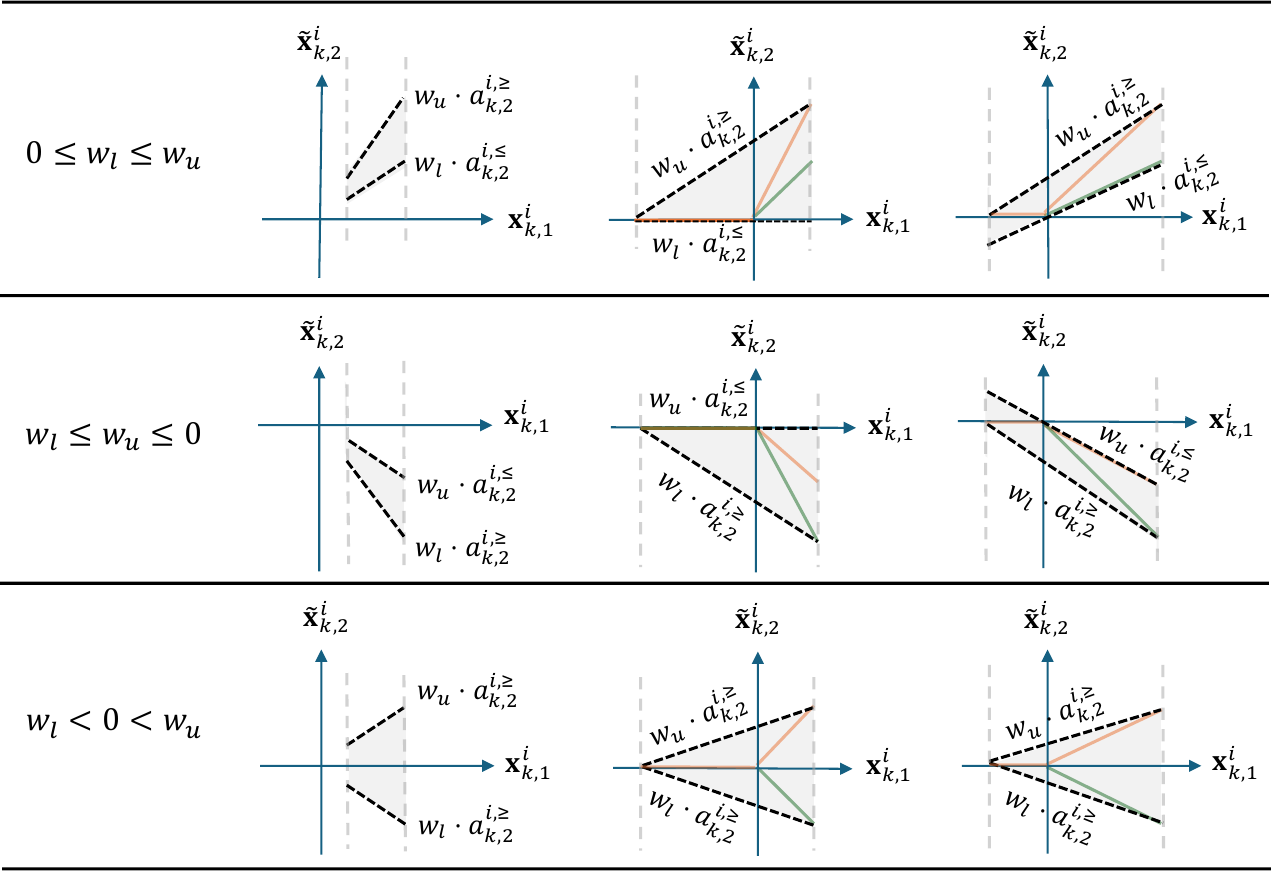}
    \caption{An illustration of the abstract transformer for the symbolic-weighted ReLU function $\overrightarrow{\bs{W}}^{i+1}_{j,k}\cdot \text{ReLU}(\bs{x}^i_{k,1})$, where the first column defines the range (interval) of values $[w_l, w_u]$ for $\overrightarrow{\bs{W}}^{i+1}_{j,k}$. The second column gives the abstraction when $l^i_{k,1}>0$ for the abstract element $\bs{a}^i_{k,1}$ of $\bs{x}^i_{k,1}$, and the third (resp. fourth) column shows the abstraction of when $(l^i_{k,1} <0<u^i_{k,1}) \wedge (l^i_{k,1}+u^i_{k,1}\le 0)$ (resp. $(l^i_{k,1} < 0 < u^i_{k,1}) \wedge (l^i_{k,1}+u^i_{k,1}>0)$).}
    \Description{An illustration of the abstract transformer for symbolic-weighted ReLU functions.}

    \label{fig:allCases}
\end{figure*}

An illustration of the above abstract transformer for the weighted-ReLU function can be found in Figure~\ref{fig:allCases}.

\begin{theorem}\label{the:deepPolyR_weight}
    The weighted ReLU abstract transformer for neuron $\tilde{\bs{x}}^i_{k,2}$ in Figure~\ref{fig:modQNN} i) is sound
    and preserves the invariant $\gamma (\bs{a}^{i,*}_{k,2}) \subseteq [\tilde{l}^i_{k,2}, \tilde{u}^i_{k,2}]$. \qed
\end{theorem}

Finally, we can apply the affine abstract transformers introduced in \deepPoly (cf. Section~\ref{sec:naiveMethod}) to the remaining affine transformations within $\bs{x}^{i+1}_{j,1}$ in equation~(\ref{eq:symWeight}), i.e., $\sum_{t\in[n_i]\backslash k} \bs{W}^{i+1}_{j,t}\bs{x}^i_{t,2}+\bs{b}^{i+1}_j$. Hence, both soundness and the domain invariant are preserved in our affine abstract transformers on equation~(\ref{eq:symWeight}) considering symbolic weight parameters.

\smallskip
\noindent
\major{\bf Symbolic weights on input neurons.}
\major{Consider $\bs{x}^2_{j,1}=\sum_{t\in [n_{2}]\backslash k}\bs{W}^{2}_{j,t}\bs{x}^1_{t}+\overrightarrow{\bs{W}}^{2}_{j,k}\bs{x}^1_k+\bs{b}^{2}_j$ with a symbolic weight $\overrightarrow{\bs{W}}^{2}_{j,k}$ connected to the input neuron $\bs{x}^1_k$. Let $[x_l,x_u]$ be the input range of the neuron $\bs{x}^1_k$, then the abstract domain for $\bs{x}^2_{j,1}$ is
$\bs{a}^{2,\ast}_{j,1} = \langle \tilde{a}^{2,\le}_{j,1}, \tilde{a}^{2,\ge}_{j,1}, \tilde{l}^{2}_{j,1}, \tilde{u}^{2}_{j,1}\rangle$ for $\bs{x}^{2}_{j,1}=\sum_{t\in [n_{2}]\backslash k}\bs{W}^{2}_{j,t}\bs{x}^1_{t}+\overrightarrow{\bs{W}}^{2}_{j,k}\bs{x}^1_k+\bs{b}^{2}_j$ with $\tilde{a}^{2,\le}_{j,1}$ and $\tilde{a}^{2,\ge}_{j,1}$ set as follows:
\[
    \tilde{a}^{2,\le}_{j,1}= \sum_{t\in [n_{2}]\backslash k}\bs{W}^{2}_{j,t}\bs{x}^1_{t} + \bs{b}^2_j + \kappa^\le \bs{x}^1_k-\eta, \quad \tilde{a}^{2,\ge}_{j,1}= \sum_{t\in [n_{2}]\backslash k}\bs{W}^{2}_{j,t}\bs{x}^1_{t} + \bs{b}^2_j + \kappa^\ge \bs{x}^1_k+\eta
\]where if $0\le x_l$, then $\{\kappa^\le = w_l$, $ \kappa^\ge = w_u$, $\eta=0\}$; if $x_u\le 0$, then $\{\kappa^\le =w_u$, $\kappa^\ge = w_l$, $\eta =0\}$; Otherwise, $\{\kappa^\le=\frac{w_l x_u -w_u x_l}{x_u-x_l}$, $\kappa^\ge=\frac{w_u x_u -w_l x_l}{x_u-x_l}$, $\eta=\frac{x_u x_l}{x_u - x_l}(w_l-w_u)\}$.}

\major{$\tilde{l}^{2}_{j,1}$ and $\tilde{u}^{2}_{j,1}$ can be determined with corresponding lower/upper bounds computation methods (cf. Section~\ref{sec:naiveMethod}). Intuitively, $\kappa^\le \bs{x}^i_k -\eta$ (resp. $\kappa^\ge \bs{x}^i_k +\eta$) expresses the lower (resp. upper) boundary of the abstract domain of the weighted input neuron $\overrightarrow{\bs{W}}^{2}_{j,k}\bs{x}^1_{k}$.}

\begin{theorem}\label{the:input}
    The abstract transformer for symbolic weighted input neuron $\bs{x}^2_{j,2}$ is sound and preserves the invariant $\gamma(\bs{a}^{2,\ast}_{j,1})\subseteq [\tilde{l}^2_{j,1}, \tilde{u}^2_{j,1}]$. \hfill\qed
\end{theorem}

\subsubsection{Affine abstract transformer for symbolic biases.}
Similar to the affine transformer for symbolic weights, for all nodes other than $\bs{x}^{i+1}_{j,1}$, we adopt the abstract transformers from \deepPoly. Our abstract transformations can be compositionally applied to settings involving multiple symbolic parameters.

Consider a symbolic bias parameter $\overrightarrow{\bs{b}}^{i+1}_j$ constrained by an interval range $[w_l,w_u]$. Then, the updated neuron function is $\bs{x}^{i+1}_{j,1}=\sum_{t\in[n_i]}\bs{W}^{i+1}_{j,t}\bs{x}^{i}_{t,2}+\overrightarrow{\bs{b}}^{i+1}_j$. Then, we define the abstract element 
$\bs{a}^{i+1}_{j,1}=\langle a^{i+1,\le}_{j,1}, a^{i+1,\ge}_{j,1},  l^{i+1}_{j,1}, u^{i+1}_{j,1}\rangle$ of neuron $\bs{x}^{i+1}_{j,1}$ as follows:
\[
      a^{i+1,\le}_{j,1}= \sum_{t\in[n_i]}\bs{W}^{i+1}_{j,t}\bs{x}^{i}_{t,2} + w_l, \quad  a^{i+1,\ge}_{j,1}= \sum_{t\in[n_i]}\bs{W}^{i+1}_{j,t}\bs{x}^{i}_{t,2} + w_u 
\] 
where $l^{i+1}_{j,1}$ and $u^{i+1}_{j,1}$ can be determined with corresponding lower/upper bounds computation methods (cf. Section~\ref{sec:naiveMethod}). 


\begin{theorem}\label{the:deepPolyR_bias}
    The affine abstract transformer for symbolic biases preserves both soundness and the invariant. \qed
\end{theorem}


\noindent
{\bf Other abstract transformers.}
In this work, for other network functions, such as the ReLU function and the maxpool operator, we directly adopt the corresponding abstract transformers from \deepPoly. Hence, \symPoly is sound.

\subsection{ Details of Function {\sf BFA\_RA}}\label{sec:bfa_ra}
In this section, we give the implementation details of our reachability analysis procedure {\sf BFA\_RA}. 
For any parameter $w$, we can always determine its concrete interval $[w_l,w_u]$ where any value obtained by flipping up to $\nn$ bits will invariably be contained. Based on this interval, given a QNN $\mN$, an input region $\mI$, and a target class $g$, we can use \symPoly to perform reachability analysis on a modified network $\mN^\ast$, where the concrete parameter $w$ is substituted with a symbolic parameter $\overrightarrow{w}$ constrained by the interval range $[w_l,w_u]$. Then, the analysis will yield two results: i) \emph{Proved}, indicating that the adversary cannot compromise $\mN$ (e.g., all inputs from the input region $\mI$ will be classified as the same class $g$) by altering at most $\nn$ bits in the two's complement representation of the parameter $w$; or ii) \emph{Unknown}, meaning that we cannot confirm the security of $w$ against such bit-flip attacks and there may exist some inputs within the region $\mI$ that will be misclassified by the attacked network into classes other than $g$.

We remark that not all bits of a parameter are equally important concerning BFAs. 
For example, given a quantized parameter $[v_Q;v_{Q-1};\cdots;v_1]$, flipping the signal bit $v_Q$ would always produce the most deviation than flipping any other bit $v_i$ (for $i\in[Q-1]$) since the former always results in the largest parameter interval by adding or subtracting $2^{Q-1}$ onto the original parameter value. Hence, if we consider all the bits at one time, we need to verify the neural network with a large interval for the parameter, and likely return \emph{Unknown}. 
Moreover, according to the abstract transformers defined for symbolic weight parameters in Section~\ref{sec:symWeight}, when the weight interval $[w_l,w_u]$ satisfies $w_l<0<w_u$, the abstract transformer for the $\text{ReLU}_w$ function would lead to a looser over-approximation compared to the other two settings ($w_l\ge 0$ or $w_u\le 0$).  
Hence, given a symbolic parameter with a constrained interval range $[w_l,w_u]$, to enhance the precision of our reachability analysis result, we first partition the parameter interval into two sub-intervals characterized by uniformly signed parameter values, either entirely positive or entirely negative. Then, we perform reachability analysis using \symPoly separately for each sub-interval. Note that, this division addresses the significant over-approximation precision loss that occurs when a symbolic weight parameter has a lower bound and upper bound with differing signs. 
An example illustrating how such an interval partition enhances the abstraction precision is given in Appendix~\ref{sec:app_ip_eff}.
Moreover, for parameter intervals that are too wide to be proved by \symPoly, we introduce a binary search method, which splits the parameter interval at its midpoint and independently verifies each resulting smaller interval iteratively.

\begin{algorithm}[t]
    \footnotesize
    \SetKwProg{myproc}{Proc}{}{}

    \myproc{{\sf BFA\_RA}{$(\mN, \mI, g, w, \nn)$}}{
        Compute two minimal sub-intervals $[w^+_l,w^+_u]$, $[w^-_l,w^-_u]$ for $w$ such that i) $w^+_l\ge 0 \wedge w^-_u\le 0$, and ii) $[w^+_l,w^+_u]\cup [w^-_l,w^-_u]$ consists of all possible values of $w$ that can be derived by flipped at most $\nn$ bits\;
        
        \If{{\sf binary\_RA}$(\mN,\mI, g, w, w^+_l,w^+_u)=\texttt{Unknown}$}   {
            \Return{ \texttt{Unknown} }\;

        }
        \If{{\sf binary\_RA}$(\mN,\mI, g, w, w^-_l,w^-_u)=\texttt{Unknown}$}   {
            \Return{ \texttt{Unknown} }\;

        }
        \Return{\texttt{Proved}}\;
    }

    \vspace{2mm}
    
    \myproc{{\sf binary\_RA}{$(\mN, \mI, g, w, w_l,w_u)$}}{

        \If{$\symPoly(\mN,\mI, g, w, w_l,w_u)=\texttt{Proved}$}{
            \Return{ \texttt{Proved}}\;
        }
        \If{$w_l==w_u$}{
            \Return{$\symPoly(\mN,\mI, g, w, w_l,w_u)$}\;
        }
        
        Split $[w_l,w_u]$ at the midpoint and get two minimal sub-intervals $[w_l, w_u']$, $[w_l',w_u]$ such that $[w_l, w_u'] \cup [w_l',w_u]$ encompasses all potential flipped values of $w$ as that in $[w_l,w_u]$\; 
        
        \If{{\sf binary\_RA}{$(\mN, \mI, g, w, w_l,w_u')=\texttt{Unknown}$ } }   {
            \Return{ \texttt{Unknown} }\;


        }\ElseIf{ {\sf binary\_RA}{$(\mN, \mI, g, w, w_l',w_u)=\texttt{ Unknown}$ } }{
            
            \Return{ \texttt{Unknown}}\;
            
        }\Else{
            \Return{\texttt{Proved}}\;
        }
        
    }
    
    \caption{{\sf BFA\_RA} function}\label{alg:verifyPolyR}

\end{algorithm}

The details of function {\sf BFA\_RA} can be found in Algorithm~\ref{alg:verifyPolyR}, where $\symPoly(\mN,\mI,g,w, w_l,w_u)$ represents that we conduct reachability analysis via \symPoly on the network $\mN$ equipped with a symbolic parameter $w$ constrained by the interval $[w_l,w_u]$ with respect to the input region $\mI$ and output class $g$.
%
The algorithm works as follows. 
Given a neural network $\mN$, an input region $\mI$, a target class $g$, an attacked parameter $w$, and the bit flipping maximum number $\nn$, we first compute two intervals $[w^+_l,w^+_u]$ and $[w^-_l,w^-_u]$ for $w$ such that i) $w^+_l\ge 0$ and $w^-_u\le 0$, and ii) these intervals are designed to encompass any values obtained by flipping up to $\nn$ bits in the two's complement representation of $w$, which can be done as follows: 

\begin{itemize}
    \item If $w\ge 0$: $w^+_l$ (resp. $w^+_u$) is obtained by flipping the most significant $\nn$ bits via $1\rightarrow 0$ (resp. $0\rightarrow 1$), and $w^-_l$ (resp. $w^-_u$) is obtained by flipping the signal bit and the most significant $\nn-1$ bits via $1\rightarrow 0$ (resp. $0\rightarrow 1$);
    \item If $w<0$: $w^+_l$ (resp. $w^+_u$) is obtained by flipping the signal bits via $1\rightarrow 0$ and the most significant $\nn-1$ bits via $1\rightarrow 0$ (resp. $0\rightarrow 1$), and $w^-_l$ (resp. $w^-_u$) is obtained by the most significant $\nn$ bits via $1\rightarrow 0$ (resp. $0\rightarrow 1$).
\end{itemize}

For parameter intervals that are too wide to be verified by \symPoly (cf. line 13), we introduce a binary search method, which splits the parameter interval at its midpoint and independently verifies each resultant smaller interval iteratively. Smaller intervals are generally more likely to yield \texttt{Proved} results, thus enhancing the overall effectiveness and precision of our reachability analysis, which has been confirmed by our experiments (cf. Section~\ref{sec:RQ1}).

\subsection{ Details of Function {\sf BFA\_MILP}}\label{sec:milp}
If {\sf BFA\_RA} fails to prove the BFA-tolerant robustness property, we then encode the verification problem as an equivalent MILP problem w.r.t the set of unproved parameters as follows. 

\vspace{2mm}
\noindent
{\bf Encoding of input regions.}
We consider input regions that are expressible by polyhedra in this work, and they can be directly encoded by linear constraints. For example, for an input region defined by $L_\infty$-norm $\mI^r_u=\{\bs{x}\in\mathbb{R}^n \mid || \bs{x}-\bs{u}||_\infty \le r\}$, we can use the following constraint set $\Theta^\mI$ to encode the input condition $\phi(\bs{x}):=\bs{x}\in\mI^r_\bs{u}$:
\[
\Theta^\mI=\{ \text{max}(\bs{u}_i-r,0)\le \bs{x}_i \le \text{min}(\bs{u}_i+r,1) \mid i\in [n]\}
\]

For input regions defined by the cartesian product of intervals $\mI^{\bs{l}\times\bs{u}}=\{\bs{x}\in \mathbb{R}^n\mid \bs{x}_i \in [\bs{l}_i, \bs{u}_j]\}$, we can use the following constraint set to encode the input condition $\phi(\bs{x}):=\bs{x}\in\mI^{\bs{l}\times \bs{u}}$:
\[
\Theta^\mI=\{ \bs{l}_i \le \bs{x}_i \le \bs{u}_i \mid i\in [n]\}
\]

\vspace{2mm}
\noindent
{\bf Encoding of output properties.}
Let $\bs{y}$ denote the output vector of $\mN^\rho$ given any attack vector $\rho$. We encode the output condition, i.e., $\psi(\bs{y}):=\text{argmax}(\bs{y}) \neq \text{argmax}(\mN(\bs{u}))=g\in[s]$ into the following set of constraints based on a set of Boolean variables $\{\eta_{i}\mid i\in[s]\backslash g\}$:
\begin{itemize}
    \item If $i<g$: then $\bs{y}_i\ge \bs{y}_g \Leftrightarrow \eta_i=1$ which can be encoded as 
    $\Theta_{i,0}^g=\{\bs{y}_g + \textbf{M}\cdot (\eta_i-1) \le \bs{y}_i, \bs{y}_i < \bs{y}_g + \textbf{M}\cdot \eta_i\}$;
    \item If $i> g$: then $\bs{y}_i> \bs{y}_g \Leftrightarrow \eta_i=1$ which can be encoded as 
    $\Theta_{i,1}^g=\{\bs{y}_g + \textbf{M}\cdot (\eta_i-1) < \bs{y}_i, \bs{y}_i < \bs{y}_g + \textbf{M}\cdot \eta_i\}$.
\end{itemize}
Intuitively, $\Theta_{i,0}^g$ (resp. $\Theta_{i,1}^g$) ensures that the $i$-th entry of the output vector $\bs{y}$ for $i<g$ (resp. $i>g$) is no less than (resp. larger than) the $g$-th entry iff $\eta_i=1$. As a result, $\text{argmax}(\bs{y})\neq g$ iff the set of constraints $\Theta_g=\bigcup_{i<g}\Theta^g_{i,0} \cup \bigcup_{i>g}\Theta^g_{i,1}\cup\{\sum_{i\in[s]\backslash g} \eta \ge 1\}$ holds.

\vspace{2mm}
\noindent
{\bf Encoding of neural networks under BFAs.} 
Next, we present the MILP encoding of the neural network under BFAs based on the intermediate analysis results from function {\sf BFA\_RA}. 

Let $\xi=\{w_1,w_2,\cdots,w_m\}$ denote the vulnerable parameters set obtained from our reachability analysis (cf. Algorithm~\ref{alg:overall}). Each parameter is quantized by $Q$ bits. The adversary can only attack one parameter from the parameter set $\xi$ and flip $\nn$ bit at most on it. We first traverse all possible $(1,\nn)$-attack vectors for each parameter $w_i \in \xi$ ($1\le i\le|\xi|$), and get the flipped value set $F_{w_i}=\{w_i^1, w_i^2, \cdot, w_i^{\mathfrak{K}}\}$ for each $w_i$, where $\mathfrak{K}= \sum_{i=1}^\nn \binom{Q}{i}$. Then, we use the following constraint set $\Theta_\xi^{\nn}$ to encode the parameters in $\xi$, where $\delta^j_i$ for $i\in|\xi|$ and $j\in\mathfrak{K}$ are binary variables:
\begin{center}
    $\Theta_{\xi}^\nn= \left\{
        \begin{array}{c}
        
             \tilde{w}_i = w_i + (w_i^1 - w_i) \delta_i^1 + (w_i^2 -w_i) \delta_i^2 + \cdots + (w_i^\mathfrak{K}-w_i) \delta_i^\mathfrak{K}, \\
                    
            i\in |\xi|,  \qquad \sum_{i=1}^{|\xi|} \sum_{j=1}^{\mathfrak{K}} \delta_i^j = 1 
        
        \end{array}\right\}$

\end{center}

Intuitively, for each parameter $w_i$, the binary variable $\delta^j_i=1$ indicates that the adversary attacks the parameter $w_i$ and alters it into a new value $w_i^j$. If $\sum_{j=1}^\mathfrak{K} \delta_i^j=0$, then it means that the adversary does not attack the parameter $w_i$. The constraint $\sum_{i=1}^{|\xi|} \sum_{j=1}^{\mathfrak{K}} \delta_i^j = 1 $ ensures that only one parameter is altered, while no more than $\nn$ bits are subject to modification.

Then, we follow the existing MILP encoding method~\cite{LomuscioM17} to encode $\mN^\rho$ into a set of mixed-integer linear constraints $\Theta_{\mN^\rho}$, where for each vulnerable parameter $w_i\in\xi$, we use $\tilde{w}_i$ in $\Theta_{\xi}^\nn$ to replace $w_i$ in the encoding of the affine function. 
Finally, the BFA-tolerant robustness verification problem is equivalent to the solving of the constraint set: $\Theta_P= \Theta_{\mN^\rho} \cup \Theta_{\xi}^\nn \cup \Theta^\mI \cup \Theta_{g}$.

\begin{theorem}\label{the:complete}
    $\mN\models^\rho_{\mm,\nn}$ holds iff $\Theta_P$ is unsatisfiable.  \hfill \qed
\end{theorem}


\minor{Overall, the complexity of {\sf BFA\_RA} is polynomial in the network size when $\mm=1$, whereas {\sf BFA\_MILP} remains NP-complete even when $\mm=1$.}

\subsection{\major{Extension to Other Networks}}
\major{This work primarily focuses on feedforward neural networks with ReLU activations. In this section, we demonstrate the extensibility of our framework to other networks, including those with sigmoid or tanh activations and architectures incorporating convolutional layers.}

\subsubsection{\major{Other activation functions}} \major{Following the idea of symbolic weights on hidden neurons in Section~\ref{sec:symWeight} and the abstract transformers proposed in \deepPoly for sigmoid and tanh, for an activation function $g(x)$ that is continuous and twice-differentiable such that 
 the first derivative $g^\prime(x)>0$ and the second derivative $g^{\prime\prime}\ge 0 \Leftrightarrow x\le 0$, we also construct an additional node $\tilde{\bs{x}}^i_{k,2}$ (the same as in Figure~\ref{fig:modQNN}) and study its abstract domain according to 
$\tilde{\bs{x}}^i_{k,2}=\overrightarrow{\bs{W}}^{i+1}_{j,k} \cdot g(\bs{x}^i_{k,1})$.
The corresponding abstract transformers for Sigmoid and Tanh considering the node $\tilde{\bs{x}}^i_{k,2}$ are given in Table~\ref{tab:log_AbT}. For the other network functions with constant parameters, we can reuse the corresponding abstract transformers from \deepPoly directly.}

\begin{table}[t]
    \centering
    \caption{\major{The abstract domain $\bs{a}^{i,\ast}_{k,2}=\langle  \tilde{a}^{i,\le}_{k,2}, \tilde{a}^{i,\ge}_{k,2}, \tilde{l}^i_{k,2}, \tilde{u}^i_{k,2} \rangle$ of $\tilde{\bs{x}}^i_{k,2}=\overrightarrow{\bs{W}}^{i+1}_{j,k}\cdot g(\bs{x}^i_{k,1})$, where $l^i_{k,1}$ and $u^i_{k,1}$ are the lower and upper bounds of $\bs{x}^i_{k,1}$, $\kappa = \frac{g(u^i_{k,1})-g(l^i_{k,1})}{u^i_{k,1}-l^i_{k,1}}$, and $\kappa^\prime = \text{min}(g^\prime (l^i_{k,1}), g^\prime(u^i_{k,1}))$}}
    \setlength{\tabcolsep}{2pt}
    \scalebox{0.8}{
    \begin{tabular}{c|c|c|c}
        \toprule
         $g(x)$ & Bounds of $\bs{x}^i_{k,1}$ & $w_l\ge 0$ & $w_u\le 0$  \\ \midrule

        & $l^i_{k,1} \ge 0$ &  \tabincell{l}{$\tilde{a}^{i,\le}_{k,2}=w_l g(l^i_{k,1})+w_l\kappa (\bs{x}^i_{k,1}-l^i_{k,1})$ \\ $\tilde{a}^{i,\ge}_{k,2}= w_u g(u^i_{k,1})+w_u\kappa^\prime (\bs{x}^i_{k,1}-u^i_{k,1})$ \\ $\tilde{l}^{i}_{k,2}= w_l g(l^i_{k,1})$, $\tilde{u}^i_{k,2}= w_u g(u^i_{k,1})$} & 
        \tabincell{l}{$\tilde{a}^{i,\le}_{k,2}= w_l g(u^i_{k,1})+w_l\kappa^\prime (\bs{x}^i_{k,1}-u^i_{k,1})$ \\ $\tilde{a}^{i,\ge}_{k,2}=w_u g(l^i_{k,1})+w_u\kappa (\bs{x}^i_{k,1}-l^i_{k,1})$ \\ $\tilde{l}^{i}_{k,2}= w_l g(u^i_{k,1})$, $\tilde{u}^i_{k,2}= w_u g(l^i_{k,1})$} \\  \cline{2-4}

        $\text{Sigmoid}(x)$ & $u^i_{k,1} \le 0$ & \tabincell{l}{$\tilde{a}^{i,\le}_{k,2}= w_l g(l^i_{k,1})+w_l\kappa^\prime (\bs{x}^i_{k,1}-l^i_{k,1})$ \\ $\tilde{a}^{i,\ge}_{k,2}=w_u g(u^i_{k,1})+w_u\kappa (\bs{x}^i_{k,1}-u^i_{k,1})$ \\ $\tilde{l}^{i}_{k,2}= w_l g(l^i_{k,1})$, $\tilde{u}^i_{k,2}= w_u g(u^i_{k,1})$} & \tabincell{l}{$\tilde{a}^{i,\le}_{k,2}= w_l g(u^i_{k,1})+w_l\kappa (\bs{x}^i_{k,1}-u^i_{k,1})$ \\ $\tilde{a}^{i,\ge}_{k,2}=w_u g(l^i_{k,1})+w_u\kappa^\prime (\bs{x}^i_{k,1}-l^i_{k,1})$ \\ $\tilde{l}^{i}_{k,2}= w_l g(u^i_{k,1})$, $\tilde{u}^i_{k,2}= w_u g(l^i_{k,1})$} \\ \cline{2-4}

        & $l^i_{k,1} < 0 < u^i_{k,1}$ &  \tabincell{l}{$\tilde{a}^{i,\le}_{k,2}= w_l g(l^i_{k,1})+w_l\kappa^\prime (\bs{x}^i_{k,1}-l^i_{k,1})$ \\ $\tilde{a}^{i,\ge}_{k,2}=w_u g(u^i_{k,1})+w_u\kappa^\prime (\bs{x}^i_{k,1}-u^i_{k,1})$\\ $\tilde{l}^{i}_{k,2}= w_l g(l^i_{k,1})$, $\tilde{u}^i_{k,2}= w_u g(u^i_{k,1})$} & 
         \tabincell{l}{$\tilde{a}^{i,\le}_{k,2}= w_l g(u^i_{k,1})+w_l\kappa^\prime (\bs{x}^i_{k,1}-u^i_{k,1})$ \\ $\tilde{a}^{i,\ge}_{k,2}=w_u g(l^i_{k,1})+w_u\kappa^\prime (\bs{x}^i_{k,1}-l^i_{k,1})$\\ $\tilde{l}^{i}_{k,2}= w_l g(u^i_{k,1})$, $\tilde{u}^i_{k,2}= w_u g(l^i_{k,1})$} \\  \midrule

         & $l^i_{k,1} \ge 0$ &  \tabincell{l}{$\tilde{a}^{i,\le}_{k,2}= w_l g(l^i_{k,1})+w_l\kappa (\bs{x}^i_{k,1}-l^i_{k,1})$ \\ $\tilde{a}^{i,\ge}_{k,2}=w_u g(u^i_{k,1})+w_u\kappa^\prime (\bs{x}^i_{k,1}-u^i_{k,1})$ \\ $\tilde{l}^{i}_{k,2}= w_l g(l^i_{k,1})$, $\tilde{u}^i_{k,2}= w_u g(u^i_{k,1})$} & 
        \tabincell{l}{$\tilde{a}^{i,\le}_{k,2}= w_l g(u^i_{k,1})+w_l\kappa^\prime (\bs{x}^i_{k,1}-u^i_{k,1})$ \\ $\tilde{a}^{i,\ge}_{k,2}=w_u g(l^i_{k,1})+w_u\kappa (\bs{x}^i_{k,1}-l^i_{k,1})$ \\ $\tilde{l}^{i}_{k,2}= w_l g(u^i_{k,1})$, $\tilde{u}^i_{k,2}= w_u g(l^i_{k,1})$} \\  \cline{2-4}

        $\text{Tanh}(x)$ & $u^i_{k,1} \le 0$ & \tabincell{l}{$\tilde{a}^{i,\le}_{k,2}= w_u g(l^i_{k,1})+w_u\kappa^\prime (\bs{x}^i_{k,1}-l^i_{k,1})$ \\ $\tilde{a}^{i,\ge}_{k,2}=w_l g(u^i_{k,1})+w_l\kappa (\bs{x}^i_{k,1}-u^i_{k,1})$ \\ $\tilde{l}^{i}_{k,2}= w_u g(l^i_{k,1})$, $\tilde{u}^i_{k,2}= w_l g(u^i_{k,1})$} & \tabincell{l}{$\tilde{a}^{i,\le}_{k,2}= w_u g(u^i_{k,1})+w_u\kappa (\bs{x}^i_{k,1}-u^i_{k,1})$ \\ $\tilde{a}^{i,\ge}_{k,2}=w_l g(l^i_{k,1})+w_l\kappa^\prime (\bs{x}^i_{k,1}-l^i_{k,1})$ \\ $\tilde{l}^{i}_{k,2}= w_u g(u^i_{k,1})$, $\tilde{u}^i_{k,2}= w_l g(l^i_{k,1})$} \\ \cline{2-4}

        & $l^i_{k,1} < 0 < u^i_{k,1}$ &  \tabincell{l}{$\tilde{a}^{i,\le}_{k,2}= w_u g(l^i_{k,1})+w_l\kappa^\prime (\bs{x}^i_{k,1}-l^i_{k,1})$ \\ $\tilde{a}^{i,\ge}_{k,2}=w_u g(u^i_{k,1})+w_l\kappa^\prime (\bs{x}^i_{k,1}-u^i_{k,1})$\\ $\tilde{l}^{i}_{k,2}= w_u g(l^i_{k,1})$, $\tilde{u}^i_{k,2}= w_u g(u^i_{k,1})$} & 
         \tabincell{l}{$\tilde{a}^{i,\le}_{k,2}= w_l g(u^i_{k,1})+w_u\kappa^\prime (\bs{x}^i_{k,1}-u^i_{k,1})$ \\ $\tilde{a}^{i,\ge}_{k,2}=w_l g(l^i_{k,1})+w_u\kappa^\prime (\bs{x}^i_{k,1}-l^i_{k,1})$\\ $\tilde{l}^{i}_{k,2}= w_l g(u^i_{k,1})$, $\tilde{u}^i_{k,2}= w_l g(l^i_{k,1})$} \\
         
        \bottomrule
    \end{tabular}}
    \label{tab:log_AbT}
\end{table}

\begin{theorem}\label{the:sigmoid}
    \major{Both the weighted Sigmoid and the weighted Tanh abstract transformers are sound and preserve the invariant $\gamma(\bs{a}^{i,\ast}_{k,2})\subseteq[\tilde{l}^i_{k,2}, \tilde{u}^i_{k,2}]$. \hfill \qed}
\end{theorem}

\major{For the MILP encoding of other activation functions, the piecewise linear approximation can be employed to encode the sigmoid and tanh functions using linear constraints. We argue that such an approximation-based MILP encoding approach is sound, however, not incomplete,
Therefore, for these logistic activation functions, we can only claim that
our approach is sound but incomplete, and limit our focus to the {\sf BFA\_RA} component.}


\subsubsection{\major{Other network architectures}}
\begin{figure*}[t]
	\centering
        \subfigure[A convolutional layer example under BFAs.]{\label{fig:conv_BFA}
		\begin{minipage}[b]{0.44\textwidth}
			\includegraphics[width=1.0\textwidth]{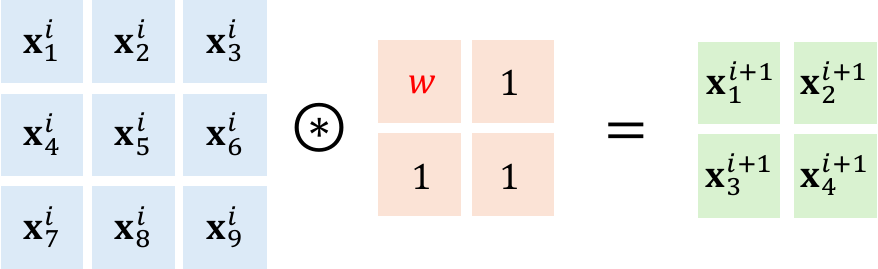}
		\end{minipage}	
	}\hspace{6mm}
        \subfigure[The transformed affine layer under BFAs.]{\label{fig:conv2aff_BFA}
		\begin{minipage}[b]{0.44\textwidth}
			\includegraphics[width=1.0\textwidth]{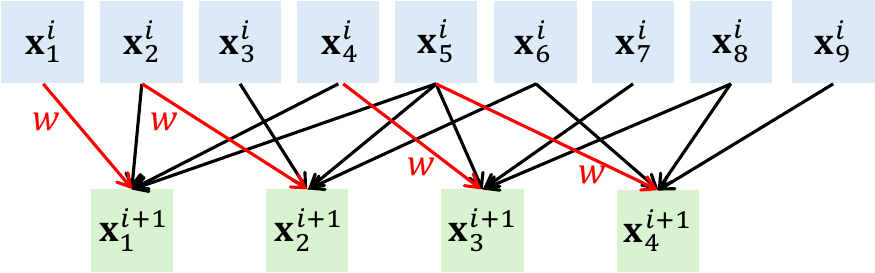}
		\end{minipage}	
	}\vspace{-3mm}
        \caption{\major{An example of affine transformation from a convolutional computation under bit-flip attacks. Given a $3\times 3$ feature layer $\bs{x}^i$ and a $2\times 2$ convolution filter with one parameter attacked (denoted as $w$) and the other three parameters to be 1 in Figure~\ref{fig:conv_BFA}, we can always get an equivalent affine layer as shown in Figure~\ref{fig:conv2aff_BFA}.}}\vspace{-2mm}
        \Description{The affine transformation from a convolutional computation under bit-flip attacks.}
    \label{fig:conv2aff}
\end{figure*}

\major{This work focuses on feed-forward network architectures, however, our approach can be generalized to shared-parameter architectures—such as convolutional networks—without additional technical challenges. Figure~\ref{fig:conv2aff} illustrates how a convolutional layer subjected to bit-flip attacks on the parameter $w$ can be transformed into an equivalent affine layer, to which \tool can be directly applied. Note that, in Figure~\ref{fig:conv2aff_BFA}, although multiple copies are under bit-flip attack, they share the same parameter $w$ in the original convolutional layer, and consequently, the attack effect is identical. Therefore, no additional combinatorial explosion occurs and the computational complexity remains equivalent to the case when $\mm=1$.} 

\major{Although no additional technical challenges occurs, \tool may suffer from significant abstraction precision loss on CNNs, compared to DNNs, due to multiple abstractions in weighted activation function and input neurons even when $\mm=1$ (in contrast to only single abstraction in feedforward networks), both contributing to higher loss than in cases without symbolic weighting.}

\subsubsection{\major{Other quantization schemes and network precisions}}
\major{\tool can be adapted to support other quantization schemes. For instance, when addressing a mixed-precision quantization scheme, only the weight interval associated with each symbolic parameter under bit-flip attacks needs to be adjusted for the BFA\_RA procedure, while no modifications are required for BFA\_MILP. }

\major{For floating-point neural networks (FPNNs), parameters are typically stored as IEEE 754 32-bit single-precision floating-point numbers, where flipping the exponent bit can cause drastic value changes (e.g., altering 0.078125 to $1.25\times 2^{124}$). \tool can be adapted to FPNNs by adjusting the parameter interval derived from bit-flipping, but the performance remains uncertain and is left for future work. 
Indeed, studies~\cite{1bitallyouneed,liyes} have shown that FPNNs are highly vulnerable to BFAs. Given this inherent vulnerability, we argue that verifying BFAs on QNNs is more reasonable, as FPNNs can be almost always compromised. Instead of formal verification, demonstrating vulnerabilities through attacks or testing is a more practical and insightful approach for FPNNs.}

\smallskip
\noindent
{\bf Clarification.}  Our method, \tool, is primarily designed to: i) effectively and efficiently prove desirable property of a given neural network under BFA and ii) or identify a relatively tight superset of vulnerable parameters that must be protected to avoid BFA. Note that the latter allows existing model integrity protection methods to be applied in a more cost-effectively way, i.e., by protecting only the identified vulnerable parameters (e.g., 0.01\% of parameters in a network) rather than all parameters. Additionally, we consider the robustness of input regions against BFA, which is more interesting yet more challenging than the robustness of individual inputs. Note that it is virtually impossible to enumerate and verify all attack vectors for an input region and attackers may use \tool to identify critical bits/parameters to flip as well as an individual input.


\section{Implementation and Evaluation}\label{sec:exp}

To validate the effectiveness of our method, we aim to answer the following research questions:

\begin{enumerate}[label=\textbf{RQ\arabic*. },itemindent=*,leftmargin=*,itemsep=0pt]
    \item How effective and efficient is {\sf BFA\_RA} for providing the sound verification result on potential attack vectors, compared with the naive baseline method (cf. Section~\ref{sec:naiveMethod})?
    
    \item Can the absence of BFAs be verified with a conclusive result for a specific network using \tool, and how effective is the MILP-based method for providing a sound and complete verification result, as a complementary approach to {\sf BFA\_RA}?
    
    \item How efficient and effective is \tool for verifying the absence of BFAs on larger networks with various activation functions?
\end{enumerate}

\noindent
{\bf Implementation.} We implemented our verification method as an end-to-end tool \tool with Gurobi~\cite{Gurobi} as the back-end MILP solver. The \symPoly \minor{component} is built upon GPUPoly~\cite{gpupoly}, an open-source GPU implementation~\cite{muellch2023eth} of \deepPoly. The quantization follows the scheme mentioned in Section~\ref{sec:quant}. During the network inference procedure and the verification process in \tool, quantized (fixed-point) parameters are stored in the IEEE 754~\cite{FPIEEE754} floating-point number format for arithmetic operation. The floating-point number soundness flag in the implementation~\cite{muellch2023eth} of GPUPoly \cite{gpupoly} is turned on for both \deepPoly and \symPoly.

\smallskip
\noindent
{\bf Datasets.} We use MNIST~\cite{MNIST} and ACAS Xu~\cite{julian2019deep} as the datasets in our experiments. MNIST contains 60,000 grayscale handwritten digits (from 0 to 9) of the size of $28\times 28$. ACAS Xu is a safety-critical system designed to provide collision avoidance advisories for unmanned aircraft.

\smallskip
\noindent
{\bf Networks.} For the MNIST dataset, we train 12$\times$2 QNNs following the post-training quantization scheme~\cite{gholami2021survey,li2024investigating} on the MNIST~\cite{MNIST} dataset, which is a common practice in prior research to improve the robustness against bit-flip attack \cite{HARDeNN,randomDNN}. We evaluate 12 architectures with varying model sizes and 2 quantization bit-widths $Q\in\{4,8\}$, using ReLU activations by default. The details of the QNNs are listed in Tables~\ref{tab:bench_A} and \ref{tab:bench_B}. The first row shows the architecture of each QNN, where $x$blk\_$y$ means that the network has $x$ hidden layers with each hidden layer containing $y$ neurons. Row 2 shows the number of parameters in these networks and Rows 3-4 give the accuracy of these networks under different quantization bit-width, i.e., $Q=4$ and $Q=8$.
Moreover, we consider two additional networks of architecture $3\text{blk}\_100$ with Sigmoid and Tanh activation functions.

For the ACAS Xu dataset, although the authors in~\cite{KBDJK17} provide 45 neural networks trained on this dataset along with 10 safety properties. We find that few of these properties can be proved via \deepPoly on these networks. Hence, in this work, we adopt retrained ones instead of the original networks from~\cite{longNew} as our benchmark. These retrained 45 networks adopt the same architecture as~\cite{KBDJK17}, i.e., 6blk\_50, and maintain comparable accuracy to the original networks (86.6\% on average). These networks output a score for five different actions: clear-of-conflict (COC), weak left (WL), weak right (WR), strong left (SL), and strong right (SR). Based on these, we built 45 QNNs following a post-training quantization scheme, setting the quantization bit-width as $Q=8$.


    


\begin{table*}[t]
    \centering
    \caption{QNNs obtained via quantization-aware training on MNIST with small network architectures.}\label{tab:bench_A}
    \setlength{\tabcolsep}{3pt}
    \scalebox{0.88}{
    \begin{tabular}{c|c|c|c|c|c|c|c}
    \toprule
    \multicolumn{2}{c|}{ArchSmall} & 3blk\_10 & 3blk\_30 & 3blk\_50 & 5blk\_10 & 5blk\_30 & 5blk\_50 \\ \midrule

    \multicolumn{2}{c|}{\# Param.} & 8.28k & 26.62k & 47.36k & 8.5k & 28.48k & 52.46k \\ \midrule
    
    & $Q=4$ & 81.09\% & 94.40\% & 95.96\% & 86.80\% & 95.63\% & 96.44\% \\
    \multirow{-2}*{Acc.} & $Q=8$ & 92.83\% & 96.38\% & 96.38\% & 92.87\% & 96.95\% & 97.20\%  \\

    \bottomrule
    \end{tabular}}
\end{table*}

\begin{table*}[t]
    \centering
    \caption{QNNs obtained via post-training quantization on MNIST with large network architectures.}\label{tab:bench_B}
    \setlength{\tabcolsep}{3pt}
    \scalebox{0.88}{
    \begin{tabular}{c|c|c|c|c|c|c|c|c|c}
    \toprule
    \multicolumn{2}{c|}{ArchLarge} & 3blk\_100& 3blk\_100$^\text{sigmoid}$ & 3blk\_100$^\text{tan}$ & 3blk\_512 & 3blk\_1024 & 5blk\_100 & 5blk\_512 & 5blk\_1024  \\ \midrule

    \multicolumn{2}{c|}{\# Param.} & 109.7k & 109.7k & 109.7k & 1,195k & 3,962k & 129.9k & 1,720k & 6,061k \\ \midrule
    
    & $Q=4$ 
      & 97.03\%  
      & 97.24\%  
      & 97.43\%  
      & 97.85\%  
      & 97.72\%  
      & 97.31\%  
      & 97.62\%  
      & 97.97\%  
      \\ 
    \multirow{-2}*{Acc.} 
    & $Q=8$ 
      & 97.53\%  
      & 97.71\%  
      & 97.56\%  
      & 98.18\%  
      & 97.84\%  
      & 97.39\%  
      & 97.74\%  
      & 98.06\%  
      \\
      
    \bottomrule
    \end{tabular}}
\end{table*}

\smallskip
\noindent
{\bf Experimental setup.} For the BFA-tolerant robustness verification problem defined in Section~\ref{sec:proDef}, we randomly selected 20 inputs from the MNIST dataset for each network. We considered 3 different attack radii, $r\in\{0,2,4\}$, for each input, resulting in 60 input regions for each network. Note that all these input regions are robust to the corresponding QNNs until the bit-flip attacks.
For the ACAS Xu benchmark, we test all 45 QNNs on the 10 properties and select the successfully proved 55 network-property pairs as our benchmarks. The details are given in Table \ref{tab:bench_C} in Appendix~\ref{sec:app-bench}.
We set the maximum number of bit flips as $\nn\in \{1,2,4\}$. 
Unless otherwise noted, each BFA\_RA task is conducted on an NVIDIA Tesla V100 accelerator and each BFA\_MILP task is conducted with 30 threads on a computer equipped with AMD EPYC 7742 64-core processors. The time limit for each verification task is 1 hour by default, \minor{considering the large number of tasks (thousands) in the subsequent experiments.}



\subsection{The Effectiveness and Efficiency of {\sf BFA\_RA}}\label{sec:RQ1}
To answer \textbf{RQ1}, for each network listed in Tables~\ref{tab:bench_A} and~\ref{tab:bench_B}, we randomly select 100 weight parameters and up to 100 bias parameters per layer for manipulation by the attacker, considering the huge amount of parameters in these networks.
For each verification task of network with architecture $x$blk\_$y$ and quantized by $Q$ bits, there are $K\cdot\sum_{i=1}^{\nn}{Q \choose i}$ $(1,\nn)$-attack vectors, where $K=100(x+1)+x\min(y,100)+10$. In total, we have $28\times 60\times 3=5040$ verification tasks (28 networks, 60 input regions per network, and 3 different values of $\nn$) for MNIST.


\begin{table}[t]
    \centering
    \caption{\minor{Verification results of {\sf BFA\_RA}, {\sf BFA\_RA} w.o binary search, and the naive method. Each entry shows
    the proportion of parameters that are proved to be safe/unknown by two compared methods. For example, the entry in the top left corner indicates that when $r=0, \nn=1$, there are 98.55\% parameters across all verification tasks on all QNNs that are both proved as safe by the naive method and the {\sf BFA\_RA} method.
    The bottom left corner entry indicates 0.03\% of parameters are proven as unknown by the naive method but proved as safe by BFA\_RA, considering 8-bit quantization and robustness radius 4.}}
    \label{tab:RA_effect}
    \setlength{\tabcolsep}{2pt}
{
    \scalebox{0.85}{
\begin{tabular}{cc|ccc|ccc|ccc|ccc}
\toprule
&  & \multicolumn{6}{c|}{\sf BFA\_RA} & \multicolumn{6}{c}{\sf BFA\_RA w.o. Binary Search}  \\ 
\cline{3-14}
\multicolumn{2}{c|}{$Q=4$} 
& \multicolumn{3}{c|}{\#Safe\_Paras} 
& \multicolumn{3}{c|}{\#Unknown\_Paras} 
& \multicolumn{3}{c|}{\#Safe\_Paras} 
& \multicolumn{3}{c}{\#Unknown\_Paras} \\

& & $\nn=1$  & $\nn=2$ & $\nn=4$  & $\nn=1$ & $\nn=2$  & $\nn=4$
& $\nn=1$  & $\nn=2$ & $\nn=4$  & $\nn=1$ & $\nn=2$  & $\nn=4$  \\ 
\midrule

\multirow{3}{*}{\makecell{\minor{Naive Method}\\ \minor{\#Safe\_Paras}}}    & $r=0$ 
& 98.55\% & 98.13\% & 98.07\%
& 0.00\% & 0.00\% & 0.00\%  
& 98.51\% & 98.00\% & 97.89\% 
& 0.04\% & 0.14\% & 0.17\% \\

 & $r=2$
& 97.98\% & 97.48\% & 97.42\%
& 0.00\% & 0.00\% & 0.00\% 
& 97.92\% & 97.33\% & 97.22\%
& 0.06\% & 0.16\% & 0.20\% \\

                & $r=4$
& 95.94\% & 95.19\% & 95.06\%
& 0.00\% & 0.00\% & 0.00\% 
& 95.70\% & 94.71\% & 94.51\%
& 0.24\% & 0.48\% & 0.55\% \\
\midrule

\multirow{3}{*}{\makecell{\minor{Naive Method}\\ \minor{\#Unknown\_Paras}}}    & $r=0$ 
& 0.00\% & 0.00\% & 0.00\%
& 1.45\% & 1.87\% & 1.93\% 
& 0.00\% & 0.00\% & 0.00\%
& 1.45\% & 1.87\% & 1.93\% \\ 

& $r=2$
& 0.00\% & 0.01\% & 0.01\%
& 2.02\% & 2.51\% & 2.57\% 
& 0.00\% & 0.00\% & 0.00\%
& 2.02\% & 2.52\% & 2.58\% \\ 

                & $r=4$
& 0.02\% & 0.06\% & 0.08\%
& 4.05\% & 4.75\% & 4.85\% 
& 0.02\% & 0.04\% & 0.04\%
& 4.05\% & 4.77\% & 4.89\% \\ 
\bottomrule
\bottomrule

&  & \multicolumn{6}{c|}{\sf BFA\_RA} & \multicolumn{6}{c}{\sf BFA\_RA w.o. Binary Search}  \\ 
\cline{3-14}
\multicolumn{2}{c|}{$Q=8$} 
& \multicolumn{3}{c|}{\#Safe\_Paras} 
& \multicolumn{3}{c|}{\#Unknown\_Paras} 
& \multicolumn{3}{c|}{\#Safe\_Paras} 
& \multicolumn{3}{c}{\#Unknown\_Paras} \\
        
& & $\nn=1$  & $\nn=2$ & $\nn=4$  & $\nn=1$ & $\nn=2$  & $\nn=4$
& $\nn=1$  & $\nn=2$ & $\nn=4$  & $\nn=1$ & $\nn=2$  & $\nn=4$  \\ 
\midrule

\multirow{3}{*}{\makecell{\minor{Naive Method}\\ \minor{\#Safe\_Paras}}}    & $r=0$

& 99.03\% & 98.59\% & 98.41\%
& 0.00\% & 0.00\% & 0.00\% 
& 99.02\% & 98.54\% & 98.33\%
& 0.01\% & 0.06\% & 0.08\% \\ 

  & $r=2$
& 98.41\% & 97.89\% & 97.66\%
& 0.00\% & 0.00\% & 0.00\% 
& 98.39\% & 97.83\% & 97.57\%
& 0.02\% & 0.06\% & 0.09\% \\

                & $r=4$
& 96.19\% & 95.43\% & 95.11\%
& 0.00\% & 0.00\% & 0.00\% 
& 95.97\% & 94.95\% & 94.53\%
& 0.21\% & 0.48\% & 0.59\% \\
\midrule

\multirow{3}{*}{\makecell{\minor{Naive Method}\\ \minor{\#Unknown\_Paras}}}    & $r=0$
& 0.00\% & 0.00\% & 0.00\%
& 0.97\% & 1.41\% & 1.59\% 
& 0.00\% & 0.00\% & 0.00\%
& 0.97\% & 1.41\% & 1.59\% \\

& $r=2$
& 0.00\% & 0.00\% & 0.01\%
& 1.59\% & 2.11\% & 2.33\% 
& 0.00\% & 0.00\% & 0.01\%
& 1.59\% & 2.11\% & 2.34\% \\

                & $r=4$
& 0.03\% & 0.08\% & 0.11\%
& 3.78\% & 4.49\% & 4.78\% 
& 0.02\% & 0.05\% & 0.06\%
& 3.79\% & 4.52\% & 4.82\% \\
\bottomrule
\end{tabular}
}
}

\end{table}

\begin{table}[t]
    \centering
    \caption{The computation time (in GPU hours) of the three methods.}
    \scalebox{0.88}{
    \begin{tabular}{c|c|c|c}
        \toprule
           & Naive Method & {\sf BFA\_RA} & {\sf BFA\_RA} w.o. Binary Search  \\ \midrule
          $Q=4$ & {$\approx$72.9h} & {$\approx $14.2h} &  {$\approx$13.6h} \\ \midrule
        $Q=8$ & {$\approx $488.2h} & {$\approx $14.8h} & {$\approx $14.0h} \\ 

        \bottomrule
    \end{tabular}}
    \label{tab:RA_time}
\end{table}

We compare the performance of {\sf BFA\_RA} with the naive method mentioned in Section~\ref{sec:naiveMethod}. Recall that, given a parameter, the naive method performs one reachability analysis for each $(1,\nn)$-attack vector
to check robustness, whereas {\sf BFA\_RA} performs one reachability analysis for all the possible $(1,\nn)$-attack vectors. Thus, {\sf BFA\_RA} is expected to reduce execution time significantly. To evaluate the trade-off between efficiency and effectiveness in {\sf BFA\_RA} compared to the naive method, we analyze the effectiveness loss/gain in Table~\ref{tab:RA_effect} and the efficiency gain of {\sf BFA\_RA} in Table~\ref{tab:RA_time}, \minor{where
\begin{itemize}
   \item Parameters proved as safe by the naive method but unknown by BFA\_RA indicate an \textbf{effectiveness loss} of {\sf BFA\_RA}, i.e., row \#Safe\_Paras and column \#Unknown\_Paras.
    \item Parameters proved as safe by both methods demonstrate the \textbf{effectiveness maintenance} of {\sf BFA\_RA},
    i.e., row \#Safe\_Paras and column \#Safe\_Paras.
    \item Parameters proved as safe by {\sf BFA\_RA} but unknown by the naive method represent an \textbf{effectiveness gain} of {\sf BFA\_RA}, i.e., row \#Unknown\_Paras and column \#Safe\_Paras.
    \item Parameters proved as unknown by both methods reveals the \textbf{limitations} of reachability analysis by both methods, i.e., row \#Unknown\_Paras and column \#Unknown\_Paras.
\end{itemize}}
Furthermore, to assess the effectiveness of the binary search strategy proposed in Section~\ref{sec:bfa_ra} for {\sf BFA\_RA}, we implement a variant that excludes binary search, referred to as {\sf BFA\_RA} w.o. Binary Search, where {\sf binary\_RA}$(\mN,\mI, g, w,w^+_l,w^+_u)$ (resp. {\sf binary\_RA}$(\mN,\mI,g,w,w^-_l,w^-_u)$) at line 3 (resp. line 5) in Algorithm~\ref{alg:verifyPolyR} is replaced by $\symPoly(\mN,\mI,g,w,w^+_l,w^+_u)$ (resp. $\symPoly(\mN,\mI,g,w, w^-_l,w^-_u)$). The corresponding experimental results are also given in Tables~\ref{tab:RA_effect} and \ref{tab:RA_time}.


\subsubsection{\minor{Effectiveness of {\sf BFA\_RA}}} 
By analyzing the experimental results reported in Table~\ref{tab:RA_effect}, we observe that {\sf BFA\_RA} consistently achieve an effectiveness gain over the naive method across all settings, including various radii of input perturbation $r$ and maximal numbers of bit to flip $\nn$. Specifically, in cases where the naive method reports parameters as unknown (\#Unknown\_Paras), {\sf BFA\_RA} successfully verifies additional parameters as safe (\#Safe\_Paras), albeit with a relatively modest gain (up to 0.11\% on average). \major{Indeed, it is reasonable that {\sf BFA\_RA} achieves a relatively small effectiveness gain. In the worst case, {\sf BFA\_RA} performs a binary search over each potential flipped weight value, similar to exhaustive traversal, making it at least as effective as the naive method. However, unlike the naive method, {\sf BFA\_RA} initially treats each symbolic weight as a range and then partitions it using a binary approach. This introduced weight range may affect the reachability analysis of non-input neurons, leading to different abstract elements, i.e., distinct value domains obtained for each neuron between the naive method and \symPoly. In this setting, if a counterexample (a neuron value leading to a successful BFA) falls within the abstract element domain obtained by \deepPoly in the naive method but not within that of \symPoly, then {\sf BFA\_RA} may exhibit an effectiveness gain. However, we argue that such cases should be rare, leading to a limited overall effectiveness gain of {\sf BFA\_RA} over the naive method.}

Furthermore, we also find that the binary-search-free variant of {\sf BFA\_RA} demonstrates a reduction in effectiveness compared to {\sf BFA\_RA}, as certain parameters that are proved as safe by both the naive method and {\sf BFA\_RA} are proved as unknown by the variant. 
For example, when $Q=8$, $r=4$, and $\nn=4$, a total of 0.59\% of the parameters that are verified as safe by both the naive method and {\sf BFA\_RA} are proved as unknown by the binary-search-free variant of {\sf BFA\_RA}.

\vspace{1mm}
\noindent
\setlength{\fboxsep}{3pt}
\setlength{\fboxrule}{1pt}
\fcolorbox{gray!90}{gray!05}{%
    \parbox{0.97\columnwidth}{
        \textbf{Result 1:} {\sf BFA\_RA} demonstrates an effectiveness gain over both the naive method and its binary-search-free variant, albeit with a relatively modest improvement.
    }
}
\vspace{1mm}

By comparing the experimental results between the naive method and binary-search-free variant of {\sf BFA\_RA}, we find that only a small proportion of parameters (up to 0.55\% for $Q=4$ and 0.59\% for $Q=8$) are verified as safe by the naive method but remain unknown by the binary-search-free variant. Recall that the binary-search-free variant performs reachability analysis based on two intervals $[w^+_l,w^+_u]$ and $[w^-_l,w^-_u]$ (cf. line 2 in Algorithm~\ref{alg:verifyPolyR}), to approximate the reachability analysis result under bit-flip attack. The observed comparison results indicate that the abstract domain proposed in \symPoly effectively captures the impact of bit-flip operations on the corresponding parameters with high accuracy.  

\vspace{1mm}
\noindent
\setlength{\fboxsep}{3pt}
\setlength{\fboxrule}{1pt}
\fcolorbox{gray!90}{gray!05}{%
    \parbox{0.97\columnwidth}{
        \textbf{Result 2:} The abstract domain proposed in \symPoly is relatively accurate in approximating the bit-flip operations.
    }
}

\subsubsection{Efficiency of {\sf BFA\_RA}}
\begin{figure*}[t]
	\centering
        
        \subfigure[$(Q,\nn)=(8,1)$.]{\label{fig:query_8_1}
		\begin{minipage}[b]{0.3\textwidth}
			\includegraphics[width=1.0\textwidth]{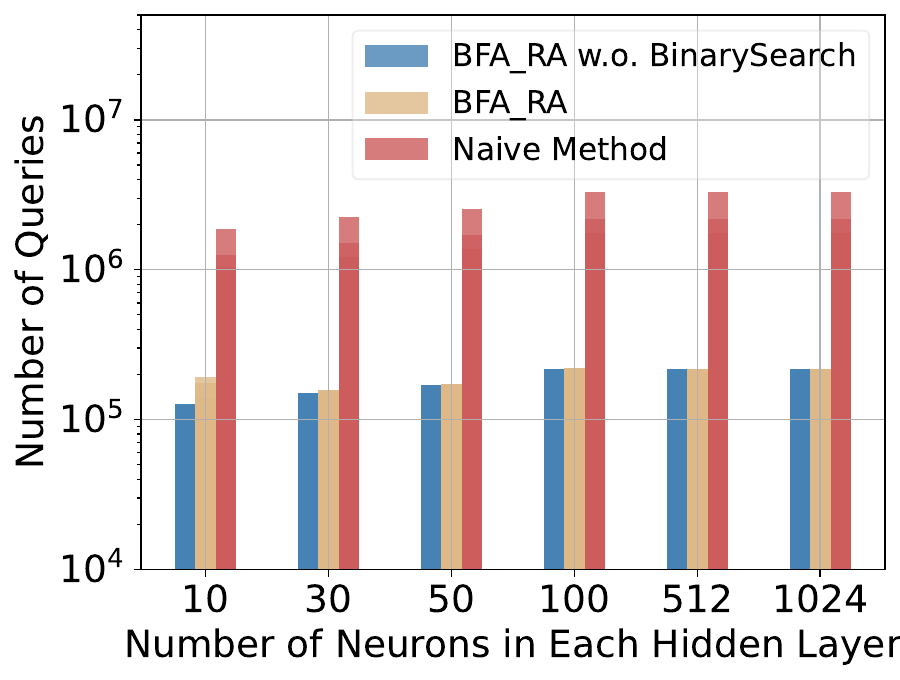}
		\end{minipage}	
	}
        \subfigure[$(Q,\nn)=(8,2)$.]{\label{fig:query_8_2}
		\begin{minipage}[b]{0.3\textwidth}
			\includegraphics[width=1.0\textwidth]{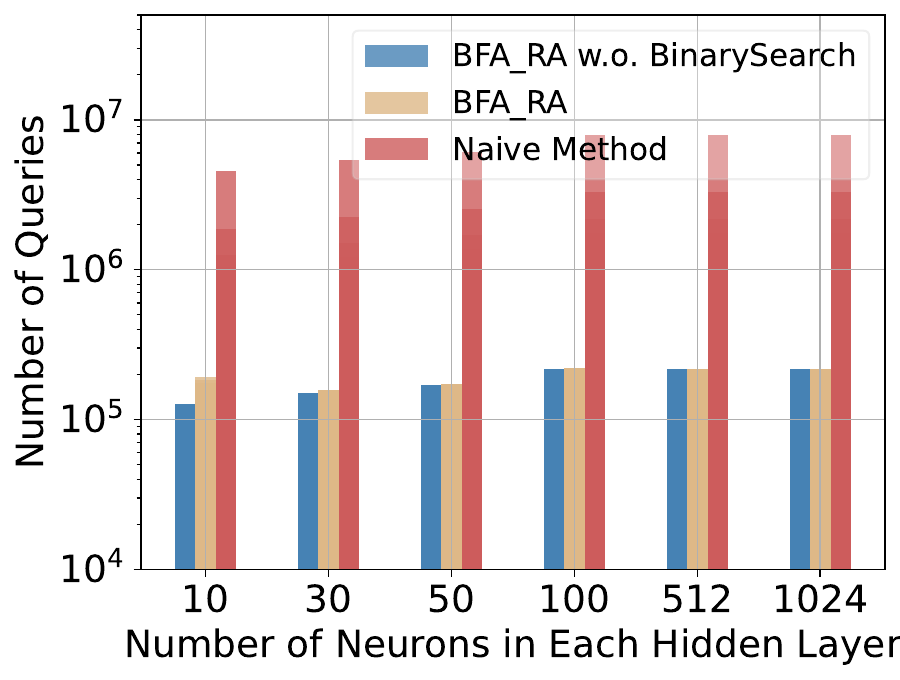}
		\end{minipage}	
	}
        \subfigure[$(Q,\nn)=(8,4)$.]{\label{fig:query_8_4}
		\begin{minipage}[b]{0.3\textwidth}
			\includegraphics[width=1.0\textwidth]{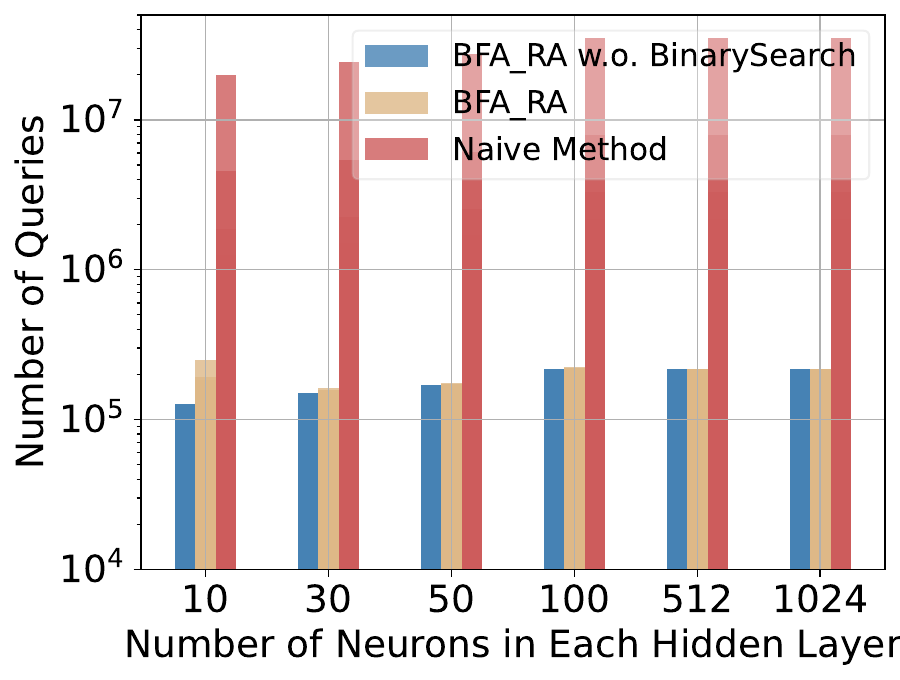}
		\end{minipage}	
	}
        
	\caption{The total number of queries with respect to \deepPoly or \symPoly in the three methods  when $Q=8$.}
	\Description{The total number of queries with respect to \deepPoly or \deepPolyR in the three methods when $Q=8$.}
    \label{fig:queryNum}
\end{figure*}

By comparing the computation time of {\sf BFA\_RA} and the naive method, as given in Table~\ref{tab:RA_time}, we find that {\sf BFA\_RA} consumes significantly less time than the naive method (up to 30x faster). It is noteworthy that the execution time for each query of \deepPoly and \symPoly is nearly identical. Therefore, the number of queries serves as a critical determinant of the overall efficiency of various methods. To illustrate it, we show the total number of queries invoked by the two methods in Figure~\ref{fig:queryNum} (we take $Q=8$ for example) and it can be observed that the naive method invokes an enormous amount of queries of \deepPoly, attributable to the fact that there are up to $K\cdot\sum_{i=1}^{\nn}{Q \choose i}$ queries for each verification task.
On the other hand, although the binary search strategy \minor{slightly} increases the number of \symPoly queries, as shown in Figure~\ref{fig:queryNum}, the execution time of {\sf BFA\_RA} remains comparable to that of its binary-search-free variant (cf. Table~\ref{tab:RA_time}).

\vspace{1mm}
\noindent
\setlength{\fboxsep}{3pt}
\setlength{\fboxrule}{1pt}
\fcolorbox{gray!90}{gray!05}{%
    \parbox{0.97\columnwidth}{
        \textbf{Result 3:} {\sf BFA\_RA} is significantly more efficient than the naive method, achieving up to a 30x speedup. Moreover, it demonstrates comparable efficiency to its binary-search-free variant. 
    }
}


\subsection{Verifying BFAs with \tool}\label{sec:exp_tool_small}
To answer \textbf{RQ2}, in this section, we use \tool to verify the BFA-tolerant robustness property across all small networks outlined in Table~\ref{tab:bench_A}  and all properties listed in Table~\ref{tab:bench_C}. This results in a total of $12\times 60\times 3=2160$ (12 networks, 60 input regions per network, and 3 different values of $\nn$) verification tasks for the MNIST dataset and $55\times 3 =165$ (55 network-property pairs and 3 different values of $\nn$) verification tasks for the ACAS Xu dataset. It is important to note that, for each verification task, we assume that all model parameters, including weights and biases, are vulnerable to bit-flip attacks. Furthermore, we consider each attack to affect only a single parameter at a time, with the attacker potentially altering up to $\nn\in\{1,2,4\}$ bits per attack.
In the following, we define a task as successfully proved by {\sf BFA\_RA} when all parameters in the corresponding network are proved as safe by {\sf BFA\_RA}. Additionally, we consider a task as successfully solved by \tool when it is either proved by {\sf BFA\_RA} or proved/falsified by {\sf BFA\_MILP}.

\begin{table}[t]
    \centering
    \caption{Verification results of \tool on ACAS Xu}\label{tab:tool_acas}
    \setlength{\tabcolsep}{2pt}
    \scalebox{0.8}{
    \begin{tabular}{c|c|c|c|c|c|c|c}
    \toprule
      &  \multicolumn{2}{c|}{{\sf BFA\_RA}} & \multicolumn{2}{c|}{{\sf BFA\_MILP}} & \multicolumn{2}{c|}{AvgTime(s)} & \\ \cline{2-7}
     \multirow{-2}*{Property} & \#Safe\_Paras & \#Proved & \#Proved & \#Falsified & {\sf BFA\_RA} & {\sf BFA\_MILP}& \multirow{-2}*{\#TO} \\ \midrule \rowcolor{gray!20}

     Prop\_3\_WL & 99.8\% & 0  & 0  & 24  & 356.0 & 4.2  & 0 \\ 
     Prop\_3\_WR & 99.9\%  & 0  & 1 & 26  & 354.7  & 4.0  & 0\\ \rowcolor{gray!20}
     Prop\_3\_SL & 99.9\%  & 0 & 0 & 23  & 354.8  & 44.2 & 1 \\
     Prop\_3\_SR & 99.7\%  & 0 & 0  & 27  & 357.5  & 5.9  & 0 \\ \rowcolor{gray!20}
     Prop\_5\_SR & 98.0\%  & 0  & 0  & 20  & 381.2  & 45.3 & 1 \\
     Prop\_10\_COC & 99.5\%  & 11 & 2  & 13  & 365.2  & 227.7  & 16\\
     
    \bottomrule
    \end{tabular}}
\end{table}


     

\begin{figure}[t]
	\centering
	\subfigure[Vulnerable weights.]{\label{fig:acas_weight}
		\begin{minipage}[b]{0.35\textwidth}
			\includegraphics[width=1.0\textwidth]{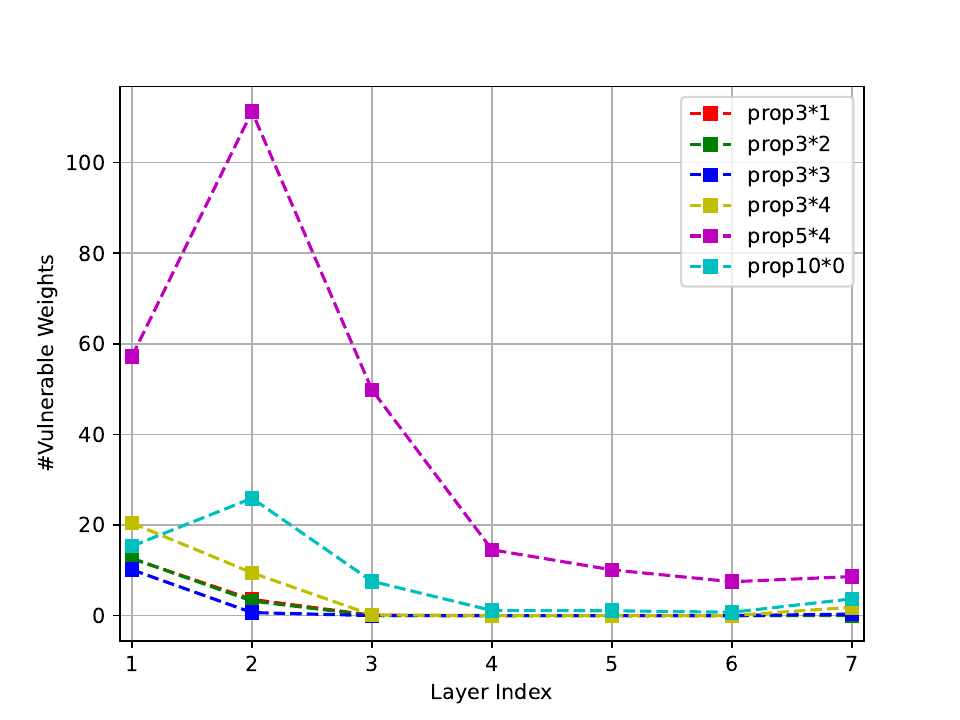}
		\end{minipage}	
	}\hspace{5mm}
        \subfigure[Vulnerable biases.]{\label{fig:acas_bias}
		\begin{minipage}[b]{0.35\textwidth}
			\includegraphics[width=1.0\textwidth]{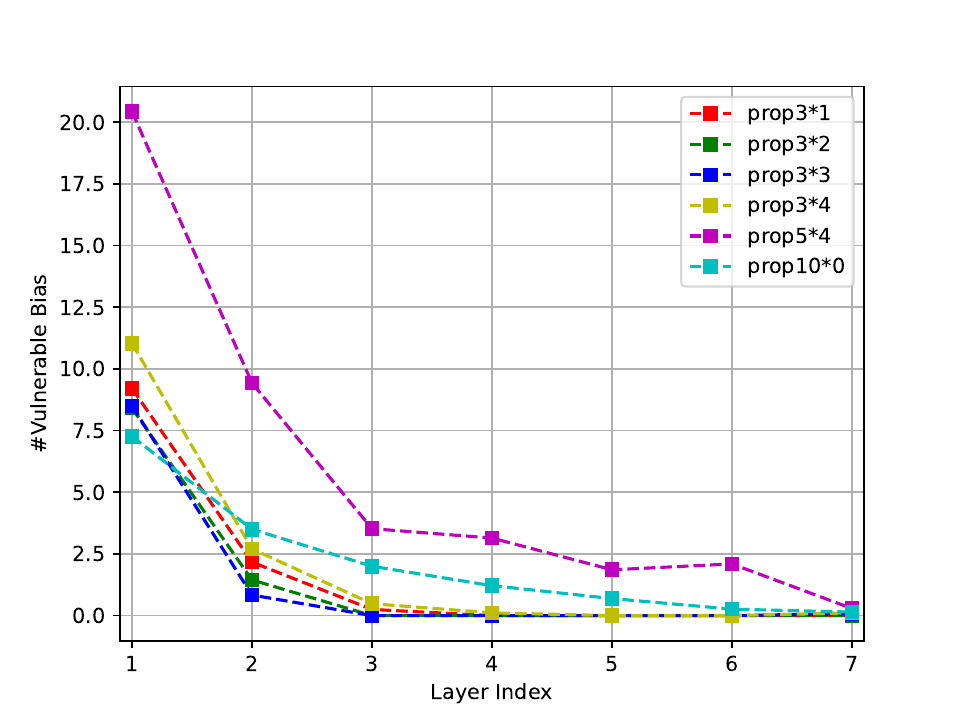}
		\end{minipage}	
	}
        
	\caption{The number of vulnerable parameters detected by {\sf BFA\_RA} in each layer on ACAS Xu.}\vspace{-2mm}
	\Description{The distribution of vulnerable parameters on ACAS Xu.}
    \label{fig:alloPara_acas}
\end{figure}

\subsubsection{ACAS Xu.}\label{sec:exp_tool_small_acas} The results for ACAS Xu are shown in Table~\ref{tab:tool_acas} and Figure~\ref{fig:alloPara_acas}. In Table~\ref{tab:tool_acas}, Column 1 shows the property verified. Column 2 shows the average proportion of parameters that are proved to be safe by {\sf BFA\_RA} across all corresponding networks and three different values of $\nn$. Column 3 gives the number of verification tasks that can be successfully verified by {\sf BFA\_RA}. Columns 4 and 5 display the verification results by {\sf BFA\_MILP}. Columns 6 and 7 give the average computation time for the two methods, and the last column gives the number of verification tasks that run out of time within 1 hour.
We can observe that for ACAS Xu, \tool successfully solved 147 tasks, with 18 tasks running out of time within 1 hour. Among all of these, {\sf BFA\_RA} proves 11 tasks independently. \major{Note that when {\sf BFA\_RA} w.o. binary search is used instead of {\sf BFA\_RA}, the total number of proved tasks via pure reachability analysis decreases by 2, and the total number of timeout tasks (by {\sf BFA\_MILP}) increases by 7. It is because that binary search strategy enables {\sf BFA\_RA} to consistently obtain a tighter value range for each vulnerable parameter, leading to a more compact MILP model (or a reduced solution space) for the {\sf BFA\_MILP} procedure and improving the overall efficiency. Detailed experimental results are presented in Table~\ref{tab:tool_acas_wo_bs} in Appendix~\ref{sec:app_exp}.}

Figures~\ref{fig:acas_weight} and~\ref{fig:acas_bias} show the detailed distribution of vulnerable weights and biases across all 7 non-input layers within the ACAS Xu networks, respectively. On average, we find that the proportion of vulnerable parameters in the earlier layers of the ACAS Xu networks is higher than that in the later layers. This observation suggests that enhanced protection measures should be prioritized for the parameters in the preceding layers to effectively mitigate the impact of bit-flip attacks. One possible reason behind this phenomenon is that the earlier layers of the ACAS Xu networks play a crucial role in feature extraction, making their parameters more susceptible to perturbations caused by BFAs. Since these layers directly influence the representations propagated through the network, any disruption in their parameters can have a cascading effect on the overall network performance, thereby increasing their vulnerability.

\begin{table}[t]
    \centering
    \caption{Verification results of \tool on MNIST for small networks when $(Q,r,\nn)=(8,0,1)$.}\label{tab:tool_mnist_small}
    \setlength{\tabcolsep}{3pt}
    \scalebox{0.85}{
    \begin{tabular}{c|c|c|c|c|c|c|c}
    \toprule
      &  \multicolumn{2}{c|}{{\sf BFA\_RA}} & \multicolumn{2}{c|}{{\sf BFA\_MILP}} & \multicolumn{2}{c|}{AvgTime(s)} & \\ \cline{2-7}
     \multirow{-2}*{Network} & \#Safe\_Paras & \#Proved & \#Proved & \#Falsified & {\sf BFA\_RA} & {\sf BFA\_MILP} & \multirow{-2}*{\#TO} \\ \midrule \rowcolor{gray!20}

    3blk\_10 & 99.7\% & 0 & 0 & 20 & 30.4 & 0.2 & 0  \\ 
    3blk\_30 & 99.9\% & 11 & 0 & 9 & 103.6 & 0.4 & 0 \\ \rowcolor{gray!20}
    3blk\_50 & 99.9\% & 19 & 1 & 0 & 203.6 & 0.7 & 0 \\

    5blk\_10 & 99.2\% & 0 & 0 & 20 & 47.2 & 0.4 & 0 \\ \rowcolor{gray!20}
    5blk\_30 & 99.9\% & 17 & 3 & 0 & 171.9 & 0.4 & 0 \\
    5blk\_50 & 99.9\% & 0 & 0 & 20 & 352.2 & 0.8 & 0 \\
     
    \bottomrule
    \end{tabular}}\vspace{-2mm}
\end{table} 

\begin{figure}[t]
	\centering
	\subfigure[Vulnerable Weights.]{\label{fig:mnist_weight}
		\begin{minipage}[b]{0.35\textwidth}
			\includegraphics[width=1.0\textwidth]{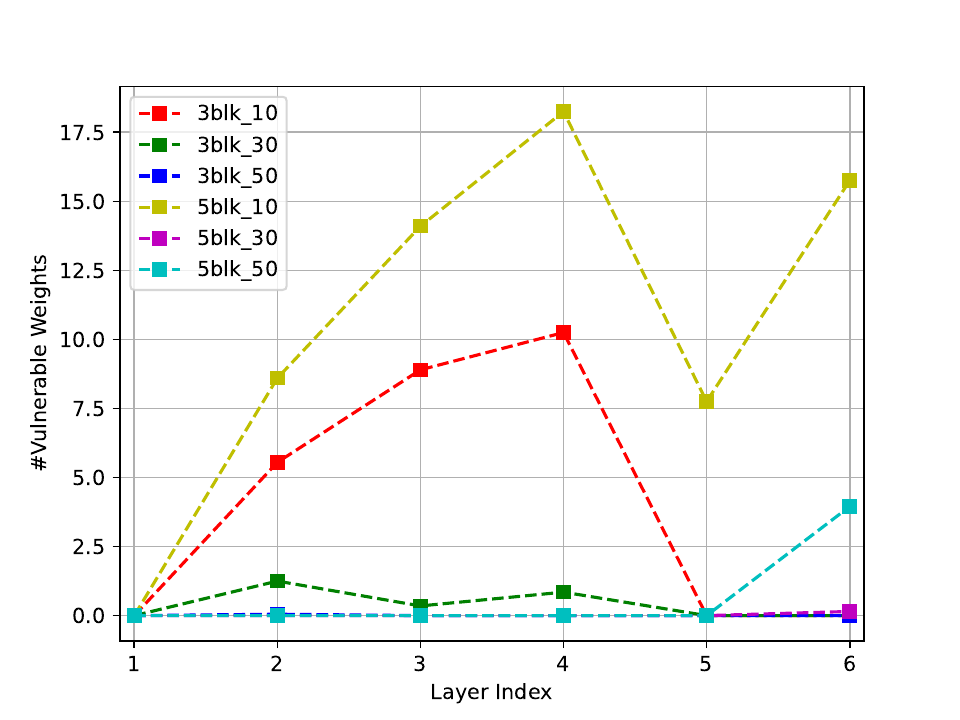}
		\end{minipage}	
	}\hspace{5mm}
        \subfigure[Vulnerable Biases.]{\label{fig:mnist_bias}
		\begin{minipage}[b]{0.35\textwidth}
			\includegraphics[width=1.0\textwidth]{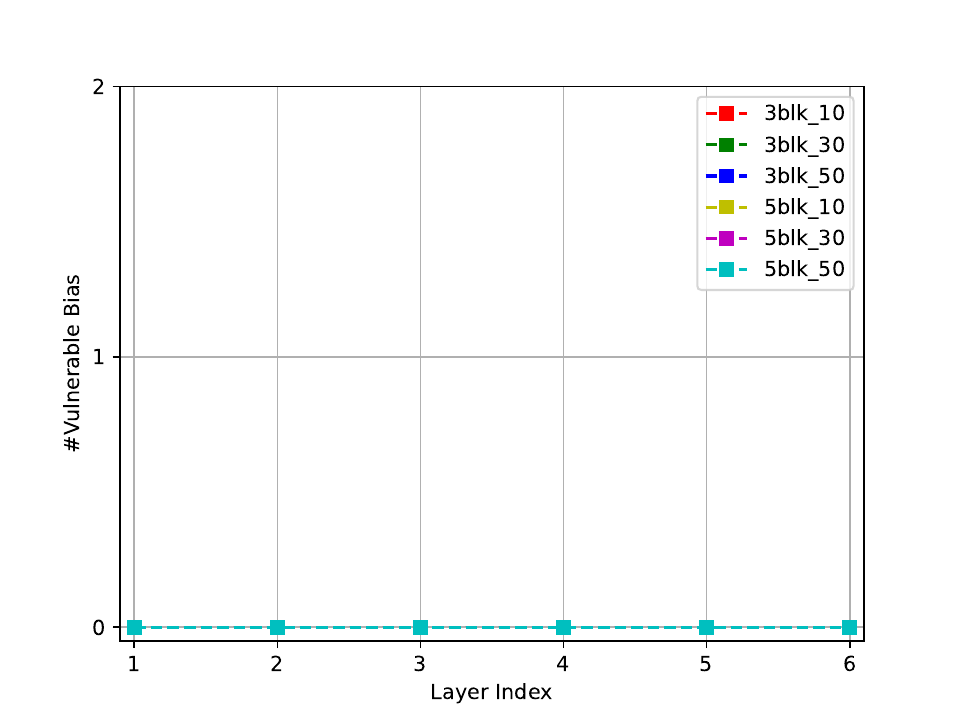}
		\end{minipage}	
	}
        
	\caption{The distribution of vulnerable parameters in each layer on MNIST when $(Q,r,\nn)=(8,0,1)$.}\vspace{-2mm}
	\Description{The distribution of vulnerable parameters in each layer on MNIST dataset when $(Q,r,\nn)=(8,0,1)$.}
    \label{fig:alloPara_mnist}
\end{figure}

\subsubsection{MNIST}
For the MNIST benchmark, \tool successfully verifies the absence of BFAs for all 2160 tasks across all 12 QNNs. Due to space limitations,  
rather than presenting the average results across all possible values of $Q$, $r$, and $\nn$, this section provides detailed verification results for a specific configuration of $(Q,r,\nn)=(8,0,1)$ as an illustrative example. 

The results are presented in Table~\ref{tab:tool_mnist_small} and Figure~\ref{fig:alloPara_mnist}. 
In Table~\ref{tab:tool_mnist_small}, the first column lists the network architecture, while the remaining columns display the same types of results as those shown in Table~\ref{tab:tool_acas}. From Table~\ref{tab:tool_mnist_small}, we can observe that for small networks, {\sf BFA\_RA} achieves relatively high precision, as the majority of tasks (69 out of 73) that fail to be proved by {\sf BFA\_RA} are indeed not robust to bit-flip attacks and subsequently falsified by {\sf BFA\_MILP}.

Figures~\ref{fig:mnist_weight} and~\ref{fig:mnist_bias} give the average number of vulnerable weights and biases in each layer within the 8 networks, respectively. We observe a distinct phenomenon compared to ACAS Xu: in these small networks, the parameters in the middle layers exhibit greater vulnerability to bit-flip attacks. This suggests that greater attention should be given to protecting the middle layer to mitigate the impact of BFAs effectively.
One possible explanation is that given the high input dimensions in these small networks, middle layers play a crucial role in transforming extracted features into high-level representations with fewer redundancy mechanisms to compensate for errors, making them more susceptible to BFAs. 
Furthermore, we find that the bias parameters of these small QNNs for MNIST exhibit significantly greater robustness against BFAs compared to those of QNNs for ACAS Xu.

\vspace{2mm}
\noindent
\setlength{\fboxsep}{3pt}
\setlength{\fboxrule}{1pt}
\fcolorbox{gray!90}{gray!05}{%
    \parbox{0.97\columnwidth}{
        \textbf{Result 4:} \tool can verify the absence of the bit-flip attacks, either prove the BFA-freeness or return a counter-example, and {\sf BFA\_MILP} is effective as a complementary method to {\sf BFA\_RA.}
    }
}
\vspace{2mm}

\subsection{\tool on Larger Networks with Various Activation Functions}\label{sec:exp_tool_large}

To answer RQ3, in this subsection, we evaluate the performance of \tool on the larger networks listed in Table~\ref{tab:bench_B} 
resulting in a total of 
\minor{$16 \times 60 \times 3=2880$ (16 networks, 60
input regions per network, and 3 different values of $\nn$)} verification tasks. Note that, although {\sf BFA\_RA} is more efficient (in polynomial time to the network size) compared to the MILP-based method (which is NP-hard), it is still possible for the MILP-based method to effectively and efficiently verify BFA-tolerant robustness properties when the size of the input region and the vulnerable parameter set, i.e., $|\xi|$ in Algorithm~\ref{alg:overall}, are limited.

\begin{table}[t]
    \centering
    \caption{\major{Verification results of \tool on the MNIST dataset for large networks, where each network undergoes 360 verification tasks in total.}}\label{tab:tool_mnist_diffAct}
    \setlength{\tabcolsep}{3pt}
    \scalebox{0.85}{
    \begin{tabular}{c|c|c|c|c|c}
    \toprule
      \multirow{2}*{} &  {\sf BFA\_RA} & \multicolumn{2}{c|}{{\sf BFA\_MILP}} & \multirow{2}*{\#TO} & \multirow{2}*{\#Solved}\\ \cline{2-4} 
     & \#Proved & \#Proved & \#Falsified & \\ \midrule \rowcolor{gray!20}
    3blk\_100 & 302 & 14 & 10 & 34 & 360 \\ 
    3blk\_100$^\text{sigmoid}$ & 347 & N/A & N/A & N/A & 347 \\ \rowcolor{gray!20}
    3blk\_100$^\text{tanh}$ & 311 & N/A & N/A & N/A & 311 \\
    5blk\_100 & 290 & 6 & 10 & 54 & 306\\ 
    \bottomrule
    \end{tabular}}
\end{table}

\begin{table}[t]
    \centering
    \caption{Detailed verification results of \tool on 3blk\_100 and 5blk\_100 with $Q\in\{4,8\}$.}\label{tab:tool_large_1}
    \setlength{\tabcolsep}{2pt}
    \scalebox{0.8}{
    \begin{tabular}{c|c|c|c|c|c|c|c|c|c}
    \toprule
     &  &  & \multicolumn{2}{c|}{{\sf BFA\_RA}} & \multicolumn{2}{c|}{{\sf BFA\_MILP}} & \multicolumn{2}{c|}{AvgTime(s)} & \\ \cline{4-9}
     \multirow{-2}*{ } & \multirow{-2}*{r} & \multirow{-2}*{$\nn$ } & \#Safe\_Paras & \#Proved & \#Proved & \#Falsified & {\sf BFA\_RA} & {\sf BFA\_MILP} & \multirow{-2}*{\#TO} \\ \midrule 

     &   & 1 & 100.0\% & 40 & 0 & 0 & 731.9 & 0 & 0\\\rowcolor{gray!20}
     \cellcolor{white} & 0 & 2 & 100.0\% & 40 & 0 & 0 & 749.8 & 0 & 0\\ 
     &   & 4 & 100.0\% & 40 & 0 & 0 & 761.9 & 0 & 0\\ \cline{2-10}

     &   & 1 & 100.0\% & 40 & 0 & 0 & 2031.5 & 0 & 0\\\rowcolor{gray!20}
     \cellcolor{white} $Q=4$ & 2 & 2 & 99.9\% & 38 & 1 & 1 & 2035.3 & 6.5 & 0\\ 
     &   & 4 & 99.9\% & 38 & 1 & 1 & 2035.7 & 6.8 & 0\\ \cline{2-10}

     &   & 1 & 99.8\% & 29 & 3 & 0 & 2035.4 & 685.4 & 8 \\ \rowcolor{gray!20}
     \cellcolor{white}  & 4 & 2 & 99.8\% & 19 & 3 & 3 & 2064.1 & 449.6 & 15 \\ 
     &   & 4 & 99.8\% & 19 & 3 & 2 & 2038.3 & 783.0 & 16 \\ 

    \bottomrule
    \bottomrule

    & &  & \multicolumn{2}{c|}{{\sf BFA\_RA}} & \multicolumn{2}{c|}{{\sf BFA\_MILP}} & \multicolumn{2}{c|}{AvgTime(s)} & \\ \cline{4-9}
     \multirow{-2}*{} &  \multirow{-2}*{r} & \multirow{-2}*{$\nn$ } & \#Safe\_Paras & \#Proved & \#Proved & \#Falsified & {\sf BFA\_RA} & {\sf BFA\_MILP} & \multirow{-2}*{\#TO} \\ \midrule

      &  & 1 & 100.0\% & 40 & 0 & 0 & 722.2 & 0 & 0\\ \rowcolor{gray!20}
    \cellcolor{white}  & 0 & 2 & 100.0\% & 40 & 0 & 0 & 745.0 & 0 & 0\\ 
     &   & 4 & 99.9\% & 39 & 0 & 1 & 754.2 & 1.6 & 0\\ \cline{2-10}

     &   & 1 & 99.9\% & 38 & 1 & 1 & 2032.5 & 2.8 & 0\\ \rowcolor{gray!20}
     \cellcolor{white} $Q=8$ & 2 & 2 & 99.9\% & 37 & 1 & 2 & 2038.6 & 6.9 & 0\\ 
     &   & 4 & 99.9\% & 37 & 0 & 3 & 2046.4 & 2.6 & 0\\ \cline{2-10}

     &   & 1 & 99.6\% & 22 & 3 & 1 & 2048.4 & 854.0 & 14 \\ \rowcolor{gray!20}
     \cellcolor{white} & 4 & 2 & 99.5\% & 18 & 3 & 2 & 2091.6 & 1482.3 & 17 \\ 
     &   & 4 & 99.4\% & 18 & 1 & 3 & 2173.7 & 297.7 & 18 \\
    
    \bottomrule
    \end{tabular}}\vspace{-2mm}
\end{table}

We observe that, with the exceptions of networks under the  3blk\_100 and 5blk\_100 architectures,
all other large networks listed in Table~\ref{tab:bench_B} cannot be successfully verified by \tool within 1 hour. \minor{For instance, a verification task for the network of architecture 3blk\_512\_512\_512 with $(Q,r,\nn)=(8,2,1)$ requires approximately 14 hours to complete.} Such failure is largely attributable to the challenges posed by i) the vast number of potential attack vectors and ii) the substantial model sizes when utilizing the MILP-based method, mirroring the issues found in existing MILP-based verification techniques in the context of vanilla network verification problems~\cite{GuyKatz2017ReluplexAE,huang2024towards}.

We report the experimental results in Tables~\ref{tab:tool_mnist_diffAct} and~\ref{tab:tool_large_1}.
%
\major{We find that \tool successfully solves the majority of verification tasks for the four networks in Table~\ref{tab:tool_mnist_diffAct}. For networks with logistic-based activations (3blk\_100$^\text{sig}$ and 3blk\_100$^\text{tanh}$), although \tool can only provide sound verification results by exclusively utilizing {\sf BFA\_RA}, \tool is still able to solve most tasks within the given time limit.} 
\minor{We also find that, compared to small networks, larger networks appear to exhibit greater robustness against BFAs. Specifically, when $r=0$ (cf. Table~\ref{tab:tool_large_1}), almost all BFA-tolerant properties can be successfully verified by {\sf BFA\_RA} solely (except for one property when $(Q,r,\nn)=(8,0,4)$), indicating enhanced resistance to bit-flip attacks within larger network architectures. Moreover, we observe that networks quantized with a lower bit-width tend to exhibit greater robustness against BFAs. This suggests that reduced bit-width quantization may inherently increase the difficulty of executing bit-flip attacks, a finding that aligns with the existing work~\cite{he2020defending}}.

\vspace{2mm}
\noindent
\setlength{\fboxsep}{3pt}
\setlength{\fboxrule}{1pt}
\fcolorbox{gray!90}{gray!05}{%
    \parbox{0.97\columnwidth}{
        \textbf{Result 5:} \tool demonstrates generalizability across various activation functions and scales to $8$-bit QNNs with a 5blk\_100 architecture, completing verification of BFAs involving up to four bit flips within 1 hour.
    }
}
\vspace{2mm}

\begin{figure}
    \centering
    \includegraphics[width=0.6\linewidth]{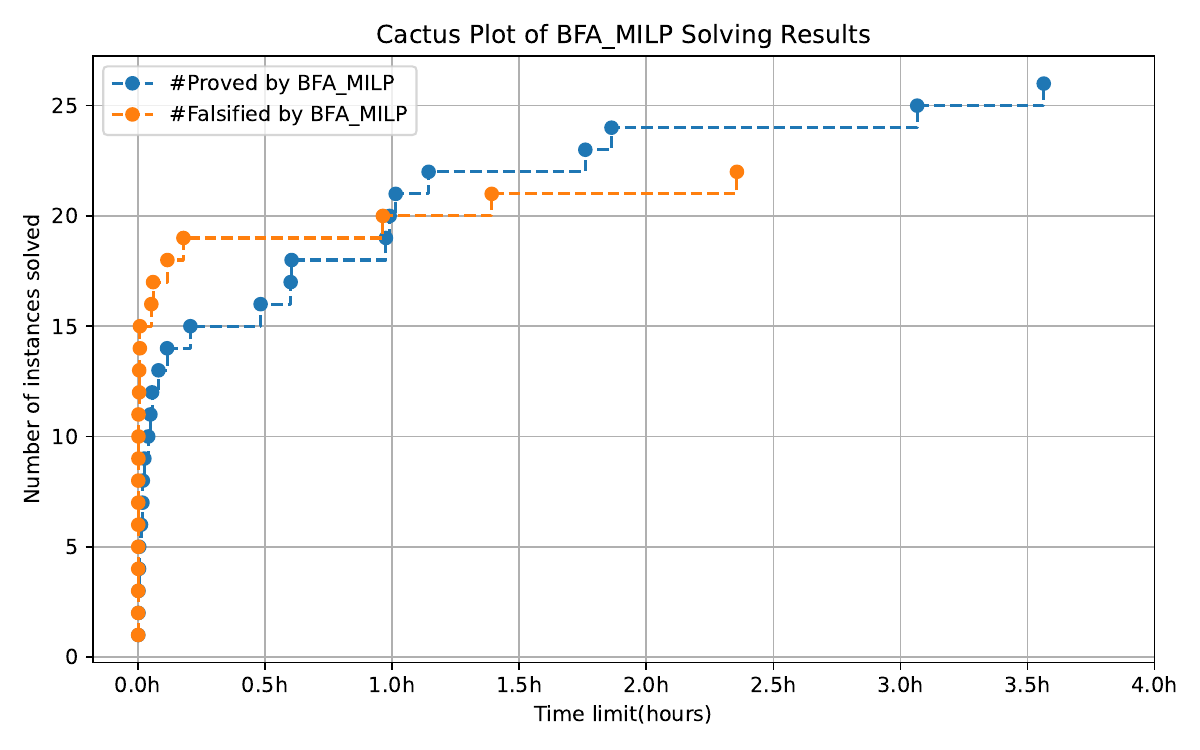}
    \caption{\major{The impact of time limit on the number of solved tasks by {\sf BFA\_MILP} for experiments in Table \ref{tab:tool_large_1}.}}
    \label{fig:table10_cactus}
\end{figure}

\major{For the unsolved tasks in Table~\ref{tab:tool_large_1}, we further investigate how the number of verified tasks evolves with extended time limits (up to 4 hours). The results are illustrated in Figure~\ref{fig:table10_cactus}. We observe that as the time limit increases, both the number of proved and the number of falsified tasks by {\sf BFA\_MILP} increase,  with the number of proved tasks exhibiting a more significant growth.} 

\section{Related Work}\label{sec:relatedwork}
In this section, we discuss the \minor{existing works closely related} to the contributions of this paper.
\vspace{2mm}

\noindent
{\bf Verification of QNNs.}
In the literature, quantization is broadly categorized into two types~\cite{gholami2021survey}: parameter-only quantization and quantization applied to both parameters and activations, leading to significant differences in verification methodologies. 
For parameter-only quantization, existing white-box DNN verification methods ~\cite{KBDJK17,LiLHY0ZXH20,betaCrown,GuoWZZSW21,backVerify22} can be applied directly, while
primarily leverage constraint-solving or abstraction. A constraint-solving-based method reduces the verification problem into 
SMT/MILP solving~\cite{KBDJK17,KatzHIJLLSTWZDK19,ChengNR17,PrabhakarA19,fischetti2018deep}. While sound and complete, they often suffer from scalability limitations. To improve efficiency, various abstraction-based methods are proposed, such as computing a conservative bound of the output range based on different abstract domains~\cite{GMDTCV18,LiLHY0ZXH20,SGMPV18,SGPV19} or obtaining abstract neural networks, rendering them more suitable for verification~\cite{AshokHKM20,ZZCSCL22,ElboherGK20,OBKatz22,LXSSXM22}. \minor{A key distinction between \symPoly and other symbolic or polyhedral abstraction-based approaches in neural network verification is that existing methods abstract only neuron value ranges with fixed parameters, focusing solely on input interval propagation. In contrast, \symPoly extends abstraction to both neuron and parameter value ranges, enabling simultaneous propagation of both input and weight intervals.}
For QNNs where both parameters and activations are quantized, existing techniques primarily rely on constraint solving~\cite{GiacobbeHL20,scaleQNN21,huang2023towards,zhang2023qebverif} or BDD~\cite{BDD4BNN,zhang2023precise}, mainly for robustness properties. 
\vspace{2mm}

\noindent
{\bf Bit-flip attacks and defense of neural networks.}
DNNs are notably vulnerable to BFAs, where a single bit alternation can cause severe performance degradation~\cite{HONG_USENIX19,SURVEY,liyes}. To mitigate this, QNNs have been explored as a more resilient alternative. Building on the foundational work by Rakin et al.~\cite{BFAICCV19}, a variety of attack technologies specifically designed for QNNs have then been investigated~\cite{DONTKNOCK,PRACTICALATTACK,SPARSEBFA,glitchinjection}. These attacks primarily manipulate bits in non-volatile memory, affecting mainly the weights and occasionally the biases (e.g., DRAM), as well as introducing faults into certain neuron activation functions. Common BFA methodologies on networks include the Rowhammer attack~\cite{ROWHAMMER,BFAICCV19,DONTKNOCK}, clock glitching attack~\cite{PRACTICALATTACK}, Voltage Frequency Scaling (VFS) attack~\cite{VFSBFA}, and lase injection attack~\cite{ovrvwlaser}. Notably, the Rowhammer, VFS, and laser inject attacks primarily manipulate the binary representations of weights and biases stored in memory~\cite{BFAICCV19,1bitallyouneed,DONTKNOCK,SPARSEBFA,VFSBFA}, while the clock glitching attack specifically disrupts the functionality of the activation functions~\cite{PRACTICALATTACK,glitchinjection}.

The primary objective of defensive techniques is to enhance accuracy and/or robustness in the presence of BFAs. A natural approach to achieving this is to implement countermeasures against the underlying mechanisms that cause attacks, specifically by addressing the attacks from a hardware or system architecture perspective. For example,~\cite{yauglikcci2021blockhammer} selectively throttles memory accesses that could otherwise potentially cause Rowhammer bit-flips. Error correction Code~\cite{cojocar2019exploiting,di2023copy} is also an effective defense mechanism, typically implemented by the memory controller. 
%
However, no existing defense method can provide a definitive guarantee of eliminating all potential risks posed by bit-flip attacks.
\vspace{2mm}

\noindent
\minor{{\bf Other related works.} \cite{weng2020towards} proposes a method that formulates certified weight perturbations as an optimization problem, employing a uniform $L_\infty$ norm perturbation within each layer. Their approach focuses on precision at the level of individual inputs, in contrast to ours \tool, which examines robustness against the BFAs over an input region. Another closely related study~\cite{wicker2020probabilistic} investigates probabilistic safety verification of Bayesian networks utilizing weight interval propagation to identify disjoint safe weight spaces based on weight distributions. Although a direct comparison between their work and ours is not feasible due to the differences in network types and verification tasks, their methodology aligns with our naive abstraction approach depicted in Figure~\ref{fig:TransAct}, which, as analyzed, exhibits lower abstract precision compared to the abstraction technique proposed in this work (cf. Figure~\ref{fig:TransX}).}

\section{Conclusion and Future Work}\label{sec:con}
We proposed a novel bit-flip attack verification method, \tool, for QNNs, which is sound, complete, and arguably efficient. To achieve this, we introduced \symPoly, the first abstract domain tailored for networks with symbolic parameters. We implemented \tool as an end-to-end tool and conducted thorough experiments on various benchmarks with networks of different model sizes and quantization bit-widths, demonstrating its effectiveness and efficiency. 

While \symPoly may not represent the theoretically optimal abstract transformer for convex relaxations of weighted activation functions, it achieves optimal when restricting static abstractions to using only two linear constraints per neuron. 
Moreover, in terms of convex relaxation, the optimal abstraction transformer may not significantly enhance \symPoly, supported by the observation in~\cite{salman2019convex} that there is an inherent \emph{barrier} to tight relaxation-based verification methods. 
However, \tool could be integrated with complementary verification techniques, such as Branch-and-Bound for ReLU splitting~\cite{bunel2018unifiedviewpiecewiselinear} and optimizable lower bounds~\cite{xu2021fastcompleteenablingcomplete}, to improve the verification precision and scalability. We consider these promising extensions for future work.

\minor{Our verification approach targets bit-flip attacks on a single parameter. Extending such threat models to attack multiple parameters simultaneously, e.g., two out of $n$ parameters (i.e., $\mm=2$), would involve a straightforward modification of Algorithm~\ref{alg:overall} by traversing all $\binom{n}{2}=\frac{n\cdot(n-1)}{2}$ two-parameter combinations in the for-loop at line 3, leading to exponential computation growth. Hence, although the abstract domain \symPoly can handle multiple symbolic parameters simultaneously, how to efficiently and effectively partition all $\binom{\nn}{2}$ combinations into groups for abstraction-refinement poses a significant and non-trivial challenge, which is also a key focus in future work.}
  

\section{Data-Availability Statement}
The source code of our tool and benchmarks are available at \cite{BFAVerifier}. 

\begin{acks}
This study was supported by the 
Strategic Priority Research Program of CAS (Award ID: XDA0320101), 
the Ministry of Education, Singapore under its Academic Research Fund Tier 2 (T2EP20222-0037), and the Ministry of Education, Singapore under its Academic Research Fund Tier 3 (MOET32020-0003).	
Any opinions, findings, conclusions, or recommendations expressed in this material are those of the author(s) and do not reflect the views of the Ministry of Education, Singapore. 
\end{acks}

\bibliographystyle{ACM-Reference-Format}

\bibliography{yd-bibfile,ydbfa-bibfile}


\begin{thebibliography}{91}


\ifx \showCODEN    \undefined \def \showCODEN     #1{\unskip}     \fi
\ifx \showDOI      \undefined \def \showDOI       #1{#1}\fi
\ifx \showISBNx    \undefined \def \showISBNx     #1{\unskip}     \fi
\ifx \showISBNxiii \undefined \def \showISBNxiii  #1{\unskip}     \fi
\ifx \showISSN     \undefined \def \showISSN      #1{\unskip}     \fi
\ifx \showLCCN     \undefined \def \showLCCN      #1{\unskip}     \fi
\ifx \shownote     \undefined \def \shownote      #1{#1}          \fi
\ifx \showarticletitle \undefined \def \showarticletitle #1{#1}   \fi
\ifx \showURL      \undefined \def \showURL       {\relax}        \fi
\providecommand\bibfield[2]{#2}
\providecommand\bibinfo[2]{#2}
\providecommand\natexlab[1]{#1}
\providecommand\showeprint[2][]{arXiv:#2}

\bibitem[FPI(2019)]%
        {FPIEEE754}
 \bibinfo{year}{2019}\natexlab{}.
\newblock \showarticletitle{IEEE Standard for Floating-Point Arithmetic}.
\newblock \bibinfo{journal}{\emph{IEEE Std 754-2019 (Revision of IEEE 754-2008)}} (\bibinfo{year}{2019}), \bibinfo{pages}{1--84}.
\newblock
\urldef\tempurl%
\url{https://doi.org/10.1109/IEEESTD.2019.8766229}
\showDOI{\tempurl}


\bibitem[Ashok et~al\mbox{.}(2020)]%
        {AshokHKM20}
\bibfield{author}{\bibinfo{person}{Pranav Ashok}, \bibinfo{person}{Vahid Hashemi}, \bibinfo{person}{Jan Kret{\'{\i}}nsk{\'{y}}}, {and} \bibinfo{person}{Stefanie Mohr}.} \bibinfo{year}{2020}\natexlab{}.
\newblock \showarticletitle{DeepAbstract: Neural Network Abstraction for Accelerating Verification}. In \bibinfo{booktitle}{\emph{Proceedings of the 18th International Symposium on Automated Technology for Verification and Analysis}}. \bibinfo{pages}{92--107}.
\newblock


\bibitem[Barenghi et~al\mbox{.}(2012)]%
        {BarenghiBKN12}
\bibfield{author}{\bibinfo{person}{Alessandro Barenghi}, \bibinfo{person}{Luca Breveglieri}, \bibinfo{person}{Israel Koren}, {and} \bibinfo{person}{David Naccache}.} \bibinfo{year}{2012}\natexlab{}.
\newblock \showarticletitle{Fault Injection Attacks on Cryptographic Devices: Theory, Practice, and Countermeasures}.
\newblock \bibinfo{journal}{\emph{Proc. {IEEE}}} \bibinfo{volume}{100}, \bibinfo{number}{11} (\bibinfo{year}{2012}), \bibinfo{pages}{3056--3076}.
\newblock


\bibitem[BFAVerifier(2025)]%
        {BFAVerifier}
\bibfield{author}{\bibinfo{person}{BFAVerifier}.} \bibinfo{year}{2025}\natexlab{}.
\newblock \bibinfo{howpublished}{\url{https://github.com/zhangyedi/BFAVerifier}}.
\newblock


\bibitem[Biham and Shamir(1997)]%
        {BihamS97}
\bibfield{author}{\bibinfo{person}{Eli Biham} {and} \bibinfo{person}{Adi Shamir}.} \bibinfo{year}{1997}\natexlab{}.
\newblock \showarticletitle{Differential Fault Analysis of Secret Key Cryptosystems}. In \bibinfo{booktitle}{\emph{Proceedings of the 17th Annual International Cryptology Conference}}. \bibinfo{pages}{513--525}.
\newblock


\bibitem[Boneh et~al\mbox{.}(1997)]%
        {BonehDL97}
\bibfield{author}{\bibinfo{person}{Dan Boneh}, \bibinfo{person}{Richard~A. DeMillo}, {and} \bibinfo{person}{Richard~J. Lipton}.} \bibinfo{year}{1997}\natexlab{}.
\newblock \showarticletitle{On the Importance of Checking Cryptographic Protocols for Faults (Extended Abstract)}. In \bibinfo{booktitle}{\emph{Proceeding of the International Conference on the Theory and Application of Cryptographic Techniques}}. \bibinfo{pages}{37--51}.
\newblock


\bibitem[Breier et~al\mbox{.}(2018)]%
        {PRACTICALATTACK}
\bibfield{author}{\bibinfo{person}{Jakub Breier}, \bibinfo{person}{Xiaolu Hou}, \bibinfo{person}{Dirmanto Jap}, \bibinfo{person}{Lei Ma}, \bibinfo{person}{Shivam Bhasin}, {and} \bibinfo{person}{Yang Liu}.} \bibinfo{year}{2018}\natexlab{}.
\newblock \showarticletitle{Practical Fault Attack on Deep Neural Networks}. In \bibinfo{booktitle}{\emph{Proceedings of the 2018 ACM SIGSAC Conference on Computer and Communications Security}} \emph{(\bibinfo{series}{CCS '18})}. \bibinfo{publisher}{Association for Computing Machinery}, \bibinfo{address}{New York, NY, USA}, \bibinfo{pages}{2204–2206}.
\newblock
\showISBNx{9781450356930}
\urldef\tempurl%
\url{https://doi.org/10.1145/3243734.3278519}
\showDOI{\tempurl}


\bibitem[Bu et~al\mbox{.}(2022)]%
        {DZBS21}
\bibfield{author}{\bibinfo{person}{Lei Bu}, \bibinfo{person}{Zhe Zhao}, \bibinfo{person}{Yuchao Duan}, {and} \bibinfo{person}{Fu Song}.} \bibinfo{year}{2022}\natexlab{}.
\newblock \showarticletitle{Taking Care of the Discretization Problem: A Comprehensive Study of the Discretization Problem and a Black-Box Adversarial Attack in Discrete Integer Domain}.
\newblock \bibinfo{journal}{\emph{IEEE Transactions on Dependable and Secure Computing}} \bibinfo{volume}{19}, \bibinfo{number}{5} (\bibinfo{year}{2022}), \bibinfo{pages}{3200--3217}.
\newblock
\urldef\tempurl%
\url{https://doi.org/10.1109/TDSC.2021.3088661}
\showDOI{\tempurl}


\bibitem[Bunel et~al\mbox{.}(2018)]%
        {bunel2018unifiedviewpiecewiselinear}
\bibfield{author}{\bibinfo{person}{Rudy Bunel}, \bibinfo{person}{Ilker Turkaslan}, \bibinfo{person}{Philip H.~S. Torr}, \bibinfo{person}{Pushmeet Kohli}, {and} \bibinfo{person}{M.~Pawan Kumar}.} \bibinfo{year}{2018}\natexlab{}.
\newblock \bibinfo{title}{A Unified View of Piecewise Linear Neural Network Verification}.
\newblock
\newblock
\showeprint[arxiv]{1711.00455}~[cs.AI]
\urldef\tempurl%
\url{https://arxiv.org/abs/1711.00455}
\showURL{%
\tempurl}


\bibitem[Chen et~al\mbox{.}(2021)]%
        {CCFDZSL20}
\bibfield{author}{\bibinfo{person}{Guangke Chen}, \bibinfo{person}{Sen Chen}, \bibinfo{person}{Lingling Fan}, \bibinfo{person}{Xiaoning Du}, \bibinfo{person}{Zhe Zhao}, \bibinfo{person}{Fu Song}, {and} \bibinfo{person}{Yang Liu}.} \bibinfo{year}{2021}\natexlab{}.
\newblock \showarticletitle{Who is Real {Bob}? Adversarial Attacks on Speaker Recognition Systems}. In \bibinfo{booktitle}{\emph{Proceedings of the 42nd IEEE Symposium on Security and Privacy}}.
\newblock


\bibitem[Chen et~al\mbox{.}(2024)]%
        {ChenZS24}
\bibfield{author}{\bibinfo{person}{Guangke Chen}, \bibinfo{person}{Yedi Zhang}, {and} \bibinfo{person}{Fu Song}.} \bibinfo{year}{2024}\natexlab{}.
\newblock \showarticletitle{{SLMIA-SR:} Speaker-Level Membership Inference Attacks against Speaker Recognition Systems}. In \bibinfo{booktitle}{\emph{31st Annual Network and Distributed System Security Symposium}}.
\newblock


\bibitem[Chen et~al\mbox{.}(2023)]%
        {ChenZZS23}
\bibfield{author}{\bibinfo{person}{Guangke Chen}, \bibinfo{person}{Yedi Zhang}, \bibinfo{person}{Zhe Zhao}, {and} \bibinfo{person}{Fu Song}.} \bibinfo{year}{2023}\natexlab{}.
\newblock \showarticletitle{{QFA2SR:} Query-Free Adversarial Transfer Attacks to Speaker Recognition Systems}. In \bibinfo{booktitle}{\emph{32nd {USENIX} Security Symposium}}, \bibfield{editor}{\bibinfo{person}{Joseph~A. Calandrino} {and} \bibinfo{person}{Carmela Troncoso}} (Eds.). \bibinfo{pages}{2437--2454}.
\newblock


\bibitem[Cheng et~al\mbox{.}(2017)]%
        {ChengNR17}
\bibfield{author}{\bibinfo{person}{Chih{-}Hong Cheng}, \bibinfo{person}{Georg N{\"{u}}hrenberg}, {and} \bibinfo{person}{Harald Ruess}.} \bibinfo{year}{2017}\natexlab{}.
\newblock \showarticletitle{Maximum Resilience of Artificial Neural Networks}. In \bibinfo{booktitle}{\emph{Proceedings of the 15th International Symposium on Automated Technology for Verification and Analysis (ATVA)}}. \bibinfo{pages}{251--268}.
\newblock


\bibitem[Cojocar et~al\mbox{.}(2019)]%
        {cojocar2019exploiting}
\bibfield{author}{\bibinfo{person}{Lucian Cojocar}, \bibinfo{person}{Kaveh Razavi}, \bibinfo{person}{Cristiano Giuffrida}, {and} \bibinfo{person}{Herbert Bos}.} \bibinfo{year}{2019}\natexlab{}.
\newblock \showarticletitle{Exploiting correcting codes: On the effectiveness of ecc memory against rowhammer attacks}. In \bibinfo{booktitle}{\emph{2019 IEEE Symposium on Security and Privacy (SP)}}. IEEE, \bibinfo{pages}{55--71}.
\newblock


\bibitem[Di~Dio et~al\mbox{.}(2023)]%
        {di2023copy}
\bibfield{author}{\bibinfo{person}{Andrea Di~Dio}, \bibinfo{person}{Koen Koning}, \bibinfo{person}{Herbert Bos}, {and} \bibinfo{person}{Cristiano Giuffrida}.} \bibinfo{year}{2023}\natexlab{}.
\newblock \showarticletitle{Copy-on-Flip: Hardening ECC Memory Against Rowhammer Attacks.}. In \bibinfo{booktitle}{\emph{NDSS}}.
\newblock


\bibitem[Dong et~al\mbox{.}(2023)]%
        {1bitallyouneed}
\bibfield{author}{\bibinfo{person}{J. Dong}, \bibinfo{person}{H. Qiu}, \bibinfo{person}{Y. Li}, \bibinfo{person}{T. Zhang}, \bibinfo{person}{Y. Li}, \bibinfo{person}{Z. Lai}, \bibinfo{person}{C. Zhang}, {and} \bibinfo{person}{S. Xia}.} \bibinfo{year}{2023}\natexlab{}.
\newblock \showarticletitle{One-bit Flip is All You Need: When Bit-flip Attack Meets Model Training}. In \bibinfo{booktitle}{\emph{2023 IEEE/CVF International Conference on Computer Vision (ICCV)}}. \bibinfo{pages}{4665--4675}.
\newblock


\bibitem[Dong et~al\mbox{.}(2021)]%
        {dong2021survey}
\bibfield{author}{\bibinfo{person}{Shi Dong}, \bibinfo{person}{Ping Wang}, {and} \bibinfo{person}{Khushnood Abbas}.} \bibinfo{year}{2021}\natexlab{}.
\newblock \showarticletitle{A survey on deep learning and its applications}.
\newblock \bibinfo{journal}{\emph{Computer Science Review}}  \bibinfo{volume}{40} (\bibinfo{year}{2021}), \bibinfo{pages}{100379}.
\newblock


\bibitem[Dumont et~al\mbox{.}(2021)]%
        {ovrvwlaser}
\bibfield{author}{\bibinfo{person}{Mathieu Dumont}, \bibinfo{person}{Pierre-Alain Moëllic}, \bibinfo{person}{Raphael Viera}, \bibinfo{person}{Jean-Max Dutertre}, {and} \bibinfo{person}{Rémi Bernhard}.} \bibinfo{year}{2021}\natexlab{}.
\newblock \showarticletitle{An Overview of Laser Injection against Embedded Neural Network Models}. In \bibinfo{booktitle}{\emph{2021 IEEE 7th World Forum on Internet of Things (WF-IoT)}}. \bibinfo{pages}{616--621}.
\newblock
\urldef\tempurl%
\url{https://doi.org/10.1109/WF-IoT51360.2021.9595075}
\showDOI{\tempurl}


\bibitem[Elboher et~al\mbox{.}(2020)]%
        {ElboherGK20}
\bibfield{author}{\bibinfo{person}{Yizhak~Yisrael Elboher}, \bibinfo{person}{Justin Gottschlich}, {and} \bibinfo{person}{Guy Katz}.} \bibinfo{year}{2020}\natexlab{}.
\newblock \showarticletitle{An Abstraction-Based Framework for Neural Network Verification}. In \bibinfo{booktitle}{\emph{Proceedings of the 32nd International Conference on Computer Aided Verification}}. \bibinfo{pages}{43--65}.
\newblock


\bibitem[Fischetti and Jo(2018)]%
        {fischetti2018deep}
\bibfield{author}{\bibinfo{person}{Matteo Fischetti} {and} \bibinfo{person}{Jason Jo}.} \bibinfo{year}{2018}\natexlab{}.
\newblock \showarticletitle{Deep neural networks and mixed integer linear optimization}.
\newblock \bibinfo{journal}{\emph{Constraints}} \bibinfo{volume}{23}, \bibinfo{number}{3} (\bibinfo{year}{2018}), \bibinfo{pages}{296--309}.
\newblock


\bibitem[Gehr et~al\mbox{.}(2018)]%
        {GMDTCV18}
\bibfield{author}{\bibinfo{person}{Timon Gehr}, \bibinfo{person}{Matthew Mirman}, \bibinfo{person}{Dana Drachsler{-}Cohen}, \bibinfo{person}{Petar Tsankov}, \bibinfo{person}{Swarat Chaudhuri}, {and} \bibinfo{person}{Martin~T. Vechev}.} \bibinfo{year}{2018}\natexlab{}.
\newblock \showarticletitle{{AI$^2$:} Safety and Robustness Certification of Neural Networks with Abstract Interpretation}. In \bibinfo{booktitle}{\emph{Proceedings of the 2018 {IEEE} Symposium on Security and Privacy}}. \bibinfo{pages}{3--18}.
\newblock


\bibitem[Gholami et~al\mbox{.}(2022)]%
        {gholami2021survey}
\bibfield{author}{\bibinfo{person}{Amir Gholami}, \bibinfo{person}{Sehoon Kim}, \bibinfo{person}{Zhen Dong}, \bibinfo{person}{Zhewei Yao}, \bibinfo{person}{Michael~W Mahoney}, {and} \bibinfo{person}{Kurt Keutzer}.} \bibinfo{year}{2022}\natexlab{}.
\newblock \showarticletitle{A survey of quantization methods for efficient neural network inference}.
\newblock In \bibinfo{booktitle}{\emph{Low-Power Computer Vision}}. \bibinfo{publisher}{Chapman and Hall/CRC}, \bibinfo{pages}{291--326}.
\newblock


\bibitem[Giacobbe et~al\mbox{.}(2020)]%
        {GiacobbeHL20}
\bibfield{author}{\bibinfo{person}{Mirco Giacobbe}, \bibinfo{person}{Thomas~A. Henzinger}, {and} \bibinfo{person}{Mathias Lechner}.} \bibinfo{year}{2020}\natexlab{}.
\newblock \showarticletitle{How Many Bits Does it Take to Quantize Your Neural Network?}. In \bibinfo{booktitle}{\emph{Proceedings of the 26th International Conference on Tools and Algorithms for the Construction and Analysis of Systems (TACAS)}}. \bibinfo{pages}{79--97}.
\newblock
\urldef\tempurl%
\url{https://doi.org/10.1007/978-3-030-45237-7\_5}
\showDOI{\tempurl}


\bibitem[Gong et~al\mbox{.}(2019)]%
        {DSQ}
\bibfield{author}{\bibinfo{person}{Ruihao Gong}, \bibinfo{person}{Xianglong Liu}, \bibinfo{person}{Shenghu Jiang}, \bibinfo{person}{Tianxiang Li}, \bibinfo{person}{Peng Hu}, \bibinfo{person}{Jiazhen Lin}, \bibinfo{person}{Fengwei Yu}, {and} \bibinfo{person}{Junjie Yan}.} \bibinfo{year}{2019}\natexlab{}.
\newblock \showarticletitle{Differentiable Soft Quantization: Bridging Full-Precision and Low-Bit Neural Networks}. In \bibinfo{booktitle}{\emph{Proceedings of the {IEEE/CVF} International Conference on Computer Vision ({ICCV})}}. \bibinfo{pages}{4851--4860}.
\newblock
\urldef\tempurl%
\url{https://doi.org/10.1109/ICCV.2019.00495}
\showDOI{\tempurl}


\bibitem[Guo et~al\mbox{.}(2021)]%
        {GuoWZZSW21}
\bibfield{author}{\bibinfo{person}{Xingwu Guo}, \bibinfo{person}{Wenjie Wan}, \bibinfo{person}{Zhaodi Zhang}, \bibinfo{person}{Min Zhang}, \bibinfo{person}{Fu Song}, {and} \bibinfo{person}{Xuejun Wen}.} \bibinfo{year}{2021}\natexlab{}.
\newblock \showarticletitle{Eager Falsification for Accelerating Robustness Verification of Deep Neural Networks}. In \bibinfo{booktitle}{\emph{Proceedings of the 32nd {IEEE} International Symposium on Software Reliability Engineering}}. \bibinfo{pages}{345--356}.
\newblock


\bibitem[Gurobi(2018)]%
        {Gurobi}
\bibfield{author}{\bibinfo{person}{Gurobi}.} \bibinfo{year}{2018}\natexlab{}.
\newblock \bibinfo{title}{A most powerful mathematical optimization solver}.
\newblock \bibinfo{howpublished}{\url{https://www.gurobi.com/}}.
\newblock


\bibitem[Han et~al\mbox{.}(2016)]%
        {HanMD15}
\bibfield{author}{\bibinfo{person}{Song Han}, \bibinfo{person}{Huizi Mao}, {and} \bibinfo{person}{William~J. Dally}.} \bibinfo{year}{2016}\natexlab{}.
\newblock \showarticletitle{Deep Compression: Compressing Deep Neural Network with Pruning, Trained Quantization and Huffman Coding}. In \bibinfo{booktitle}{\emph{Proceedings of the 4th International Conference on Learning Representations}}.
\newblock


\bibitem[He et~al\mbox{.}(2020)]%
        {he2020defending}
\bibfield{author}{\bibinfo{person}{Zhezhi He}, \bibinfo{person}{Adnan~Siraj Rakin}, \bibinfo{person}{Jingtao Li}, \bibinfo{person}{Chaitali Chakrabarti}, {and} \bibinfo{person}{Deliang Fan}.} \bibinfo{year}{2020}\natexlab{}.
\newblock \showarticletitle{Defending and harnessing the bit-flip based adversarial weight attack}. In \bibinfo{booktitle}{\emph{Proceedings of the IEEE/CVF Conference on Computer Vision and Pattern Recognition}}. \bibinfo{pages}{14095--14103}.
\newblock


\bibitem[Henzinger et~al\mbox{.}(2021)]%
        {scaleQNN21}
\bibfield{author}{\bibinfo{person}{Thomas~A. Henzinger}, \bibinfo{person}{Mathias Lechner}, {and} \bibinfo{person}{Dorde Zikelic}.} \bibinfo{year}{2021}\natexlab{}.
\newblock \showarticletitle{Scalable Verification of Quantized Neural Networks}. In \bibinfo{booktitle}{\emph{Proceedings of the 35th {AAAI} Conference on Artificial Intelligence ({AAAI})}}. \bibinfo{pages}{3787--3795}.
\newblock
\urldef\tempurl%
\url{https://doi.org/10.1609/AAAI.V35I5.16496}
\showDOI{\tempurl}


\bibitem[Hong et~al\mbox{.}(2019)]%
        {HONG_USENIX19}
\bibfield{author}{\bibinfo{person}{Sanghyun Hong}, \bibinfo{person}{Pietro Frigo}, \bibinfo{person}{Yigitcan Kaya}, \bibinfo{person}{Cristiano Giuffrida}, {and} \bibinfo{person}{Tudor Dumitras}.} \bibinfo{year}{2019}\natexlab{}.
\newblock \showarticletitle{Terminal Brain Damage: Exposing the Graceless Degradation in Deep Neural Networks Under Hardware Fault Attacks}. In \bibinfo{booktitle}{\emph{28th USENIX Security Symposium (USENIX Security 19)}}. \bibinfo{publisher}{USENIX Association}, \bibinfo{address}{Santa Clara, CA}, \bibinfo{pages}{497--514}.
\newblock
\showISBNx{978-1-939133-06-9}


\bibitem[Huang et~al\mbox{.}(2023)]%
        {huang2023towards}
\bibfield{author}{\bibinfo{person}{Pei Huang}, \bibinfo{person}{Haoze Wu}, \bibinfo{person}{Yuting Yang}, \bibinfo{person}{Ieva Daukantas}, \bibinfo{person}{Min Wu}, \bibinfo{person}{Yedi Zhang}, {and} \bibinfo{person}{Clark Barrett}.} \bibinfo{year}{2023}\natexlab{}.
\newblock \showarticletitle{Towards Efficient Verification of Quantized Neural Networks}.
\newblock \bibinfo{journal}{\emph{arXiv preprint arXiv:2312.12679}} (\bibinfo{year}{2023}).
\newblock


\bibitem[Huang et~al\mbox{.}(2024)]%
        {huang2024towards}
\bibfield{author}{\bibinfo{person}{Pei Huang}, \bibinfo{person}{Haoze Wu}, \bibinfo{person}{Yuting Yang}, \bibinfo{person}{Ieva Daukantas}, \bibinfo{person}{Min Wu}, \bibinfo{person}{Yedi Zhang}, {and} \bibinfo{person}{Clark Barrett}.} \bibinfo{year}{2024}\natexlab{}.
\newblock \showarticletitle{Towards Efficient Verification of Quantized Neural Networks}. In \bibinfo{booktitle}{\emph{Proceedings of the AAAI Conference on Artificial Intelligence}}, Vol.~\bibinfo{volume}{38}. \bibinfo{pages}{21152--21160}.
\newblock


\bibitem[Jacob et~al\mbox{.}(2018)]%
        {JacobKCZTHAK18}
\bibfield{author}{\bibinfo{person}{Benoit Jacob}, \bibinfo{person}{Skirmantas Kligys}, \bibinfo{person}{Bo Chen}, \bibinfo{person}{Menglong Zhu}, \bibinfo{person}{Matthew Tang}, \bibinfo{person}{Andrew~G. Howard}, \bibinfo{person}{Hartwig Adam}, {and} \bibinfo{person}{Dmitry Kalenichenko}.} \bibinfo{year}{2018}\natexlab{}.
\newblock \showarticletitle{Quantization and Training of Neural Networks for Efficient Integer-Arithmetic-Only Inference}. In \bibinfo{booktitle}{\emph{Proceedings of the {IEEE} Conference on Computer Vision and Pattern Recognition ({CVPR})}}. \bibinfo{pages}{2704--2713}.
\newblock


\bibitem[Julian et~al\mbox{.}(2019)]%
        {julian2019deep}
\bibfield{author}{\bibinfo{person}{Kyle~D Julian}, \bibinfo{person}{Mykel~J Kochenderfer}, {and} \bibinfo{person}{Michael~P Owen}.} \bibinfo{year}{2019}\natexlab{}.
\newblock \showarticletitle{Deep neural network compression for aircraft collision avoidance systems}.
\newblock \bibinfo{journal}{\emph{Journal of Guidance, Control, and Dynamics}} \bibinfo{volume}{42}, \bibinfo{number}{3} (\bibinfo{year}{2019}), \bibinfo{pages}{598--608}.
\newblock


\bibitem[Katz et~al\mbox{.}(2017a)]%
        {GuyKatz2017ReluplexAE}
\bibfield{author}{\bibinfo{person}{Guy Katz}, \bibinfo{person}{Clark Barrett}, \bibinfo{person}{David~L. Dill}, \bibinfo{person}{Kyle~D. Julian}, {and} \bibinfo{person}{Mykel~J. Kochenderfer}.} \bibinfo{year}{2017}\natexlab{a}.
\newblock \showarticletitle{Reluplex: An Efficient {SMT} Solver for Verifying Deep Neural Networks}. In \bibinfo{booktitle}{\emph{Proceedings of the 29th International Conference on Computer Aided Verification}}. \bibinfo{pages}{97--117}.
\newblock


\bibitem[Katz et~al\mbox{.}(2017b)]%
        {KBDJK17}
\bibfield{author}{\bibinfo{person}{Guy Katz}, \bibinfo{person}{Clark~W. Barrett}, \bibinfo{person}{David~L. Dill}, \bibinfo{person}{Kyle Julian}, {and} \bibinfo{person}{Mykel~J. Kochenderfer}.} \bibinfo{year}{2017}\natexlab{b}.
\newblock \showarticletitle{Reluplex: An Efficient {SMT} Solver for Verifying Deep Neural Networks}. In \bibinfo{booktitle}{\emph{Proceedings of the 29th International Conference on Computer Aided Verification}}. \bibinfo{pages}{97--117}.
\newblock


\bibitem[Katz et~al\mbox{.}(2019)]%
        {KatzHIJLLSTWZDK19}
\bibfield{author}{\bibinfo{person}{Guy Katz}, \bibinfo{person}{Derek~A. Huang}, \bibinfo{person}{Duligur Ibeling}, \bibinfo{person}{Kyle Julian}, \bibinfo{person}{Christopher Lazarus}, \bibinfo{person}{Rachel Lim}, \bibinfo{person}{Parth Shah}, \bibinfo{person}{Shantanu Thakoor}, \bibinfo{person}{Haoze Wu}, \bibinfo{person}{Aleksandar Zeljic}, \bibinfo{person}{David~L. Dill}, \bibinfo{person}{Mykel~J. Kochenderfer}, {and} \bibinfo{person}{Clark~W. Barrett}.} \bibinfo{year}{2019}\natexlab{}.
\newblock \showarticletitle{The Marabou Framework for Verification and Analysis of Deep Neural Networks}. In \bibinfo{booktitle}{\emph{Proceedings of the 31st International Conference on Computer Aided Verification}}. \bibinfo{pages}{443--452}.
\newblock


\bibitem[Khalid et~al\mbox{.}(2021)]%
        {khalid2021exploiting}
\bibfield{author}{\bibinfo{person}{Faiq Khalid}, \bibinfo{person}{Muhammad~Abdullah Hanif}, {and} \bibinfo{person}{Muhammad Shafique}.} \bibinfo{year}{2021}\natexlab{}.
\newblock \showarticletitle{Exploiting vulnerabilities in deep neural networks: Adversarial and fault-injection attacks}.
\newblock \bibinfo{journal}{\emph{arXiv preprint arXiv:2105.03251}} (\bibinfo{year}{2021}).
\newblock


\bibitem[Khoshavi et~al\mbox{.}(2022)]%
        {HARDeNN}
\bibfield{author}{\bibinfo{person}{Navid Khoshavi}, \bibinfo{person}{Mohammad Maghsoudloo}, \bibinfo{person}{Arman Roohi}, \bibinfo{person}{Saman Sargolzaei}, {and} \bibinfo{person}{Yu Bi}.} \bibinfo{year}{2022}\natexlab{}.
\newblock \showarticletitle{HARDeNN: Hardware-assisted attack-resilient deep neural network architectures}.
\newblock \bibinfo{journal}{\emph{Microprocessors and Microsystems}}  \bibinfo{volume}{95} (\bibinfo{year}{2022}), \bibinfo{pages}{104710}.
\newblock


\bibitem[Kim et~al\mbox{.}(2014)]%
        {kim2014flipping}
\bibfield{author}{\bibinfo{person}{Yoongu Kim}, \bibinfo{person}{Ross Daly}, \bibinfo{person}{Jeremie Kim}, \bibinfo{person}{Chris Fallin}, \bibinfo{person}{Ji~Hye Lee}, \bibinfo{person}{Donghyuk Lee}, \bibinfo{person}{Chris Wilkerson}, \bibinfo{person}{Konrad Lai}, {and} \bibinfo{person}{Onur Mutlu}.} \bibinfo{year}{2014}\natexlab{}.
\newblock \showarticletitle{Flipping bits in memory without accessing them: An experimental study of DRAM disturbance errors}.
\newblock \bibinfo{journal}{\emph{ACM SIGARCH Computer Architecture News}} \bibinfo{volume}{42}, \bibinfo{number}{3} (\bibinfo{year}{2014}), \bibinfo{pages}{361--372}.
\newblock


\bibitem[LeCun and Cortes(2010)]%
        {MNIST}
\bibfield{author}{\bibinfo{person}{Yann LeCun} {and} \bibinfo{person}{Corinna Cortes}.} \bibinfo{year}{2010}\natexlab{}.
\newblock \bibinfo{title}{MNIST handwritten digit database}.
\newblock
\newblock


\bibitem[Lee and Chandrakasan(2022)]%
        {SPARSEBFA}
\bibfield{author}{\bibinfo{person}{Kyungmi Lee} {and} \bibinfo{person}{Anantha~P. Chandrakasan}.} \bibinfo{year}{2022}\natexlab{}.
\newblock \showarticletitle{SparseBFA: Attacking Sparse Deep Neural Networks with the Worst-Case Bit Flips on Coordinates}. In \bibinfo{booktitle}{\emph{ICASSP 2022 - 2022 IEEE International Conference on Acoustics, Speech and Signal Processing (ICASSP)}}. \bibinfo{pages}{4208--4212}.
\newblock
\urldef\tempurl%
\url{https://doi.org/10.1109/ICASSP43922.2022.9747337}
\showDOI{\tempurl}


\bibitem[Li et~al\mbox{.}(2019)]%
        {LiLYCHZ19}
\bibfield{author}{\bibinfo{person}{Jianlin Li}, \bibinfo{person}{Jiangchao Liu}, \bibinfo{person}{Pengfei Yang}, \bibinfo{person}{Liqian Chen}, \bibinfo{person}{Xiaowei Huang}, {and} \bibinfo{person}{Lijun Zhang}.} \bibinfo{year}{2019}\natexlab{}.
\newblock \showarticletitle{Analyzing Deep Neural Networks with Symbolic Propagation: Towards Higher Precision and Faster Verification}. In \bibinfo{booktitle}{\emph{Proceedings of the 26th International Symposium on Static Analysis}}. \bibinfo{pages}{296--319}.
\newblock


\bibitem[Li et~al\mbox{.}(2024a)]%
        {li2024investigating}
\bibfield{author}{\bibinfo{person}{Qun Li}, \bibinfo{person}{Yuan Meng}, \bibinfo{person}{Chen Tang}, \bibinfo{person}{Jiacheng Jiang}, {and} \bibinfo{person}{Zhi Wang}.} \bibinfo{year}{2024}\natexlab{a}.
\newblock \showarticletitle{Investigating the Impact of Quantization on Adversarial Robustness}.
\newblock \bibinfo{journal}{\emph{arXiv preprint arXiv:2404.05639}} (\bibinfo{year}{2024}).
\newblock


\bibitem[Li et~al\mbox{.}(2020)]%
        {LiLHY0ZXH20}
\bibfield{author}{\bibinfo{person}{Renjue Li}, \bibinfo{person}{Jianlin Li}, \bibinfo{person}{Cheng{-}Chao Huang}, \bibinfo{person}{Pengfei Yang}, \bibinfo{person}{Xiaowei Huang}, \bibinfo{person}{Lijun Zhang}, \bibinfo{person}{Bai Xue}, {and} \bibinfo{person}{Holger Hermanns}.} \bibinfo{year}{2020}\natexlab{}.
\newblock \showarticletitle{PRODeep: a platform for robustness verification of deep neural networks}. In \bibinfo{booktitle}{\emph{Proceedings of the 28th {ACM} Joint European Software Engineering Conference and Symposium on the Foundations of Software Engineering}}. \bibinfo{pages}{1630--1634}.
\newblock


\bibitem[Li et~al\mbox{.}(2024b)]%
        {liyes}
\bibfield{author}{\bibinfo{person}{Shaofeng Li}, \bibinfo{person}{Xinyu Wang}, \bibinfo{person}{Minhui Xue}, \bibinfo{person}{Haojin Zhu}, \bibinfo{person}{Zhi Zhang}, \bibinfo{person}{Yansong Gao}, \bibinfo{person}{Wen Wu}, {and} \bibinfo{person}{Xuemin~Sherman Shen}.} \bibinfo{year}{2024}\natexlab{b}.
\newblock \showarticletitle{Yes, One-Bit-Flip Matters! Universal DNN Model Inference Depletion with Runtime Code Fault Injection}.
\newblock  (\bibinfo{year}{2024}).
\newblock


\bibitem[Liu et~al\mbox{.}(2024)]%
        {LXSSXM22}
\bibfield{author}{\bibinfo{person}{Jiaxiang Liu}, \bibinfo{person}{Yunhan Xing}, \bibinfo{person}{Xiaomu Shi}, \bibinfo{person}{Fu Song}, \bibinfo{person}{Zhiwu Xu}, {and} \bibinfo{person}{Zhong Ming}.} \bibinfo{year}{2024}\natexlab{}.
\newblock \showarticletitle{Abstraction and Refinement: Towards Scalable and Exact Verification of Neural Networks}.
\newblock \bibinfo{journal}{\emph{{ACM} Trans. Softw. Eng. Methodol.}} \bibinfo{volume}{33}, \bibinfo{number}{5} (\bibinfo{year}{2024}), \bibinfo{pages}{129:1--129:35}.
\newblock
\urldef\tempurl%
\url{https://doi.org/10.1145/3644387}
\showDOI{\tempurl}


\bibitem[Liu et~al\mbox{.}(2023)]%
        {advwa}
\bibfield{author}{\bibinfo{person}{Liang Liu}, \bibinfo{person}{Yanan Guo}, \bibinfo{person}{Yueqiang Cheng}, \bibinfo{person}{Youtao Zhang}, {and} \bibinfo{person}{Jun Yang}.} \bibinfo{year}{2023}\natexlab{}.
\newblock \showarticletitle{Generating Robust DNN With Resistance to Bit-Flip Based Adversarial Weight Attack}.
\newblock \bibinfo{journal}{\emph{IEEE Trans. Comput.}} \bibinfo{volume}{72}, \bibinfo{number}{2} (\bibinfo{year}{2023}), \bibinfo{pages}{401--413}.
\newblock
\urldef\tempurl%
\url{https://doi.org/10.1109/TC.2022.3211411}
\showDOI{\tempurl}


\bibitem[Liu et~al\mbox{.}(2020)]%
        {glitchinjection}
\bibfield{author}{\bibinfo{person}{Wenye Liu}, \bibinfo{person}{Chip-Hong Chang}, \bibinfo{person}{Fan Zhang}, {and} \bibinfo{person}{Xiaoxuan Lou}.} \bibinfo{year}{2020}\natexlab{}.
\newblock \showarticletitle{Imperceptible misclassification attack on deep learning accelerator by glitch injection}. In \bibinfo{booktitle}{\emph{Proceedings of the 57th ACM/EDAC/IEEE Design Automation Conference}} (Virtual Event, USA) \emph{(\bibinfo{series}{DAC '20})}. \bibinfo{publisher}{IEEE Press}, Article \bibinfo{articleno}{29}, \bibinfo{numpages}{6}~pages.
\newblock
\showISBNx{9781450367257}


\bibitem[Liu et~al\mbox{.}(2017)]%
        {LiuWLX17}
\bibfield{author}{\bibinfo{person}{Yannan Liu}, \bibinfo{person}{Lingxiao Wei}, \bibinfo{person}{Bo Luo}, {and} \bibinfo{person}{Qiang Xu}.} \bibinfo{year}{2017}\natexlab{}.
\newblock \showarticletitle{Fault injection attack on deep neural network}. In \bibinfo{booktitle}{\emph{Proceedings of the {IEEE/ACM} International Conference on Computer-Aided Design}}. \bibinfo{pages}{131--138}.
\newblock


\bibitem[Lomuscio and Maganti(2017)]%
        {LomuscioM17}
\bibfield{author}{\bibinfo{person}{Alessio Lomuscio} {and} \bibinfo{person}{Lalit Maganti}.} \bibinfo{year}{2017}\natexlab{}.
\newblock \showarticletitle{An approach to reachability analysis for feed-forward {ReLU} neural networks}.
\newblock \bibinfo{journal}{\emph{CoRR}}  \bibinfo{volume}{abs/1706.07351} (\bibinfo{year}{2017}).
\newblock


\bibitem[muellch et~al\mbox{.}(2023)]%
        {muellch2023eth}
\bibfield{author}{\bibinfo{person}{muellch}, \bibinfo{person}{Gleb Makarchuk}, \bibinfo{person}{GgnDp}, \bibinfo{person}{skcho}, \bibinfo{person}{Gagandeep Singh}, \bibinfo{person}{François Serre}, \bibinfo{person}{Tobias Zimmermann}, \bibinfo{person}{Anian Ruoss}, \bibinfo{person}{Mark Müller}, \bibinfo{person}{Shachar Itzhaky}, \bibinfo{person}{Jingxuan He}, \bibinfo{person}{Haoze(Andrew) Wu}, \bibinfo{person}{Isac Andrei}, \bibinfo{person}{jorgenavas}, \bibinfo{person}{Jose Calderon}, {and} \bibinfo{person}{Jianlin Li}.} \bibinfo{year}{2023}\natexlab{}.
\newblock \bibinfo{booktitle}{\emph{eth-sri/ELINA}}.
\newblock
\urldef\tempurl%
\url{https://github.com/eth-sri/ELINA}
\showURL{%
\tempurl}


\bibitem[Musa et~al\mbox{.}(2023)]%
        {MusaHLLY23}
\bibfield{author}{\bibinfo{person}{Abubakar~Ahmad Musa}, \bibinfo{person}{Adamu Hussaini}, \bibinfo{person}{Weixian Liao}, \bibinfo{person}{Fan Liang}, {and} \bibinfo{person}{Wei Yu}.} \bibinfo{year}{2023}\natexlab{}.
\newblock \showarticletitle{Deep Neural Networks for Spatial-Temporal Cyber-Physical Systems: {A} Survey}.
\newblock \bibinfo{journal}{\emph{Future Internet}} \bibinfo{volume}{15}, \bibinfo{number}{6} (\bibinfo{year}{2023}), \bibinfo{pages}{199}.
\newblock
\urldef\tempurl%
\url{https://doi.org/10.3390/FI15060199}
\showDOI{\tempurl}


\bibitem[Mutlu and Kim(2020)]%
        {ROWHAMMER}
\bibfield{author}{\bibinfo{person}{Onur Mutlu} {and} \bibinfo{person}{Jeremie~S. Kim}.} \bibinfo{year}{2020}\natexlab{}.
\newblock \showarticletitle{RowHammer: A Retrospective}.
\newblock \bibinfo{journal}{\emph{IEEE Transactions on Computer-Aided Design of Integrated Circuits and Systems}} \bibinfo{volume}{39}, \bibinfo{number}{8} (\bibinfo{year}{2020}), \bibinfo{pages}{1555--1571}.
\newblock
\urldef\tempurl%
\url{https://doi.org/10.1109/TCAD.2019.2915318}
\showDOI{\tempurl}


\bibitem[Ostrovsky et~al\mbox{.}(2022)]%
        {OBKatz22}
\bibfield{author}{\bibinfo{person}{Matan Ostrovsky}, \bibinfo{person}{Clark~W. Barrett}, {and} \bibinfo{person}{Guy Katz}.} \bibinfo{year}{2022}\natexlab{}.
\newblock \showarticletitle{An Abstraction-Refinement Approach to Verifying Convolutional Neural Networks}.
\newblock \bibinfo{journal}{\emph{CoRR}}  \bibinfo{volume}{abs/2201.01978} (\bibinfo{year}{2022}).
\newblock


\bibitem[Pham and Sun(2022)]%
        {backVerify22}
\bibfield{author}{\bibinfo{person}{Long~H Pham} {and} \bibinfo{person}{Jun Sun}.} \bibinfo{year}{2022}\natexlab{}.
\newblock \showarticletitle{Verifying Neural Networks Against Backdoor Attacks}. In \bibinfo{booktitle}{\emph{Proceedings of the 34th International Conference on Computer Aided Verification ({CAV})}}. \bibinfo{pages}{171--192}.
\newblock
\urldef\tempurl%
\url{https://doi.org/10.1007/978-3-031-13185-1\_9}
\showDOI{\tempurl}


\bibitem[Pham and Sun(2024)]%
        {longNew}
\bibfield{author}{\bibinfo{person}{Long~H Pham} {and} \bibinfo{person}{Jun Sun}.} \bibinfo{year}{2024}\natexlab{}.
\newblock \showarticletitle{Certified Continual Learning for Neural Network Regression}.
\newblock \bibinfo{journal}{\emph{arXiv preprint arXiv:2407.06697}} (\bibinfo{year}{2024}).
\newblock


\bibitem[Prabhakar and Afzal(2019)]%
        {PrabhakarA19}
\bibfield{author}{\bibinfo{person}{Pavithra Prabhakar} {and} \bibinfo{person}{Zahra~Rahimi Afzal}.} \bibinfo{year}{2019}\natexlab{}.
\newblock \showarticletitle{Abstraction based Output Range Analysis for Neural Networks}. In \bibinfo{booktitle}{\emph{Proceedings of the Annual Conference on Neural Information Processing Systems}}, \bibfield{editor}{\bibinfo{person}{Hanna~M. Wallach}, \bibinfo{person}{Hugo Larochelle}, \bibinfo{person}{Alina Beygelzimer}, \bibinfo{person}{Florence d'Alch{\'{e}}{-}Buc}, \bibinfo{person}{Emily~B. Fox}, {and} \bibinfo{person}{Roman Garnett}} (Eds.). \bibinfo{pages}{15762--15772}.
\newblock


\bibitem[Qian et~al\mbox{.}(2023)]%
        {SURVEY}
\bibfield{author}{\bibinfo{person}{Cheng Qian}, \bibinfo{person}{Ming Zhang}, \bibinfo{person}{Yuanping Nie}, \bibinfo{person}{Shuaibing Lu}, {and} \bibinfo{person}{Huayang Cao}.} \bibinfo{year}{2023}\natexlab{}.
\newblock \showarticletitle{A Survey of Bit-Flip Attacks on Deep Neural Network and Corresponding Defense Methods}.
\newblock \bibinfo{journal}{\emph{Electronics}} \bibinfo{volume}{12}, \bibinfo{number}{4} (\bibinfo{year}{2023}).
\newblock
\showISSN{2079-9292}
\urldef\tempurl%
\url{https://doi.org/10.3390/electronics12040853}
\showDOI{\tempurl}


\bibitem[Rakin et~al\mbox{.}(2022)]%
        {rakin2022deepsteal}
\bibfield{author}{\bibinfo{person}{Adnan~Siraj Rakin}, \bibinfo{person}{Md~Hafizul~Islam Chowdhuryy}, \bibinfo{person}{Fan Yao}, {and} \bibinfo{person}{Deliang Fan}.} \bibinfo{year}{2022}\natexlab{}.
\newblock \showarticletitle{Deepsteal: Advanced model extractions leveraging efficient weight stealing in memories}. In \bibinfo{booktitle}{\emph{2022 IEEE symposium on security and privacy (SP)}}. IEEE, \bibinfo{pages}{1157--1174}.
\newblock


\bibitem[Rakin et~al\mbox{.}(2019)]%
        {BFAICCV19}
\bibfield{author}{\bibinfo{person}{Adnan~Siraj Rakin}, \bibinfo{person}{Zhezhi He}, {and} \bibinfo{person}{Deliang Fan}.} \bibinfo{year}{2019}\natexlab{}.
\newblock \showarticletitle{Bit-Flip Attack: Crushing Neural Network With Progressive Bit Search}. In \bibinfo{booktitle}{\emph{Proceedings of the IEEE/CVF International Conference on Computer Vision (ICCV)}}.
\newblock


\bibitem[Rakin et~al\mbox{.}(2021)]%
        {VFSBFA}
\bibfield{author}{\bibinfo{person}{Adnan~Siraj Rakin}, \bibinfo{person}{Yukui Luo}, \bibinfo{person}{Xiaolin Xu}, {and} \bibinfo{person}{Deliang Fan}.} \bibinfo{year}{2021}\natexlab{}.
\newblock \showarticletitle{{Deep-Dup}: An Adversarial Weight Duplication Attack Framework to Crush Deep Neural Network in {Multi-Tenant} {FPGA}}. In \bibinfo{booktitle}{\emph{30th USENIX Security Symposium (USENIX Security 21)}}. \bibinfo{publisher}{USENIX Association}, \bibinfo{pages}{1919--1936}.
\newblock
\showISBNx{978-1-939133-24-3}
\urldef\tempurl%
\url{https://www.usenix.org/conference/usenixsecurity21/presentation/rakin}
\showURL{%
\tempurl}


\bibitem[Salman et~al\mbox{.}(2019)]%
        {salman2019convex}
\bibfield{author}{\bibinfo{person}{Hadi Salman}, \bibinfo{person}{Greg Yang}, \bibinfo{person}{Huan Zhang}, \bibinfo{person}{Cho-Jui Hsieh}, {and} \bibinfo{person}{Pengchuan Zhang}.} \bibinfo{year}{2019}\natexlab{}.
\newblock \showarticletitle{A convex relaxation barrier to tight robustness verification of neural networks}.
\newblock \bibinfo{journal}{\emph{Advances in Neural Information Processing Systems}}  \bibinfo{volume}{32} (\bibinfo{year}{2019}).
\newblock


\bibitem[Serre et~al\mbox{.}(2021)]%
        {gpupoly}
\bibfield{author}{\bibinfo{person}{Fran{\c c}ois Serre}, \bibinfo{person}{Christoph M{\"u}ller}, \bibinfo{person}{Gagandeep Singh}, \bibinfo{person}{Markus P{\"u}schel}, {and} \bibinfo{person}{Martin Vechev}.} \bibinfo{year}{2021}\natexlab{}.
\newblock \showarticletitle{Scaling Polyhedral Neural Network Verification on {GPU}s}. In \bibinfo{booktitle}{\emph{Proc. Machine Learning and Systems (MLSys'21)}}.
\newblock


\bibitem[Singh et~al\mbox{.}(2018)]%
        {SGMPV18}
\bibfield{author}{\bibinfo{person}{Gagandeep Singh}, \bibinfo{person}{Timon Gehr}, \bibinfo{person}{Matthew Mirman}, \bibinfo{person}{Markus P{\"{u}}schel}, {and} \bibinfo{person}{Martin~T. Vechev}.} \bibinfo{year}{2018}\natexlab{}.
\newblock \showarticletitle{Fast and Effective Robustness Certification}. In \bibinfo{booktitle}{\emph{Proceedings of the Annual Conference on Neural Information Processing Systems}}. \bibinfo{pages}{10825--10836}.
\newblock


\bibitem[Singh et~al\mbox{.}(2019)]%
        {SGPV19}
\bibfield{author}{\bibinfo{person}{Gagandeep Singh}, \bibinfo{person}{Timon Gehr}, \bibinfo{person}{Markus P{\"{u}}schel}, {and} \bibinfo{person}{Martin~T. Vechev}.} \bibinfo{year}{2019}\natexlab{}.
\newblock \showarticletitle{An abstract domain for certifying neural networks}.
\newblock \bibinfo{journal}{\emph{Proceedings of the ACM on Programming Languages (POPL)}}  \bibinfo{volume}{3} (\bibinfo{year}{2019}), \bibinfo{pages}{41:1--41:30}.
\newblock
\urldef\tempurl%
\url{https://doi.org/10.1145/3290354}
\showDOI{\tempurl}


\bibitem[Song et~al\mbox{.}(2021)]%
        {SongLCFL21}
\bibfield{author}{\bibinfo{person}{Fu Song}, \bibinfo{person}{Yusi Lei}, \bibinfo{person}{Sen Chen}, \bibinfo{person}{Lingling Fan}, {and} \bibinfo{person}{Yang Liu}.} \bibinfo{year}{2021}\natexlab{}.
\newblock \showarticletitle{Advanced evasion attacks and mitigations on practical ML-based phishing website classifiers}.
\newblock \bibinfo{journal}{\emph{Int. J. Intell. Syst.}} \bibinfo{volume}{36}, \bibinfo{number}{9} (\bibinfo{year}{2021}), \bibinfo{pages}{5210--5240}.
\newblock


\bibitem[Stutz et~al\mbox{.}(2023)]%
        {randomDNN}
\bibfield{author}{\bibinfo{person}{David Stutz}, \bibinfo{person}{Nandhini Chandramoorthy}, \bibinfo{person}{Matthias Hein}, {and} \bibinfo{person}{Bernt Schiele}.} \bibinfo{year}{2023}\natexlab{}.
\newblock \showarticletitle{Random and Adversarial Bit Error Robustness: Energy-Efficient and Secure DNN Accelerators}.
\newblock \bibinfo{journal}{\emph{IEEE Transactions on Pattern Analysis and Machine Intelligence}} \bibinfo{volume}{45}, \bibinfo{number}{3} (\bibinfo{year}{2023}), \bibinfo{pages}{3632--3647}.
\newblock
\urldef\tempurl%
\url{https://doi.org/10.1109/TPAMI.2022.3181972}
\showDOI{\tempurl}


\bibitem[Tol et~al\mbox{.}(2023)]%
        {DONTKNOCK}
\bibfield{author}{\bibinfo{person}{M. Tol}, \bibinfo{person}{S. Islam}, \bibinfo{person}{A.~J. Adiletta}, \bibinfo{person}{B. Sunar}, {and} \bibinfo{person}{Z. Zhang}.} \bibinfo{year}{2023}\natexlab{}.
\newblock \showarticletitle{Don't Knock! Rowhammer at the Backdoor of DNN Models}. In \bibinfo{booktitle}{\emph{2023 53rd Annual IEEE/IFIP International Conference on Dependable Systems and Networks (DSN)}}. \bibinfo{publisher}{IEEE Computer Society}, \bibinfo{address}{Los Alamitos, CA, USA}, \bibinfo{pages}{109--122}.
\newblock
\urldef\tempurl%
\url{https://doi.org/10.1109/DSN58367.2023.00023}
\showDOI{\tempurl}


\bibitem[Tol et~al\mbox{.}(2022)]%
        {tol2022toward}
\bibfield{author}{\bibinfo{person}{M~Caner Tol}, \bibinfo{person}{Saad Islam}, \bibinfo{person}{Berk Sunar}, {and} \bibinfo{person}{Ziming Zhang}.} \bibinfo{year}{2022}\natexlab{}.
\newblock \showarticletitle{Toward realistic backdoor injection attacks on dnns using rowhammer}.
\newblock \bibinfo{journal}{\emph{arXiv preprint arXiv:2110.07683}} (\bibinfo{year}{2022}).
\newblock


\bibitem[Tran et~al\mbox{.}(2020)]%
        {TranBXJ20}
\bibfield{author}{\bibinfo{person}{Hoang{-}Dung Tran}, \bibinfo{person}{Stanley Bak}, \bibinfo{person}{Weiming Xiang}, {and} \bibinfo{person}{Taylor~T. Johnson}.} \bibinfo{year}{2020}\natexlab{}.
\newblock \showarticletitle{Verification of Deep Convolutional Neural Networks Using ImageStars}. In \bibinfo{booktitle}{\emph{Proceedings of the International Conference on Computer Aided Verification}}. \bibinfo{pages}{18--42}.
\newblock


\bibitem[Tran et~al\mbox{.}(2019)]%
        {TranLMYNXJ19}
\bibfield{author}{\bibinfo{person}{Hoang{-}Dung Tran}, \bibinfo{person}{Diego~Manzanas Lopez}, \bibinfo{person}{Patrick Musau}, \bibinfo{person}{Xiaodong Yang}, \bibinfo{person}{Luan~Viet Nguyen}, \bibinfo{person}{Weiming Xiang}, {and} \bibinfo{person}{Taylor~T. Johnson}.} \bibinfo{year}{2019}\natexlab{}.
\newblock \showarticletitle{Star-Based Reachability Analysis of Deep Neural Networks}. In \bibinfo{booktitle}{\emph{Proceedings of the 3rd World Congress on Formal Methods}}. \bibinfo{pages}{670--686}.
\newblock


\bibitem[Walker et~al\mbox{.}(2021)]%
        {walker2021dram}
\bibfield{author}{\bibinfo{person}{Andrew~J Walker}, \bibinfo{person}{Sungkwon Lee}, {and} \bibinfo{person}{Dafna Beery}.} \bibinfo{year}{2021}\natexlab{}.
\newblock \showarticletitle{On DRAM rowhammer and the physics of insecurity}.
\newblock \bibinfo{journal}{\emph{IEEE Transactions on Electron Devices}} \bibinfo{volume}{68}, \bibinfo{number}{4} (\bibinfo{year}{2021}), \bibinfo{pages}{1400--1410}.
\newblock


\bibitem[Wang et~al\mbox{.}(2018)]%
        {WangPWYJ18}
\bibfield{author}{\bibinfo{person}{Shiqi Wang}, \bibinfo{person}{Kexin Pei}, \bibinfo{person}{Justin Whitehouse}, \bibinfo{person}{Junfeng Yang}, {and} \bibinfo{person}{Suman Jana}.} \bibinfo{year}{2018}\natexlab{}.
\newblock \showarticletitle{Formal security analysis of neural networks using symbolic intervals}. In \bibinfo{booktitle}{\emph{{USENIX} Security Symposium'18}}. \bibinfo{pages}{1599--1614}.
\newblock


\bibitem[Wang et~al\mbox{.}(2021)]%
        {betaCrown}
\bibfield{author}{\bibinfo{person}{Shiqi Wang}, \bibinfo{person}{Huan Zhang}, \bibinfo{person}{Kaidi Xu}, \bibinfo{person}{Xue Lin}, \bibinfo{person}{Suman Jana}, \bibinfo{person}{Cho-Jui Hsieh}, {and} \bibinfo{person}{J~Zico Kolter}.} \bibinfo{year}{2021}\natexlab{}.
\newblock \showarticletitle{Beta-crown: Efficient bound propagation with per-neuron split constraints for neural network robustness verification}.
\newblock \bibinfo{journal}{\emph{Advances in Neural Information Processing Systems}}  \bibinfo{volume}{34} (\bibinfo{year}{2021}).
\newblock


\bibitem[Weng et~al\mbox{.}(2020)]%
        {weng2020towards}
\bibfield{author}{\bibinfo{person}{Tsui-Wei Weng}, \bibinfo{person}{Pu Zhao}, \bibinfo{person}{Sijia Liu}, \bibinfo{person}{Pin-Yu Chen}, \bibinfo{person}{Xue Lin}, {and} \bibinfo{person}{Luca Daniel}.} \bibinfo{year}{2020}\natexlab{}.
\newblock \showarticletitle{Towards certificated model robustness against weight perturbations}. In \bibinfo{booktitle}{\emph{Proceedings of the AAAI Conference on Artificial Intelligence}}, Vol.~\bibinfo{volume}{34}. \bibinfo{pages}{6356--6363}.
\newblock


\bibitem[Wicker et~al\mbox{.}(2020)]%
        {wicker2020probabilistic}
\bibfield{author}{\bibinfo{person}{Matthew Wicker}, \bibinfo{person}{Luca Laurenti}, \bibinfo{person}{Andrea Patane}, {and} \bibinfo{person}{Marta Kwiatkowska}.} \bibinfo{year}{2020}\natexlab{}.
\newblock \showarticletitle{Probabilistic safety for bayesian neural networks}. In \bibinfo{booktitle}{\emph{Conference on uncertainty in artificial intelligence}}. PMLR, \bibinfo{pages}{1198--1207}.
\newblock


\bibitem[WikiChip(2022)]%
        {FSDChip}
\bibfield{author}{\bibinfo{person}{WikiChip}.} \bibinfo{year}{Accessed April 30, 2022}\natexlab{}.
\newblock \bibinfo{title}{FSD Chip - Tesla}.
\newblock \bibinfo{howpublished}{\url{https://en.wikichip.org/wiki/tesla_(car_company)/fsd_chip}}.
\newblock


\bibitem[Xu et~al\mbox{.}(2021)]%
        {xu2021fastcompleteenablingcomplete}
\bibfield{author}{\bibinfo{person}{Kaidi Xu}, \bibinfo{person}{Huan Zhang}, \bibinfo{person}{Shiqi Wang}, \bibinfo{person}{Yihan Wang}, \bibinfo{person}{Suman Jana}, \bibinfo{person}{Xue Lin}, {and} \bibinfo{person}{Cho-Jui Hsieh}.} \bibinfo{year}{2021}\natexlab{}.
\newblock \bibinfo{title}{Fast and Complete: Enabling Complete Neural Network Verification with Rapid and Massively Parallel Incomplete Verifiers}.
\newblock
\newblock
\showeprint[arxiv]{2011.13824}~[cs.AI]
\urldef\tempurl%
\url{https://arxiv.org/abs/2011.13824}
\showURL{%
\tempurl}


\bibitem[Ya{\u{g}}lik{\c{c}}i et~al\mbox{.}(2021)]%
        {yauglikcci2021blockhammer}
\bibfield{author}{\bibinfo{person}{A~Giray Ya{\u{g}}lik{\c{c}}i}, \bibinfo{person}{Minesh Patel}, \bibinfo{person}{Jeremie~S Kim}, \bibinfo{person}{Roknoddin Azizi}, \bibinfo{person}{Ataberk Olgun}, \bibinfo{person}{Lois Orosa}, \bibinfo{person}{Hasan Hassan}, \bibinfo{person}{Jisung Park}, \bibinfo{person}{Konstantinos Kanellopoulos}, \bibinfo{person}{Taha Shahroodi}, {et~al\mbox{.}}} \bibinfo{year}{2021}\natexlab{}.
\newblock \showarticletitle{Blockhammer: Preventing rowhammer at low cost by blacklisting rapidly-accessed dram rows}. In \bibinfo{booktitle}{\emph{2021 IEEE International Symposium on High-Performance Computer Architecture (HPCA)}}. IEEE, \bibinfo{pages}{345--358}.
\newblock


\bibitem[Yang et~al\mbox{.}(2021)]%
        {YLLHWSXZ20}
\bibfield{author}{\bibinfo{person}{Pengfei Yang}, \bibinfo{person}{Renjue Li}, \bibinfo{person}{Jianlin Li}, \bibinfo{person}{Cheng{-}Chao Huang}, \bibinfo{person}{Jingyi Wang}, \bibinfo{person}{Jun Sun}, \bibinfo{person}{Bai Xue}, {and} \bibinfo{person}{Lijun Zhang}.} \bibinfo{year}{2021}\natexlab{}.
\newblock \showarticletitle{Improving Neural Network Verification through Spurious Region Guided Refinement}. In \bibinfo{booktitle}{\emph{Proceedings of 27th International Conference on Tools and Algorithms for the Construction and Analysis of Systems (TACAS)}}, \bibfield{editor}{\bibinfo{person}{Jan~Friso Groote} {and} \bibinfo{person}{Kim~Guldstrand Larsen}} (Eds.). \bibinfo{pages}{389--408}.
\newblock


\bibitem[Yao et~al\mbox{.}(2020)]%
        {yao2020deephammer}
\bibfield{author}{\bibinfo{person}{Fan Yao}, \bibinfo{person}{Adnan~Siraj Rakin}, {and} \bibinfo{person}{Deliang Fan}.} \bibinfo{year}{2020}\natexlab{}.
\newblock \showarticletitle{DeepHammer: Depleting the intelligence of deep neural networks through targeted chain of bit flips}. In \bibinfo{booktitle}{\emph{29th USENIX Security Symposium (USENIX Security 20)}}. \bibinfo{pages}{1463--1480}.
\newblock


\bibitem[Zhang et~al\mbox{.}(2023c)]%
        {zhang2023post}
\bibfield{author}{\bibinfo{person}{Jinjie Zhang}, \bibinfo{person}{Yixuan Zhou}, {and} \bibinfo{person}{Rayan Saab}.} \bibinfo{year}{2023}\natexlab{c}.
\newblock \showarticletitle{Post-training quantization for neural networks with provable guarantees}.
\newblock \bibinfo{journal}{\emph{SIAM Journal on Mathematics of Data Science}} \bibinfo{volume}{5}, \bibinfo{number}{2} (\bibinfo{year}{2023}), \bibinfo{pages}{373--399}.
\newblock


\bibitem[Zhang et~al\mbox{.}(2024)]%
        {ZhangCSSD24}
\bibfield{author}{\bibinfo{person}{Yedi Zhang}, \bibinfo{person}{Guangke Chen}, \bibinfo{person}{Fu Song}, \bibinfo{person}{Jun Sun}, {and} \bibinfo{person}{Jin~Song Dong}.} \bibinfo{year}{2024}\natexlab{}.
\newblock \showarticletitle{Certified Quantization Strategy Synthesis for Neural Networks}. In \bibinfo{booktitle}{\emph{Proceedings of the 26th International Symposium on Formal Methods ({FM}), Part {I}}}, \bibfield{editor}{\bibinfo{person}{Andr{\'{e}} Platzer}, \bibinfo{person}{Kristin~Yvonne Rozier}, \bibinfo{person}{Matteo Pradella}, {and} \bibinfo{person}{Matteo Rossi}} (Eds.), Vol.~\bibinfo{volume}{14933}. \bibinfo{pages}{343--362}.
\newblock
\urldef\tempurl%
\url{https://doi.org/10.1007/978-3-031-71162-6\_18}
\showDOI{\tempurl}


\bibitem[Zhang et~al\mbox{.}(2023a)]%
        {zhang2023qebverif}
\bibfield{author}{\bibinfo{person}{Yedi Zhang}, \bibinfo{person}{Fu Song}, {and} \bibinfo{person}{Jun Sun}.} \bibinfo{year}{2023}\natexlab{a}.
\newblock \showarticletitle{QEBVerif: Quantization error bound verification of neural networks}. In \bibinfo{booktitle}{\emph{International Conference on Computer Aided Verification}}. Springer, \bibinfo{pages}{413--437}.
\newblock


\bibitem[Zhang et~al\mbox{.}(2021)]%
        {BDD4BNN}
\bibfield{author}{\bibinfo{person}{Yedi Zhang}, \bibinfo{person}{Zhe Zhao}, \bibinfo{person}{Guangke Chen}, \bibinfo{person}{Fu Song}, {and} \bibinfo{person}{Taolue Chen}.} \bibinfo{year}{2021}\natexlab{}.
\newblock \showarticletitle{{BDD4BNN:} {A} BDD-Based Quantitative Analysis Framework for Binarized Neural Networks}. In \bibinfo{booktitle}{\emph{Proceedings of the 33rd International Conference on Computer Aided Verification (CAV)}}. \bibinfo{pages}{175--200}.
\newblock
\urldef\tempurl%
\url{https://doi.org/10.1007/978-3-030-81685-8\_8}
\showDOI{\tempurl}


\bibitem[Zhang et~al\mbox{.}(2023b)]%
        {zhang2023precise}
\bibfield{author}{\bibinfo{person}{Yedi Zhang}, \bibinfo{person}{Zhe Zhao}, \bibinfo{person}{Guangke Chen}, \bibinfo{person}{Fu Song}, {and} \bibinfo{person}{Taolue Chen}.} \bibinfo{year}{2023}\natexlab{b}.
\newblock \showarticletitle{Precise Quantitative Analysis of Binarized Neural Networks: A {BDD}-based Approach}.
\newblock \bibinfo{journal}{\emph{ACM Transactions on Software Engineering and Methodology}} \bibinfo{volume}{32}, \bibinfo{number}{3} (\bibinfo{year}{2023}), \bibinfo{pages}{1--51}.
\newblock
\urldef\tempurl%
\url{https://doi.org/10.1145/3563212}
\showDOI{\tempurl}


\bibitem[Zhang et~al\mbox{.}(2022)]%
        {ZZCSZC22}
\bibfield{author}{\bibinfo{person}{Yedi Zhang}, \bibinfo{person}{Zhe Zhao}, \bibinfo{person}{Guangke Chen}, \bibinfo{person}{Fu Song}, \bibinfo{person}{Min Zhang}, \bibinfo{person}{Taolue Chen}, {and} \bibinfo{person}{Jun Sun}.} \bibinfo{year}{2022}\natexlab{}.
\newblock \showarticletitle{QVIP: An ILP-based Formal Verification Approach for Quantized Neural Networks}. In \bibinfo{booktitle}{\emph{Proceedings of the 37th IEEE/ACM International Conference on Automated Software Engineering (ASE)}}. \bibinfo{pages}{82:1--82:13}.
\newblock
\urldef\tempurl%
\url{https://doi.org/10.1145/3551349.3556916}
\showDOI{\tempurl}


\bibitem[Zhao et~al\mbox{.}(2024)]%
        {ZhaoCLLSWS24}
\bibfield{author}{\bibinfo{person}{Zhe Zhao}, \bibinfo{person}{Guangke Chen}, \bibinfo{person}{Tong Liu}, \bibinfo{person}{Taishan Li}, \bibinfo{person}{Fu Song}, \bibinfo{person}{Jingyi Wang}, {and} \bibinfo{person}{Jun Sun}.} \bibinfo{year}{2024}\natexlab{}.
\newblock \showarticletitle{Attack as Detection: Using Adversarial Attack Methods to Detect Abnormal Examples}.
\newblock \bibinfo{journal}{\emph{{ACM} Trans. Softw. Eng. Methodol.}} \bibinfo{volume}{33}, \bibinfo{number}{3} (\bibinfo{year}{2024}), \bibinfo{pages}{68:1--68:45}.
\newblock
\urldef\tempurl%
\url{https://doi.org/10.1145/3631977}
\showDOI{\tempurl}


\bibitem[Zhao et~al\mbox{.}(2022)]%
        {ZZCSCL22}
\bibfield{author}{\bibinfo{person}{Zhe Zhao}, \bibinfo{person}{Yedi Zhang}, \bibinfo{person}{Guangke Chen}, \bibinfo{person}{Fu Song}, \bibinfo{person}{Taolue Chen}, {and} \bibinfo{person}{Jiaxiang Liu}.} \bibinfo{year}{2022}\natexlab{}.
\newblock \showarticletitle{{CLEVEREST:} Accelerating CEGAR-based Neural Network Verification via Adversarial Attacks}. In \bibinfo{booktitle}{\emph{Proceedings of the 29th International Symposium on Static Analysis}}. \bibinfo{pages}{449--473}.
\newblock
\urldef\tempurl%
\url{https://doi.org/10.1007/978-3-031-22308-2\_20}
\showDOI{\tempurl}


\bibitem[Zhou et~al\mbox{.}(2022)]%
        {zhou2022incremental}
\bibfield{author}{\bibinfo{person}{Aojun Zhou}, \bibinfo{person}{Anbang Yao}, \bibinfo{person}{Yiwen Guo}, \bibinfo{person}{Lin Xu}, {and} \bibinfo{person}{Yurong Chen}.} \bibinfo{year}{2022}\natexlab{}.
\newblock \showarticletitle{Incremental Network Quantization: Towards Lossless CNNs with Low-precision Weights}. In \bibinfo{booktitle}{\emph{International Conference on Learning Representations}}.
\newblock


\end{thebibliography}

\appendix

\section{Overview of the ACAS Xu Benchmark}\label{sec:app-bench}
Table~\ref{tab:bench_C} gives the details of the ACAS Xu Benchmark we used throughout Section~\ref{sec:exp}. We evaluate all 45 QNNs, denoted as $N_{i,j}$ with $1\le i\le 5,1\le j\le 9$, on the 10 properties, then select the successfully proved ones, resulting in a total of $8+9+8+9+7+14=55$ network-property pairs as our benchmarks. 

\begin{table}[h]
    \centering
    \caption{Benchmarks of properties and QNNs obtained via post-quantization training for ACAS Xu.}
    \setlength{\tabcolsep}{2pt}
    \scalebox{0.8}{
    \begin{tabular}{l|l|l}
        \toprule
         Property  & Description &  Network    \\ \midrule
         Prop\_3\_WL & \tabincell{l}{If the intruder is directly ahead and is moving towards the \\ ownership, a ``Weak Left'' maneuver is advised.} & \tabincell{l} {(8): $N_{2,6},N_{2,7},N_{2,8},N_{2,9},$ \\ $N_{4,6},N_{4,7},N_{4,8},N_{4,9}$} \\ \midrule
         
         Prop\_3\_WR & \tabincell{l}{If the intruder is directly ahead and is moving towards the \\ ownership, a ``Weak Right'' maneuver is advised.} & \tabincell{l}{(9): $N_{1,6},N_{3,6},N_{3,7},N_{3,8},$ \\ $N_{3,9},N_{5,6},N_{5,7},N_{5,8},N_{5,9}$ } \\ \midrule

         Prop\_3\_SL & \tabincell{l}{If the intruder is directly ahead and is moving towards the \\ ownership, a ``Strong Left'' maneuver is advised.} & \tabincell{l}{(8): $N_{2,2},N_{2,3},N_{2,4},N_{2,5},$ \\ $N_{4,2},N_{4,3},N_{4,4},N_{4,5}$} \\ \midrule

         Prop\_3\_SR & \tabincell{l}{If the intruder is directly ahead and is moving towards the \\ ownership, a ``Strong Right'' maneuver is advised.} & \tabincell{l}{(9): $N_{3,1},N_{3,3},N_{3,4},N_{3,5},$ \\ $N_{5,1}, N_{5,2},N_{5,3},N_{5,4},N_{5,5}$} \\ \midrule
         
         Prop\_5\_SR & \tabincell{l}{If the intruder is near and approaching from the left, a \\ ``Strong Right'' maneuver is advised.} & \tabincell{l}{(7): $N_{3,1},N_{3,2},N_{3,3},N_{5,2},$ \\ $N_{5,3},N_{5,4}, N_{5,5}$} \\ \midrule

         Prop\_10\_COC & \tabincell{l}{For a far away intruder, a ``Clear of Conflict'' maneuver is \\ advised.} & \tabincell{l}{(14): $N_{1,3},N_{1,4},N_{1,5},N_{1,6}$\\ $N_{3,2},N_{3,6},N_{3,7},N_{4,1},N_{4,2},$ \\ $N_{4,4},N_{4,5},N_{5,1},N_{5,4},N_{5,6}$} \\ 
        
        \bottomrule
    \end{tabular}}
    \label{tab:bench_C}
\end{table}

\section{Missing Proofs in Sections \ref{sec:pro} and~\ref{sec:method}}

\subsection{Proof of Theorem~\ref{the:npc}}

\begin{proof}
To show that the problem of checking whether $\mN\models^\rho_{\mm,\nn}\langle \phi,\psi \rangle$ holds is in NP, we can
\begin{enumerate}
  \item {\bf Step 1:} non-deterministically guess an input $\bs{x}\in \mathbb{R}^n$ and an $(k,\nn)$-attack vector $\rho=\{(\alpha_1,P_1),\cdots, (\alpha_k,P_k)\}$ for $k\leq \mm$;
  \item {\bf Step 2:} build a new neural network $\mN^\rho$ according to the $(k,\nn)$-fault attack vector $\rho$;
  \item {\bf Step 3:} compute $\mN^\rho(\bs{x})$  by feeding the values of the input $\bs{x}$ forward through the network;
  \item {\bf Step 4:} check if both $\phi(\bs{x})$ and $\psi(\mN^\rho(\bs{x}))$ hold, and return satisfiable if both $\phi(\bs{x})$ and $\psi(\mN^\rho(\bs{x}))$ hold.
\end{enumerate}
Since Steps 2--4 can be done in polynomial time, we conclude the proof.

The NP-hardness is proved by reducing from the satisfiability problem of the vanilla neural network verification problem $\mN\models \langle \phi,\psi\rangle$ which is NP-complete~\cite{GuyKatz2017ReluplexAE}.
Consider a vanilla neural network verification problem of checking whether $\mN\models \langle \phi,\psi\rangle$ holds.
Suppose the inputs and outputs of the neural network are $\bs{x}$ and $\bs{y} = \mN(\bs{x})$, respectively.
We construct a neural network  $\mN'$ as follows:
\begin{itemize}
  \item $\mN'$ comprises $\nn+1$ copies of the network verification $\mN$ in parallel,
  \item all the copies share the same inputs $\bs{x}$,
  \item the outputs of the $i$-th copy are renamed by $\bs{y}_i$,
  \item the weights of the edges between two neurons in two different copies are $0$,  ensuring that the neurons
  in the $i$-th copy are not affected by the neurons in other copies.
\end{itemize}
Let $\psi'=\bigvee_{i=1}^{\nn+1}\psi_i$, where $\psi_i$ is obtained from the property $\psi$ by renaming
the outputs $\bs{y}$ with the outputs $\bs{y}_i$.

{\bf Claim.} \textit{For any fixed constants $\mm$ and $\nn$,
$\mN' \models^\rho_{\mm,\nn} \langle \phi',\psi'\rangle$ holds
iff $\mN \models \langle \phi,\psi\rangle$ holds.}

$(\Leftarrow)$ Suppose the vanilla neural network verification $\mN\models\langle \phi,\psi\rangle$ holds,
then for any inputs $\bs{x}\in \mathbb{R}^n$ that satisfies the pre-condition $\phi$, $\bs{y}=N(\bs{x})$ also satisfies
the post-condition $\phi$. According to the construction of $N'$, 
for any $(k,\nn)$-fault attack vector $\rho$ with $k\leq \mm$, 
there exists a copy of $N$, say
the $i$-th copy of $N$, such that the outputs $\bs{y}_i$ are the same as the outputs $\bs{y}$.
It implies that $N'(\bs{x})$ satisfies $\psi_i$, hence $\psi'$.
Thus, $\mN'\models^\rho_{\mm,\nn}\langle \phi,\psi'\rangle$ holds.

$(\Rightarrow)$
Suppose the vanilla neural network verification problem $\mN\models \langle \phi,\psi\rangle$ does not hold, then 
there exists a counterexample $\bs{x}\in \mathbb{R}^n$ such that
$\bs{x}$ satisfies the pre-condition $\phi$ but $\bs{y}=\mN(\bs{x})$ does not satisfy
the post-condition $\phi$. According to the construction of $\mN'$, 
the outputs $\mN'(\bs{x})$ of $\mN'$ under an $(\mm,0)$-fault attack vector $\rho$ (i.e., no parameters can be changed)
are $\nn+1$ copies of $\bs{y}=\mN(\bs{x})$.
Thus, $\mN'(\bs{x})$ does not satisfy $\psi'=\bigvee_{i=1}^{\nn+1}\psi_i$,
i.e., $\mN'\models^\rho_{\mm,\nn}\langle \phi,\psi'\rangle$ does not hold.
\end{proof}

\subsection{Proof of Theorem~\ref{the:deepPolyR_weight}}

\begin{proof}

We first prove the soundness of the weighted-ReLU abstract transformer. Let $\bs{a}^{i}_{k,1}=\langle a^{i,\le}_{k,1},a^{i,\ge}_{k,1},l^{i}_{k,1},u^{i}_{k,1} \rangle$ be the abstract element of $\bs{x}^i_{k,1}$ and $\bs{a}^{i}_{k,2}=\langle a^{i,\le}_{k,2},a^{i,\ge}_{k,2},l^{i}_{k,2},u^{i}_{k,2} \rangle$ be the abstract element of $\bs{x}^i_{k,2}=\text{ReLU}(\bs{x}^i_{k,1})$. 
$\gamma(\bs{a}^i_{k,1}) = \{ x\in\mathbb{R}\mid a^{i,\le}_{k,1}\le x \le a^{i,\ge}_{k,1}\}$.         
Given $\overrightarrow{\bs{W}}^{i+1}_{j,k}\in[w_l,w_u]$, we prove the soundness by considering the following 5 cases:
\begin{itemize}
        \item If $l^i_{k,1}\ge 0$: $\text{ReLU}(\bs{x}^i_{k,1})=\bs{x}^i_{k,1}$ for $\bs{x}^i_{k,1}\in\gamma(\bs{a}^i_{k,1})$ and $a^{i,\le}_{k,2}=a^{i,\ge}_{k,2}=\bs{x}^{i}_{k,1}$. Then, we have $w_l \cdot \bs{x}^i_{k,1}\le \overrightarrow{\bs{W}}^{i+1}_{j,k}\cdot \text{ReLU}(\bs{x}^i_{k,1})\le w_u\cdot \bs{x}^{i}_{k,1}$ and therefore
        \begin{displaymath}
        \begin{aligned}
            \overrightarrow{\bs{W}}^{i+1}_{j,k}\cdot \text{ReLU}(\gamma(\bs{a}^i_{k,1})) & \subseteq \{ x\in\mathbb{R} \mid w_l \cdot x' \le x \le w_u \cdot x' \wedge a^{i,\le}_{k,1} \le x'\le a^{i,\ge}_{k,1}\} \\
            & = \{ x\in\mathbb{R} \mid w_l \cdot a^{i,\le}_{k,2} \le x \le w_u\cdot a^{i,\ge}_{k,2}\} = \gamma(\bs{a}^{i,*}_{k,2})
        \end{aligned}
        \end{displaymath}
        
        \item If $u^i_{k,1}\le 0$: $\text{ReLU}(\bs{x}^i_{k,1})=0$ for $\bs{x}^i_{k,1}\in\gamma(\bs{a}^i_{k,1})$ and $a^{i,\le}_{k,2}=a^{i,\ge}_{k,2}=0$. Then, we have $0 \le \overrightarrow{\bs{W}}^{i+1}_{j,k}\cdot \text{ReLU}(\bs{x}^i_{k,1})\le 0$ and therefore 
        \begin{displaymath}
        \begin{aligned}
            \overrightarrow{\bs{W}}^{i+1}_{j,k}\cdot \text{ReLU}(\gamma(\bs{a}^i_{k,1})) & \subseteq \{ x\in\mathbb{R} \mid 0 \le x \le 0 \wedge a^{i,\le}_{k,1} \le x'\le a^{i,\ge}_{k,1}\} \\
            & = \{ x\in\mathbb{R} \mid w_l \cdot a^{i,\le}_{k,2} \le x \le w_u\cdot a^{i,\ge}_{k,2}\} = \gamma(\bs{a}^{i,*}_{k,2})
        \end{aligned}
        \end{displaymath}
        
        \item If $l^i_{k,1} u^i_{k,1}<0$: we have $\lambda \cdot \bs{x}^i_{k,1}\le \text{ReLU}(\bs{x}^i_{k,1})\le \frac{u^i_{k,1}}{u^i_{k,1}-l^i_{k,1}}(\bs{x}^i_{k,1}-l^i_{k,1})$ for $\bs{x}^i_{k,1}\in\gamma(\bs{a}^i_{k,1})$.   
        $a^{i,\le}_{k,2}=\lambda\cdot \bs{x}^{i}_{k,1}$ and $a^{i,\ge}_{k,2}=\frac{u^i_{k,1}}{u^i_{k,1}-l^i_{k,1}}(\bs{x}^i_{k,1}-l^i_{k,1})$.
        \begin{itemize}
            \item When $w_l \ge 0$: we have $w_l \cdot \lambda \cdot \bs{x}^i_{k,1}\le \overrightarrow{\bs{W}}^{i+1}_{j,k}\cdot \text{ReLU}(\bs{x}^i_{k,1})\le w_u\cdot \frac{u^i_{k,1}}{u^i_{k,1}-l^i_{k,1}}(\bs{x}^i_{k,1}-l^i_{k,1})$ and therefore
            \begin{displaymath}
            \begin{aligned}
                \overrightarrow{\bs{W}}^{i+1}_{j,k}\cdot \text{ReLU}(\gamma(\bs{a}^i_{k,1})) & \subseteq \{ x\in\mathbb{R} \mid w_l \cdot \lambda \cdot x' \le x \le w_u \cdot \frac{u^i_{k,1}}{u^i_{k,1}-l^i_{k,1}}(x'-l^i_{k,1})  \wedge a^{i,\le}_{k,1} \le x'\le a^{i,\ge}_{k,1}\} \\
                & = \{ x\in\mathbb{R} \mid w_l \cdot a^{i,\le}_{k,2} \le x \le w_u\cdot a^{i,\ge}_{k,2}\} \\
                & = \{x\in\mathbb{R} \mid \tilde{a}^{i,\le}_{k,2} \le x \le \tilde{a}^{i,\ge}_{k,2}\} = \gamma(\bs{a}^{i,*}_{k,2})
            \end{aligned}
            \end{displaymath}

            \item When $w_u \le 0$: we have $w_l \cdot \frac{u^i_{k,1}}{u^i_{k,1}-l^i_{k,1}}(\bs{x}^i_{k,1}-l^i_{k,1}) \le \overrightarrow{\bs{W}}^{i+1}_{j,k}\cdot \text{ReLU}(\bs{x}^i_{k,1})\le w_u \cdot \lambda \cdot \bs{x}^i_{k,1}$ and therefore
            \begin{displaymath}
            \begin{aligned}
                \overrightarrow{\bs{W}}^{i+1}_{j,k}\cdot \text{ReLU}(\gamma(\bs{a}^i_{k,1})) & \subseteq \{ x\in\mathbb{R} \mid w_l \cdot \frac{u^i_{k,1}}{u^i_{k,1}-l^i_{k,1}}(x'-l^i_{k,1}) \le x \le  w_u \cdot \lambda \cdot x' \wedge a^{i,\le}_{k,1} \le x'\le a^{i,\ge}_{k,1}\} \\
                & = \{ x\in\mathbb{R} \mid w_l \cdot a^{i,\ge}_{k,2} \le x \le w_u\cdot a^{i,\le}_{k,2}\} = \{x\in\mathbb{R} \mid \tilde{a}^{i,\le}_{k,2} \le x \le \tilde{a}^{i,\ge}_{k,2}\} = \gamma(\bs{a}^{i,*}_{k,2})
            \end{aligned}
            \end{displaymath}

            \item When $w_l < 0 < w_u$: if $\lambda=0$, we have $0\le \text{ReLU}(\bs{x}^i_{k,1})\le \frac{u^i_{k,1}}{u^i_{k,1}-l^i_{k,1}}(\bs{x}^i_{k,1}-l^i_{k,1})$. Hence, we have $ w_l \cdot \frac{u^i_{k,1}}{u^i_{k,1}-l^i_{k,1}}(\bs{x}^i_{k,1}-l^i_{k,1}) \le \overrightarrow{\bs{W}}^{i+1}_{j,k}\cdot \text{ReLU}(\bs{x}^i_{k,1})\le w_u \cdot \frac{u^i_{k,1}}{u^i_{k,1}-l^i_{k,1}}(\bs{x}^i_{k,1}-l^i_{k,1})$. If $\lambda=1$, we have $\bs{x}^i_{k,1} \le \text{ReLU}(\bs{x}^i_{k,1})\le \frac{u^i_{k,1}}{u^i_{k,1}-l^i_{k,1}}(\bs{x}^i_{k,1}-l^i_{k,1})$. Since $w_l<0$, we have $w_l\cdot \bs{x}^i_{k,1} \ge w_l\cdot \frac{u^i_{k,1}}{u^i_{k,1}-l^i_{k,1}}(\bs{x}^i_{k,1}-l^i_{k,1})$.
            Finally, we have $ w_l \cdot \frac{u^i_{k,1}}{u^i_{k,1}-l^i_{k,1}}(\bs{x}^i_{k,1}-l^i_{k,1}) \le \overrightarrow{\bs{W}}^{i+1}_{j,k}\cdot \text{ReLU}(\bs{x}^i_{k,1})\le w_u \cdot \frac{u^i_{k,1}}{u^i_{k,1}-l^i_{k,1}}(\bs{x}^i_{k,1}-l^i_{k,1})$ for $\lambda\in\{0,1\}$ and therefore             
            \begin{displaymath}
            \begin{aligned}
                \overrightarrow{\bs{W}}^{i+1}_{j,k}\cdot \text{ReLU}(\gamma(\bs{a}^i_{k,1})) & \subseteq \{ x\in\mathbb{R} \mid w_l \cdot \frac{u^i_{k,1}(x'-l^i_{k,1})}{u^i_{k,1}-l^i_{k,1}} \le x \le  w_u \cdot \frac{u^i_{k,1}(x'-l^i_{k,1})}{u^i_{k,1}-l^i_{k,1}} \wedge a^{i,\le}_{k,1} \le x'\le a^{i,\ge}_{k,1}\} \\
                & = \{ x\in\mathbb{R} \mid w_l \cdot a^{i,\ge}_{k,2} \le x \le w_u\cdot a^{i,\ge}_{k,2}\} = \{x\in\mathbb{R} \mid \tilde{a}^{i,\le}_{k,2} \le x \le \tilde{a}^{i,\ge}_{k,2}\} = \gamma(\bs{a}^{i,*}_{k,2})
            \end{aligned}
            \end{displaymath}
        \end{itemize}
    \end{itemize}
    
    Therefore, \textbf{our weighted-ReLU abstract transformer is sound} in all cases. We then prove that the invariant preserving. Let $\gamma (\bs{a}^{i,*}_{k,2})=\{x\in\mathbb{R}\mid \tilde{a}^{i,\le}_{k,2} \le x \le \tilde{a}^{i,\ge}_{k,2}\}$ and $l^i_{k,2}\le a^{i,\le}_{k,2} \le \bs{x}^i_{k,2} \le a^{i,\ge}_{k,2} \le u^i_{k,2}$:
    \begin{itemize}
        \item If $w_l\ge 0$, we have $ w_l \cdot l^i_{k,2} \le w_l \cdot a^{i,\le}_{k,2}$ and $ w_u \cdot a^{i,\ge}_{k,2} \le w_u \cdot u^i_{k,2}$. Then, we have $\tilde{l}^i_{k,2} = w_l \cdot l^i_{k,2} \le w_l \cdot a^{i,\le}_{k,2} = \tilde{a}^{i,\le}_{k,2} \le \tilde{\bs{x}}^i_{k,2} \le \tilde{a}^{i,\ge}_{k,2} = w_u \cdot a^{i,\ge}_{k,2} \le w_u\cdot u^i_{k,2} = \tilde{u}^i_{k,2}$;

        \item If $w_u\le 0$, we have $ w_l \cdot u^i_{k,2} \le w_l \cdot a^{i,\ge}_{k,2}$ and $ w_u \cdot a^{i,\le}_{k,2} \le w_u \cdot l^i_{k,2}$. Then, we have $\tilde{l}^i_{k,2} = w_l \cdot u^i_{k,2} \le w_l \cdot a^{i,\ge}_{k,2} = \tilde{a}^{i,\le}_{k,2} \le \tilde{\bs{x}}^i_{k,2} \le \tilde{a}^{i,\ge}_{k,2} = w_u \cdot a^{i,\le}_{k,2} \le w_u\cdot l^i_{k,2} = \tilde{u}^i_{k,2}$;

        \item If $w_l<0<w_u$, we have $ w_l \cdot u^i_{k,2} \le w_l \cdot a^{i,\ge}_{k,2}$ and $ w_u \cdot a^{i,\ge}_{k,2} \le w_u \cdot u^i_{k,2}$. Then, we have $\tilde{l}^i_{k,2} = w_l \cdot u^i_{k,2} \le w_l \cdot a^{i,\ge}_{k,2} = \tilde{a}^{i,\le}_{k,2} \le \tilde{\bs{x}}^i_{k,2} \le \tilde{a}^{i,\ge}_{k,2} = w_u \cdot a^{i,\ge}_{k,2} \le w_u\cdot u^i_{k,2} = \tilde{u}^i_{k,2}$;
    \end{itemize}
    
    Therefore, \textbf{our weight-ReLU abstract transformer preserves the invariant} in all cases, and we finish the proving.
\end{proof}


\subsection{Proof of Theorem~\ref{the:input}}
\begin{proof}
    \major{Consider $\bs{x}^2_{j,1}=\sum_{t\in [n_{2}]\backslash k}\bs{W}^{2}_{j,t}\bs{x}^1_{t}+\overrightarrow{\bs{W}}^{2}_{j,k}\bs{x}^1_k+\bs{b}^{2}_j$, we can inherit the proof of affine abstract transformer in \cite{SGPV19} directly on all input neurons $\bs{x}^i_t$ for $t\in[n_2]\backslash k$. To prove the theorem, it remains to demonstrate that the abstract transformer for the weighted input neuron $\overrightarrow{\bs{W}}^2_{j,k}\bs{x}^1_k$ is sound, i.e., to prove that the abstraction $\langle a^\le, a^\ge, l,u\rangle$ preserves all possible concrete values of $\overrightarrow{\bs{W}}^2_{j,k}\bs{x}^1_k$ with the input range of $\bs{x}^1_k$ as $[x_l,x_u]$ and value range of $\overrightarrow{\bs{W}}^2_{j,k}$ as $[w_l,w_u]$, where $a^\le = \kappa^\le \bs{x}^1_k -\eta$, $a^\ge=\kappa^\le \bs{x}^1_k +\eta$, and:
    \begin{itemize}
        \item If $x_l\ge 0$, then $\{\kappa^\le = w_l, \kappa^\ge=w_u, \eta=0\}$; 
        \item If $x_u \le 0$, then $\{\kappa^\le=w_u,\kappa^\ge=w_l,\eta=0\}$;
        \item Otherwise, $\{\kappa^\le=\frac{w_u x_u - w_l x_l}{x_u-x_l}, \kappa^\ge = \frac{w_l x_u - w_u x_l}{x_u-x_l}, \eta = \frac{x_u x_l}{x_u-x_l}(w_l-w_u)\}$.
    \end{itemize}}

    \major{Intuitively, an intuitive sound abstract transformer for the function $y=\overrightarrow{\bs{W}}^2_{j,k} \bs{x}^1_k$ considering $\bs{x}^1_k \in[x_l,x_u]$ and $\overrightarrow{\bs{W}}^2_{j,k}\in[w_l,w_u]$ is the convex quadrilateral constructed by the four vertices: $A=(x_l, w_lx_l)$, $B=(x_u,w_lx_u)$, $C=(x_l, w_ux_l)$, and $D=(x_u,w_ux_u)$. The upper and lower boundaries of the quadrilateral are shown in Table~\ref{tab:affine_bound}.}
    \begin{table}[h]
    \centering
    \caption{\major{Boundaries of the quadrilateral constructed by the four vertices: $A=(x_l, w_lx_l)$, $B=(x_u,w_lx_u)$, $C=(x_l, w_ux_l)$, and $D=(x_u,w_ux_u)$, where $\overline{AB}$ denote the boundary is the line segment of AB.}}\label{tab:affine_bound}
    \setlength{\tabcolsep}{3pt}
    \scalebox{0.85}{
    \begin{tabular}{c|c|c|c|c|c|c|c|c|c}
    \toprule
      &  \multicolumn{3}{c|}{$w_l\ge 0$} & \multicolumn{3}{c|}{$w_u \le 0$} & \multicolumn{3}{c}{$w_l < 0 <w_u$} \\ \cline{2-10}
     & $x_l \ge 0$ & $x_u \le 0$ & $x_l <0<x_u $& $x_l \ge 0$ & $x_u \le 0$ & $x_l <0<x_u $& $x_l \ge 0$ & $x_u \le 0$ & $x_l <0<x_u $ \\ \midrule 

     Upper Boundary& $\overline{CD}$ & $\overline{AB}$ & $\overline{AD}$ & $\overline{CD}$ & $\overline{AB}$ & $\overline{AD}$ & $\overline{CD}$ & $\overline{AB}$ & $\overline{AD}$\\ 
     Lower Boundary& $\overline{AB}$ & $\overline{CD}$ & $\overline{BC}$ & $\overline{AB}$ & $\overline{CD}$ & $\overline{BC}$ & $\overline{AB}$ & $\overline{CD}$ & $\overline{BC}$ \\ 
     
    \bottomrule
    \end{tabular}}
\end{table}

\major{For any $x\in[x_l,x_u]$, the segment $\overline{AB}$ is given by $y=w_l x$, $\overline{CD}$ is given by  $y=w_u x$, $\overline{AD}$ is given by $y=w_lx_l+\frac{w_ux_u-w_lx_l}{x_u-x_l}(x-x_l)$, and $\overline{BC}$ is given by $y=w_lx_u+\frac{w_lx_u-w_ux_l}{x_u-x_l}(x-x_u)$. Since these expressions fully characterize the convex hull of the function over the given input interval, the theorem is thereby proved.}
\end{proof}

\subsection{Proof of Theorem~\ref{the:deepPolyR_bias}}

\begin{proof}
    Since the abstraction loses no precision, soundness and the invariant are preserved directly.
\end{proof}

\subsection{Proof of Theorem~\ref{the:sigmoid}}

\begin{proof}

\major{We prove it by construction. }

\major{Assuming that the abstract element of $g(x)$ obtained in \deepPoly is $\langle a^\le,a^\ge, l, u\rangle$, let $\langle \tilde{a}^\le, \tilde{a}^\ge, \tilde{l},\tilde{u} \rangle$ denote the abstract element of $w\cdot g(x)$. We use $g^\prime(x)$ and $g^{\prime\prime}(x)$ to denote the first and second derivatives of $g(x)$. Next, we prove the soundness of the abstract transformer demonstrated in Table~\ref{tab:log_AbT} by construction based on the existing proof ideas on the abstract transformers of the Sigmoid and Tanh activation functions in~\cite{SGPV19}, which can be illustrated as follows:
\begin{itemize}
    \item When $l \ge 0$, $a^\le$ is given by the line segment defined by two points $(l, g(l))$ and $(u, g(u))$, i.e., the slope is $\kappa = \frac{g(u)-g(l)}{u-l}$. This is because $g$ is concave on $[l, u]$; Otherwise, $a^\le$ is given by the function defined by the point $(l, g(l))$ and a minimum slope $\kappa^\prime = \text{min}(g^\prime(l), g^\prime(u))$, i.e., $y(x)=\kappa^\prime(x-l)+g(l)$.
    This is because $g^\prime$ is non-decreasing on $(l,0]$ and decreasing on $[0, u)$, then by setting the slope as  $\kappa^\prime$, we can always guarantee that $\kappa^\prime$ is the minimum derivative for any point on $g(x)$ with $x\in[l,u]$, hence for any point $x_c\in[l,u]$, $\kappa^\prime (x-x_c)+g(x_c)$ always lies below of $g(x)$ on $[x_c,u]$. Hence, when $x_c=l$, we have $\kappa^\prime (x-l)+g(l)$ always lies below of $g(x)$ on $[l,u]$.
    \item When $u \le 0$, $a^\ge$ is given by the line segment defined by two points $(l, g(l))$ and $(u, g(u))$,  i.e., the slope is $\kappa = \frac{g(u)-g(l)}{u-l}$. This is because $g$ is convex on $[l, u]$; Otherwise, $a^\ge$ is given by the function defined by the point $(u,g(u))$ and the minimum slope $\kappa^\prime = \text{min}(g^\prime(l), g^\prime(u))$, i.e., $y(x)=\kappa^\prime(x-u)+g(u)$. This is because $g^\prime$ is non-decreasing on $(l,0]$ and decreasing on $[0, u)$, then by setting the slope as  $\kappa^\prime$, we can always guarantee that $\kappa^\prime$ is the minimum derivative for any point on $g(x)$ with $x\in[l,u]$, hence for any point $x_c\in[l,u]$, $\kappa^\prime (x-x_c)+g(x_c)$ always lies above of $g(x)$ on $[l, x_c]$. Hence, when $x_c=u$, we have $\kappa^\prime (x-u)+g(u)$ always lies above of $g(x)$ on $[l,u]$.
\end{itemize}}

\major{\bf We first prove the theorem when $g(x)=\text{Sigmoid}(x)$ by construction.} 

\major{Assuming that $a^\ge_{w_u}$ is the upper boundary of $w_u\cdot \text{Sigmoid}(x)$, then we have $a^\ge_{w_u}(x)\ge w_u\cdot \text{Sigmoid}(x) \ge w\cdot \text{Sigmoid}(x)$  with $w\in[w_l,w_u] \wedge x\in[l,u]$. Similarly, assuming $a^\le_{w_l}$ is the lower boundary of $w_l\cdot \text{Sigmoid}(x)$, then we have $a^\le_{w_l}(x)\le w_l\cdot \text{Sigmoid}(x) \le w\cdot \text{Sigmoid}(x)$  with $w\in[w_l,w_u] \wedge x\in[l,u]$. Hence, $a^\le_{w_l}$ and $a^\ge_{w_u}$ are sound lower and upper boundaries, i.e., $\tilde{a}^\le$ and $\tilde{a}^\ge$, of abstract domain $w\cdot \text{Sigmoid}(x)$ with $w\in[w_l,w_u]$ on $x\in[l,u]$, respectively. 
By following the above abstract element construction idea from~\cite{SGPV19}, we obtain $\tilde{a}^\le = a^\le_{w_l}$ and $\tilde{a}^\ge = a^\ge_{w_u}$ as follows:}
\begin{figure}[h]
    \centering
    \subfigure[$0 \le w_l \le w_u$.]{\label{fig:sig_pos}
        \begin{minipage}[b]{0.3\textwidth}
            \includegraphics[width=1.0\textwidth]{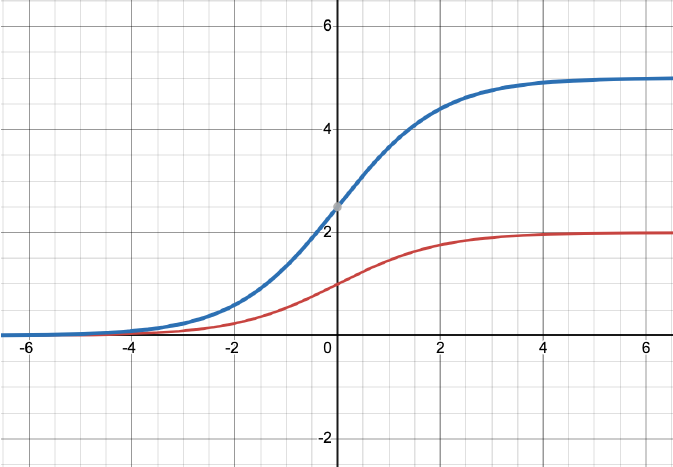}
        \end{minipage}	
    }\hspace{8mm}
        \subfigure[$w_l \le w_u \le 0$.]{\label{fig:sig_neg}
        \begin{minipage}[b]{0.3\textwidth}
            \includegraphics[width=1.0\textwidth]{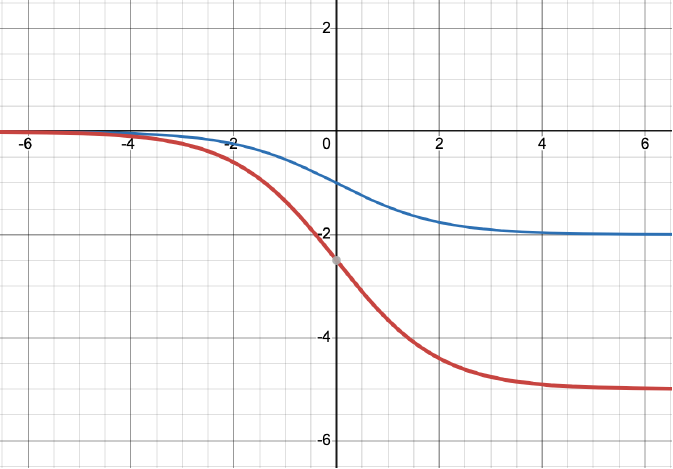}
        \end{minipage}	
    }
    \caption{$w\cdot \text{Sigmoid}(x)$ with $w\in\{w_l,w_u\}$, $w_l$ in red and $w_u$ in blue.}
    \Description{Weighted sigmoid function.}
    \label{fig:weight_sigmoid}
\end{figure}
    
\major{\begin{itemize}
        \item When $w_l\ge0$ (cf. Figure~\ref{fig:sig_pos}), we have $\tilde{l}=w_l g(l)$, $\tilde{u}=w_u g(u)$, and 
        \begin{itemize}
            \item If $l \ge 0$, $a^\le_{w_l}$ is given by the line segment defined by two points $(l, w_l g(l))$ and $(u, w_lg(u))$, i.e., $a^\le_{w_l}=w_l g(l) +\frac{w_l g(u)-w_l g(l)}{u-l}(x-l)=w_lg(l)+w_l\kappa(x-l)$; 
            Otherwise, $a^\le_{w_l}$ is given by the function defined by the point $(l, w_l g(l))$ and the minimum slope of $w_l g(x)$ on $[l,u]$, i.e., $a^\le_{w_l}= w_l g(l) + \text{min}((w_l g(x))^\prime\mid_{x=u}, (w_l g(x))^\prime\mid_{x=l})(x-l)=w_lg(l)+w_l\text{min}(g^\prime(u),g^\prime(l))(x-l) = w_lg(l)+w_l\kappa^\prime (x-l)$; 
            \item If $u \le 0$, $a^\ge_{w_u}$ is given by the line segment defined by two points $(l, w_u g(l))$ and $(u, w_u g(u))$, i.e., $a^\ge_{w_u}= w_u g(u) + \frac{w_u g(u)-w_u g(l)}{u-l}(x-u)=w_ug(u)+w_u\kappa(x-u)$; 
            Otherwise, $a^\ge_{w_u}$ is given by the function defined by the point $(u, w_u g(u))$ and the minimum slope of $w_u g(x)$ on $[l,u]$, i.e., $a^\ge_{w_u}= w_u g(u) + \text{min}((w_u g(x))^\prime\mid_{x=u}, (w_u g(x))^\prime\mid_{x=l})(x-u)=w_ug(u)+w_u\text{min}(g^\prime(u),g^\prime(l))(x-u) = w_ug(u)+w_u\kappa^\prime (x-u)$;
        \end{itemize}
        \item When $w_u\le 0$ (cf. Figure~\ref{fig:sig_neg}), we have $\tilde{l}= w_l g(u)$, $\tilde{u}=w_u g(l)$, and
        \begin{itemize}
            \item If $u\le 0$, $a^\le_{w_l}$ is given by the line segment defined by two points $(l,w_lg(l))$ and $(u,w_lg(u))$, i.e., $a^\le_{w_l}= w_l g(u) + \frac{w_l g(l)-w_l g(l)}{u-l}(x-u)=w_lg(u)+w_l\kappa(x-u)$;
            Otherwise, $a^\le_{w_l}$ is given by the function defined by the point $(u,w_l g(u))$ and the maximum slope of $w_l g(x)$ on $[l,u]$, i.e., $a^\le_{w_l}= w_l g(u) + \text{max}((w_l g(x))^\prime\mid_{x=u}, (w_u g(x))^\prime\mid_{x=l})(x-u)=w_lg(u)+w_l\text{min}(g^\prime(u),g^\prime(l))(x-u) = w_lg(u)+w_l\kappa^\prime (x-u)$; 
            \item If $l \ge 0$, $a^\ge_{w_u}$ is given by the line segment defined by two points $(l,w_ug(l))$ and $(u,w_ug(u))$, i.e., $a^\ge_{w_u}= w_u g(l) + \frac{w_u g(u)-w_u g(l)}{u-l}(x-l)=w_ug(l)+w_u\kappa(x-l)$; 
            Otherwise, $a^\ge_{w_u}$ is given by the function defined by the point $(l,w_u g(l))$ and the maximum slope of $w_u g(x)$ on $[l,u]$, i.e., $a^\ge_{w_u}= w_u g(l) + \text{max}((w_u g(x))^\prime\mid_{x=u}, (w_u g(x))^\prime\mid_{x=l})(x-l)=w_ug(l)+w_u\text{min}(g^\prime(u),g^\prime(l))(x-l) = w_ug(l)+w_u\kappa^\prime (x-l)$;
        \end{itemize}
    \end{itemize}}
    
\major{Finally, given the weighted Sigmoid function $\overrightarrow{\bs{W}}^{i+1}_{j,k}\cdot \text{Sigmoid}(\bs{x}^i_{k,1})$, we can construct and obtain its sound abstract element presented in Table~\ref{tab:log_AbT}. }

\major{\bf We next prove the theorem when $g(x)=\text{Tanh}(x)$ by construction.}
\major{Note that the sign of the value changes when crossing $x=0$ in the Tanh function. To ensure soundness, we construct the abstract element based on different values of $x$ and $w$ directly as follows:}
     
\begin{figure}[h]
    \centering
    \subfigure[$0 \le w_l \le w_u$.]{\label{fig:tanh_pos}
        \begin{minipage}[b]{0.3\textwidth}
            \includegraphics[width=1.0\textwidth]{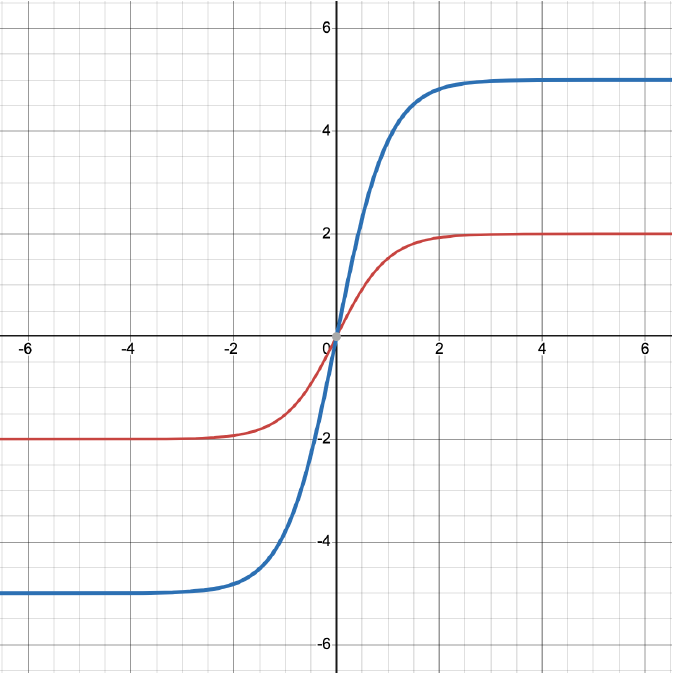}
        \end{minipage}	
    }\hspace{8mm}
        \subfigure[$w_l \le w_u \le 0$.]{\label{fig:tanh_neg}
        \begin{minipage}[b]{0.3\textwidth}
            \includegraphics[width=1.0\textwidth]{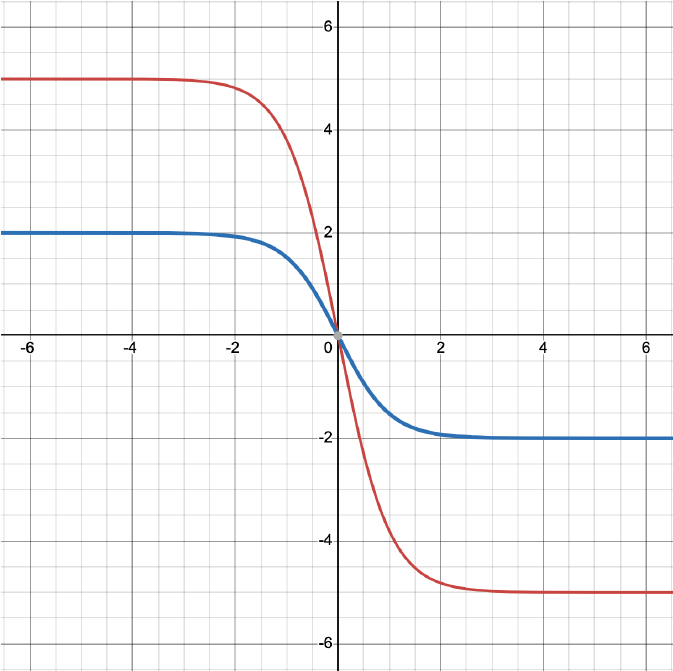}
        \end{minipage}	
    }
        
    \caption{$w\cdot \text{Tanh}(x)$ with $w\in\{w_l,w_u\}$, $w_l$ in red and $w_u$ in blue.}
    \Description{Weighted tanh function.}
    \label{fig:weight_sigmoid}
\end{figure}

\major{\begin{itemize}
    \item When $w_l\ge0$ (cf. Figure~\ref{fig:tanh_pos}):
    \begin{itemize}
        \item If $l\ge 0$, then $\tilde{l}=w_l g(l)$, $\tilde{u}=w_u g(u)$, $\tilde{a}^\le=a^\le_{w_l}$, and $\tilde{a}^\ge = a^\ge_{w_u}$, where $a^\le_{w_l} = w_l g(l) + w_l \kappa (x-l)$ and $a^\ge_{w_u} = w_u g(u) + w_u \kappa^\prime (x-u)$;
        \item If $u\le 0$, then $\tilde{l}=w_u g(l)$, $\tilde{u}=w_l g(u)$, $\tilde{a}^\le = a^\le_{w_u}$, and $\tilde{a}^\ge = a^\ge_{w_l}$, where $a^\le_{w_u} = w_u g(l) + w_u \kappa^\prime (x-l)$ and $a^\ge_{w_l}= w_l g(u) + w_l \kappa (x-u)$;
        \item If $l<0<u$, then $\tilde{l}= w_u g(l)$, $\tilde{u}=w_u g(u)$. $\tilde{a}^\le$ is given by the point $(l, w_u g(l))$ and the minimum slope of all slopes of $w_l g(x)$ and $w_u g(x)$ on $[l,u]$, i.e., $\tilde{a}^\le = w_u g(l) + \text{min} ( w_u g^\prime(x)\mid_{x=u}, w_u g^\prime(x)\mid_{x=l}, w_l g^\prime(x)\mid_{x=l}, w_l g^\prime(x)\mid_{x=u}) (x-l) = w_u g(l) + \text{min}(w_l g^\prime(x)\mid_{x=u}, w_l g^\prime(x)\mid_{x=l})(x-l) = w_u g(l)+ w_l\kappa^\prime (x-l)$; Similarly, $\tilde{a}^\ge$ is given by the point $(u, w_u g(u))$ and the same minimum slope, i.e., $\tilde{a}^\ge = w_u g(u) + w_l \kappa^\prime (x-u)$.
    \end{itemize}    
    \item When $w_u \le 0$ (cf. Figure~\ref{fig:tanh_neg}):
    \begin{itemize}
        \item If $l\ge 0$, then $\tilde{l}=w_l g(u)$, $\tilde{u}=w_u g(l)$, $\tilde{a}^\le = a^\le_{w_l}$, and $\tilde{a}^\ge=a^\ge_{w_u}$. $a^\le_{w_l}$ is given by function defined by the point $(u, w_l g(u))$ and the maximum slope $\text{max}(w_l g^\prime(l), w_l g^\prime(u))= w_l \kappa^\prime$, i.e., $\tilde{a}^\le= w_l g(u) + w_l \kappa^\prime (x-u)$. $a^\ge_{w_u}$ is given by line segment defined by two points $(l, w_u g(l))$ and $(u, w_u g(u))$, i.e., $\tilde{a}^\ge = w_u g(l) + w_u \kappa (x-l)$;
        \item If $u\le 0$, then $\tilde{l}= w_u g(u)$, $\tilde{u}=w_l g(l)$, $\tilde{a}^\le = a^\le_{w_u}$, and $\tilde{a}^\ge = a^\ge_{w_l}$. $a^\le_{w_u}$ is given by line segment by two points $(l, w_u g(l))$ and $(u, w_u g(u))$, i.e., $\tilde{a}^\le = w_u g(u) + w_u \kappa (x-u)$. $a^\ge_{w_l}$ is given by function defined by the point $(l, w_l g(l))$ and the maximum slope of $w_l g(x)$ on $[l,u]$, i.e., $\tilde{a}^\ge = w_l g(l) + w_l \kappa^\prime (x-l)$;
        \item If $l < 0 < u$, then $\tilde{l}= w_l g(u)$, $\tilde{u}=w_l g(l)$. $\tilde{a}^\le$ is given by the point $u, w_l g(u)$ and the maximum slope of all slopes of $w_l g(x)$ and $w_u g(x)$ on $[l,u]$, i.e., $\tilde{a}^\le = w_l g(u) +  \text{max}(w_l g^\prime(x)\mid_{x=u}, w_l g^\prime(x)\mid_{x=l}, w_u g^\prime(x)\mid_{x=u}, w_u g^\prime(x)\mid_{x=l}) (x-u) = w_l g(u) + \text{max}(w_u g^\prime(x)\mid_{x=u}, w_u g^\prime(x)\mid_{x=l}) (x-u)$ $ = w_l g(u) + w_u \kappa^\prime (x-u)$; Similarly, $\tilde{a}^\ge$ is given by the point $(l, w_l g(l))$ and the same maximum slope, i.e., $\tilde{a}^\ge = w_l g(l) + w_u \kappa^\prime(x-l)$.
    \end{itemize}
\end{itemize}}

\major{Finally, given the weighted Tanh function $\overrightarrow{\bs{W}}^{i+1}_{j,k}\cdot \text{Tanh}(\bs{x}^i_{k,1})$, we can construct and obtain its sound abstract element presented in Table~\ref{tab:log_AbT}.}
\end{proof}

\section{\major{Illustration of Interval Partition Effectiveness}}\label{sec:app_ip_eff}
\begin{figure}[h]
        \centering
        \subfigure[Abstraction concerning ${[w_l,w_u]}$.]{\label{fig:affine_0}
            \begin{minipage}[b]{0.15\textwidth}
                \includegraphics[width=1.0\textwidth]{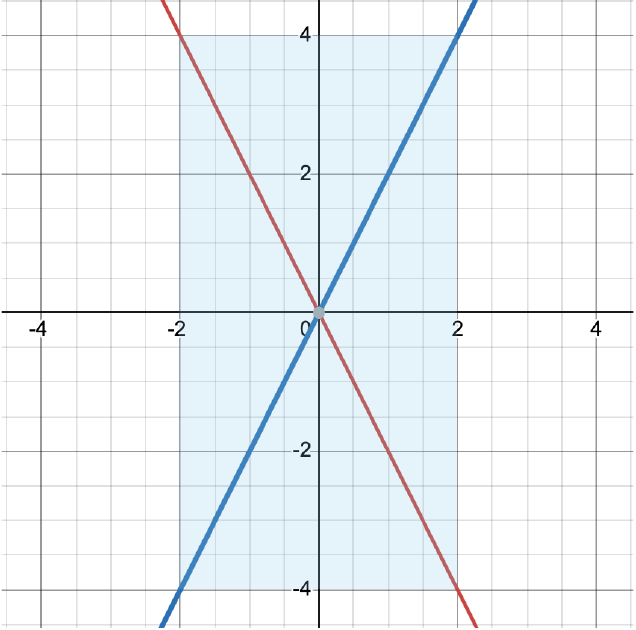}
            \end{minipage}	
        } \hspace{2mm}
        \subfigure[Abstraction concerning ${[w_l,0]}$.]{\label{fig:affine_0_1}
            \begin{minipage}[b]{0.15\textwidth}
                \includegraphics[width=1.0\textwidth]{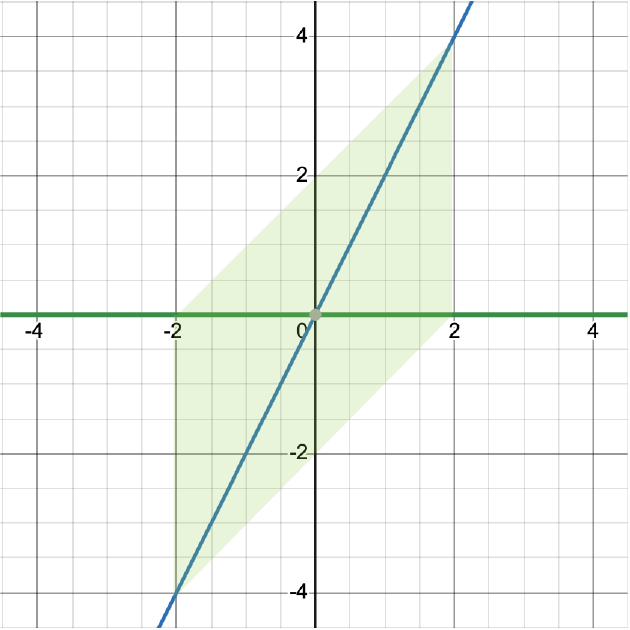}
            \end{minipage}	
        }\hspace{2mm}
        \subfigure[Abstraction concerning ${[0,w_u]}$.]{\label{fig:affine_0_2}
            \begin{minipage}[b]{0.15\textwidth}
                \includegraphics[width=1.0\textwidth]{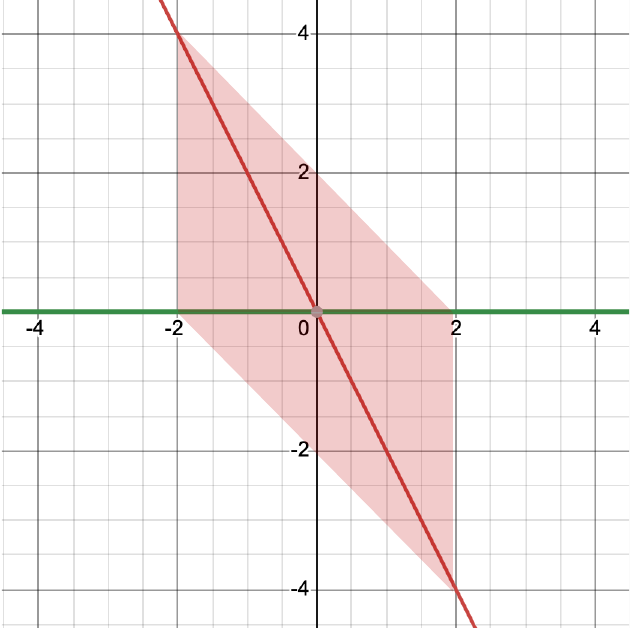}
            \end{minipage}	
        }\hspace{2mm}
        \subfigure[The union of \ref{fig:affine_0_1} and \ref{fig:affine_0_2}.]{\label{fig:affine_1}
            \begin{minipage}[b]{0.15\textwidth}
                \includegraphics[width=1.0\textwidth]{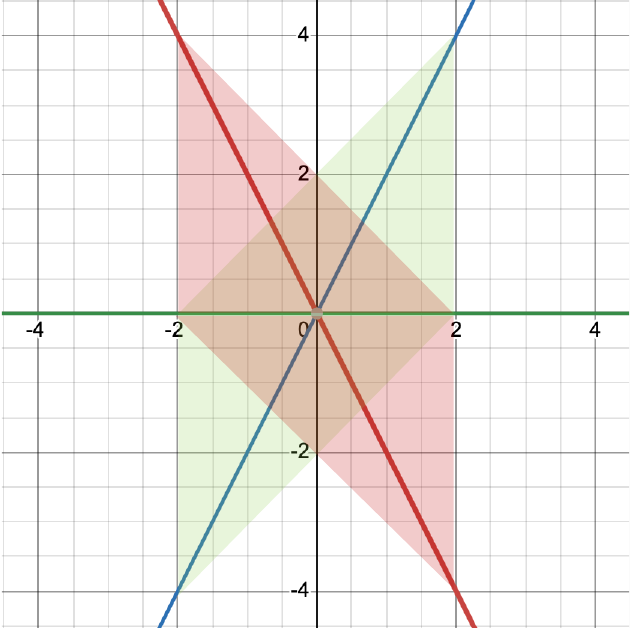}
            \end{minipage}	
        }\hspace{2mm}
        \subfigure[The abstraction ground truth.]{\label{fig:affine_2}
            \begin{minipage}[b]{0.15\textwidth}
                \includegraphics[width=1.0\textwidth]{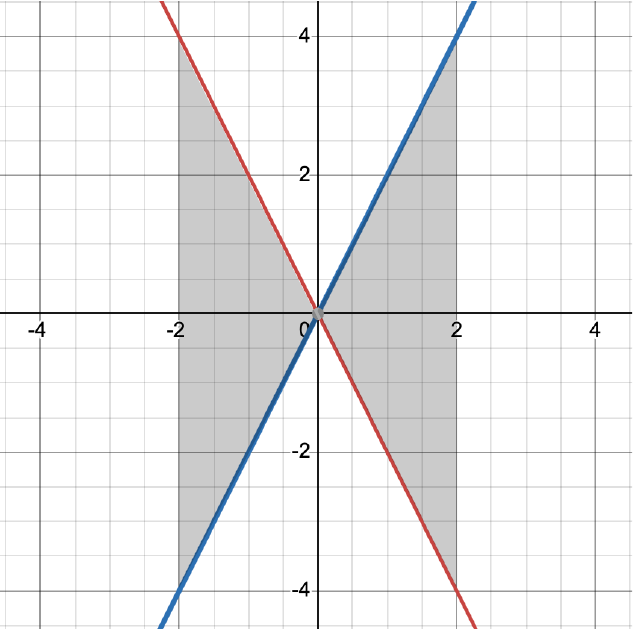}
            \end{minipage}	
        }
        \caption{\major{An illustration example explaining why interval partitions can enhance the abstraction precision when $w_l < 0 < w_u$ for the weighted input neuron $\overrightarrow{\bs{W}}^2_{j,k}\bs{x}^i_k$ with $\overrightarrow{\bs{W}}^2_{j,k}\in[w_l,w_u]$.}}
        \Description{Weighted tanh function.}
        \label{fig:aff_illu}
\end{figure}

\major{This section presents an illustrative example (see Figure~\ref{fig:aff_illu}) to explain how the interval partition, as part of the binary search strategy, can enhance abstraction precision, albeit to a limited extent. Specifically, the blue-shaded region in Figure~\ref{fig:affine_0} represents the value domain of the abstract element when abstraction is directly applied considering $w_l\le w \le w_u$ with $w_l <0<w_u$. In contrast, the green and red regions, shown in Figures~\ref{fig:affine_0_1} and \ref{fig:affine_0_2}, respectively, illustrate the value domains of the abstract elements obtained by separately applying abstraction concerning $w_l\le w \le 0$ and $0\le w\le w_u$. 
By combining the attraction results after interval partition, we can find that the final union of value domains, as depicted in Figure~\ref{fig:affine_1}, is more precise (i.e., smaller) compared to that in Figure~\ref{fig:affine_0}. This demonstrates an improvement in abstraction precision achieved through the interval partition strategy. The ground truth of the abstraction in Figure~\ref{fig:affine_2}.}

\section{Additional Experimental Results}\label{sec:app_exp}
\major{This section presents the details of the experimental results by \tool without Binary Search in {\sf BFA\_RA}, which are omitted in Section~\ref{sec:exp_tool_small} and Section~\ref{sec:exp_tool_large}.}

\begin{table}[h]
    \centering
    \caption{\major{Verification results of \tool without Binary Search in BFA\_RA on ACAS Xu.}}\label{tab:tool_acas_wo_bs}
    \scalebox{0.9}{
    \begin{tabular}{c|c|c|c|c|c|c|c}
    \toprule
      &  \multicolumn{2}{c|}{{\sf BFA\_RA}} & \multicolumn{2}{c|}{{\sf BFA\_MILP}} & \multicolumn{2}{c|}{AvgTime(s)} & \\ \cline{2-7}
     \multirow{-2}*{Property} & \#Safe\_Paras & \#Proved & \#Proved & \#Falsified & {\sf BFA\_RA} & {\sf BFA\_MILP}& \multirow{-2}*{\#TO} \\ \midrule \rowcolor{gray!20}

     Prop\_3\_WL & 99.7\% & 0 & 0 & 24 & 353.4 & 34.6 & 0\\ 
     Prop\_3\_WR & 99.8\% & 0 & 1 & 26 & 355.7 & 8.4 & 0\\ \rowcolor{gray!20}
     Prop\_3\_SL & 99.8\% & 0 & 0 & 24 & 356.5 & 15.3 & 0 \\
     Prop\_3\_SR & 99.5\% & 0 & 0 & 27 & 357.3 & 176.3 & 0\\ \rowcolor{gray!20}
     Prop\_5\_SR & 97.4\% & 0 & 0 & 18 & 379.0 & 376.3 & 3\\
     Prop\_10\_COC & 99.3\% & 9 & 1 & 10 & 361.7 & 84.7 & 22\\
     
    \bottomrule
    \end{tabular}}
\end{table}

\begin{table}[h]
    \centering
    \caption{\major{Verification results of \tool without Binary Search in BFA\_RA on MNIST for small networks when $(Q,r,\nn)=(8,0,1)$.}}\label{tab:tool_mnist_small}
    \setlength{\tabcolsep}{3pt}
    \scalebox{0.85}{
    \begin{tabular}{c|c|c|c|c|c|c|c}
    \toprule
      &  \multicolumn{2}{c|}{{\sf BFA\_RA}} & \multicolumn{2}{c|}{{\sf BFA\_MILP}} & \multicolumn{2}{c|}{AvgTime(s)} & \\ \cline{2-7}
     \multirow{-2}*{Network} & \#Safe\_Paras & \#Proved & \#Proved & \#Falsified & {\sf BFA\_RA} & {\sf BFA\_MILP} & \multirow{-2}*{\#TO} \\ \midrule \rowcolor{gray!20}

    3blk\_10 & 99.7\% & 0 & 0 & 20 & 30.8 & 0.2 & 0  \\ 
    3blk\_30 & 99.9\% & 11 & 0 & 9 & 103.4 & 0.4 & 0 \\ \rowcolor{gray!20}
    3blk\_50 & 99.9\% & 19 & 1 & 0 & 204.7 & 0.6 & 0 \\
    5blk\_10 & 99.2\% & 0 & 0 & 20 & 47.1 & 0.4 & 0 \\ \rowcolor{gray!20}
    5blk\_30 & 99.9\% & 17 & 3 & 0 & 171.4 & 0.4 & 0 \\
    5blk\_50 & 99.9\% & 0 & 0 & 20 & 349.5 & 0.8 & 0 \\
     
    \bottomrule
    \end{tabular}}
\end{table} 

\begin{table}[h]
    \centering
    \caption{\major{Detailed verification results of \tool without Binary Search in {\sf BFA\_RA} on 3blk\_100 and 5blk\_100 with $Q\in\{4,8\}$.}}\label{tab:tool_large_2}
    \setlength{\tabcolsep}{3pt}
    \scalebox{0.85}{
    \begin{tabular}{c|c|c|c|c|c|c|c|c|c}
    \toprule
     &  &  & \multicolumn{2}{c|}{{\sf BFA\_RA}} & \multicolumn{2}{c|}{{\sf BFA\_MILP}} & \multicolumn{2}{c|}{AvgTime(s)} & \\ \cline{4-9}
     \multirow{-2}*{ } & \multirow{-2}*{r} & \multirow{-2}*{$\nn$ } & \#Safe\_Paras & \#Proved & \#Proved & \#Falsified & {\sf BFA\_RA} & {\sf BFA\_MILP} & \multirow{-2}*{\#TO} \\ \midrule 

     &   & 1 & 100.0\% & 40 & 0 & 0 & 727.1 & 0 & 0\\\rowcolor{gray!20}
     \cellcolor{white} & 0 & 2 & 100.0\% & 40 & 0 & 0 & 747.7 & 0 & 0\\ 
     &   & 4 & 100.0\% & 40 & 0 & 0 & 760.8 & 0 & 0\\ \cline{2-10}

     &   & 1 & 100.0\% & 40 & 0 & 0 & 2025.1 & 0 & 0\\\rowcolor{gray!20}
     \cellcolor{white} $Q=4$ & 2 & 2 & 99.9\% & 38 & 1 & 1 & 2058.7 & 6.5 & 0\\ 
     &   & 4 & 99.9\% & 38 & 1 & 1 & 2034.2 & 6.8 & 0\\ \cline{2-10}

     &   & 1 & 99.8\% & 29 & 3 & 0 & 2037.0 & 1038.9 & 8 \\ \rowcolor{gray!20}
     \cellcolor{white}  & 4 & 2 & 99.7\% & 19 & 3 & 3 & 2044.2 & 694.5 & 15 \\ 
     &   & 4 & 99.8\% & 19 & 2 & 2 & 2032.4 & 103.0 & 17 \\ 

    \bottomrule
    \bottomrule

    & &  & \multicolumn{2}{c|}{{\sf BFA\_RA}} & \multicolumn{2}{c|}{{\sf BFA\_MILP}} & \multicolumn{2}{c|}{AvgTime(s)} & \\ \cline{4-9}
     \multirow{-2}*{} &  \multirow{-2}*{r} & \multirow{-2}*{$\nn$ } & \#Safe\_Paras & \#Proved & \#Proved & \#Falsified & {\sf BFA\_RA} & {\sf BFA\_MILP} & \multirow{-2}*{\#TO} \\ \midrule

      &  & 1 & 100.0\% & 40 & 0 & 0 & 717.6 & 0 & 0\\ \rowcolor{gray!20}
    \cellcolor{white}  & 0 & 2 & 100.0\% & 40 & 0 & 0 & 738.6 & 0 & 0\\ 
     &   & 4 & 99.9\% & 39 & 0 & 1 & 755.5 & 1.7 & 0\\ \cline{2-10}

     &   & 1 & 99.9\% & 38 & 1 & 1 & 2036.7 & 2.7 & 0\\ \rowcolor{gray!20}
     \cellcolor{white} $Q=8$ & 2 & 2 & 99.9\% & 37 & 1 & 2 & 2029.5 & 8.1 & 0\\ 
     &   & 4 & 99.9\% & 36 & 0 & 3 & 2027.9 & 4.6 & 1 \\ \cline{2-10}

     &   & 1 & 99.5\% & 22 & 3 & 1 & 2046.6 & 1419.0 & 14 \\ \rowcolor{gray!20}
     \cellcolor{white} & 4 & 2 & 99.3\% & 18 & 3 & 1 & 2036.1 & 858.4 & 18 \\ 
     &   & 4 & 99.4\% & 18 & 1 & 3 & 2067.3 & 295.6 & 18 \\
    
    \bottomrule
    \end{tabular}}
\end{table}

\end{document}